 \documentclass[aps,prb,preprint,groupedaddress,amsmath,amssymb]{revtex4}

\usepackage{graphicx}
\usepackage{amsmath}
\usepackage{dcolumn}
\usepackage{bm}

\newcommand{\bff}{\mathbf{f}}

\newcommand{\im}{\mathrm{i}}

\newcommand{\ra}{\rangle}
\newcommand{\la}{\langle}

\newcommand{\vepsilon}{\bm \epsilon}
\newcommand{\vvarepsilon}{\bm \varepsilon}
\newcommand{\vxi}{\bm \xi}
\newcommand{\vx}{\bm x}
\newcommand{\vy}{\bm y}
\newcommand{\vz}{\bm z}

\newcommand{\vf}{\bm f}

\newcommand{\vu}{\bm u}
\newcommand{\vv}{\bm v}
\newcommand{\vw}{\bm w}

\newcommand{\ii}{ \mathrm{i}}
\newcommand{\jj}{ \mathrm{j}}

\begin{document}
\title{Linear Algebra for Mueller Calculus}
\author{A. Aiello}
\author{J.P. Woerdman}
\affiliation{Huygens Laboratory, Leiden University\\
P.O.\ Box 9504, 2300 RA Leiden, The Netherlands}
\begin{abstract}
We give a self-contained exposition of some mathematical aspects
of the Mueller-Stokes formalism. In the first part we review some
basic notions of linear algebra and establish a proper notation.
In the second part we introduce the Mueller-Stokes formalism and
derive some useful mathematical relation between physical
quantities. Finally a useful decomposition theorem is reviewed.
\end{abstract}
 \maketitle
\section{Introduction and notation}
In these notes we  have collected some mathematical results that
are not easy  to find in the literature. We assume that the reader
is already knowledgeable about the Mueller-Stokes formalism. All
the results presented here can be found in the following
references:
\begin{small}
\begin{description}
  \item[[ 1]] R. W. Schmieder, ``Stokes-Algebra Formalism'',
  J. Opt. Soc. Am. {\bf 59}, 297-302 (1969).
  \item[[ 2]] S. R. Cloude, ``Group theory and polarisation
  algebra'',
  Optik {\bf 75}, 26-36 (1986).
    \item[[ 3]] K. Kim, L. Mandel, and E. Wolf,
   ``Relationship between Jones and Mueller matrices for
   random media'', J. Opt. Soc. A {\bf 4}, 433-437 (1987).
  \item[[ 4]] S. R. Cloude, ``Conditions for the physical realisability of matrix
  operators in polarimetry'', in \emph{Polarization Considerations
   for Optical Systems II}, R. A. Chipman ed., Proc. Soc.
   Photo-Opt. Instrum. Eng. {\bf 1166}, 177-185 (1989).
  \item[[ 5]] S. R. Cloude, ``Lie Groups in Electromagnetic Wave
   Propagation and Scattering'', Journal of Electromagnetic Waves
   and Applications {\bf 6}, 947-974 (1992).
    \item[[ 6]] D. G. M. Anderson and R. Barakat,
   ``Necessary and sufficient conditions for a
   Mueller matrix to be derivable from a Jones
   matrix'', J. Opt. Soc. A {\bf 11}, 2305-2319 (1994).
\end{description}
\end{small}
Different authors use different notations which makes difficult to
recognize the same result appearing on different papers. For this
reason we have tried to simplify and unify the notation by
adopting one which seems (at least to us) to be the closest to the
physics (especially the quantum  physics) of the problem. For
example,
 we have not adopted the awkward ``optical'' notation for
the Pauli matrices
\begin{small}
\begin{equation}\label{i1}
\sigma_1 = \begin{pmatrix}
  1 & 0 \\
  0 & -1
\end{pmatrix},
\qquad \sigma_2 = \begin{pmatrix}
  0 & 1 \\
  1 & 0
\end{pmatrix},
\qquad \sigma_3 = \begin{pmatrix}
  0 & -\im \\
  \im & 0
\end{pmatrix},
\end{equation}
\end{small}
but we have adopted the standard ``quantum'' notation
\begin{small}
\begin{equation}\label{i2}
\sigma_1 =\begin{pmatrix}
  0 & 1 \\
  1 & 0
\end{pmatrix} ,
\qquad \sigma_2 = \begin{pmatrix}
  0 & -\im \\
  \im & 0
\end{pmatrix}, \qquad \sigma_3 = \begin{pmatrix}
  1 & 0 \\
  0 & -1
\end{pmatrix}.
\end{equation}
\end{small}
Of course, as a consequence  of this choice, also the Stokes
parameters defined in these notes are different from the standard
``optical'' one. If with $\mathbf{E} = X \vx + Y \vy$ we denote
the electric field of an homogeneous plane wave propagating along
the axis $\vz$, then \emph{our} Stokes parameters $\{ S_0, S_1,
S_2, S_3 \}$ are defined as
\begin{equation}\label{i6}
\begin{array}{lcccccccc}
  S_0 & = & |X|^2 + |Y|^2 & = & I & = & S_0^\mathrm{BW} & = & S_0^\mathrm{H} ,\\
  S_1 & = &  X Y^* + X^*Y  & = & U & = &  S_2^\mathrm{BW} & = &  S_2^\mathrm{H},\\
  S_2 & = &  \im( X Y^* - X^*Y ) &  = &  -V & = & -S_3^\mathrm{BW}& = & -S_3^\mathrm{H} ,\\
  S_3 & = & |X|^2 - |Y|^2 & = & Q & = & S_1^\mathrm{BW}& = & - S_1^\mathrm{H},
\end{array}
\end{equation}
where the last three columns display the traditional ($\{I,Q,U,V
\}$), the ``Born-Wolf''\footnotemark \footnotetext[1]{M. Born, E.
Wolf, \textit{Principles of optics}, 7th ed., (Cambridge
University Press, Cambridge, 1999).} ($\{
S_0^\mathrm{BW},S_1^\mathrm{BW},S_2^\mathrm{BW},S_3^\mathrm{BW}
\}$), and the ``van de Hulst''\footnotemark \footnotetext[2]{H. C.
van de Hulst, \textit{Light Scattering by Small Particles}, (Dover
Publications, Inc., New York, 1981).} ($\{
S_0^\mathrm{H},S_1^\mathrm{H},S_2^\mathrm{H},S_3^\mathrm{H} \}$)
definitions of the Stokes parameters, respectively.
 It is curious to notice that this change of
notation was already suggested in the sixties [1] but it was
unadopted. The last three Stokes parameters form a Cartesian
coordinate system on the Poincar\'e sphere (Fig. 1). This change
of notation for the Stokes parameters also causes a change in the
definition of the Mueller matrix. For example, if we write the
Stokes \emph{vectors} in ours and in the ``van de Hulst'' notation
as
\begin{equation}\label{i7}
\vec{S} = \begin{pmatrix}
  S_{0} \\
 S _{1} \\
 S _{2} \\
 S _{3}
\end{pmatrix}, \qquad \vec{S}^\mathrm{H} = \begin{pmatrix}
  S_{0}^\mathrm{H} \\
 S _{1}^\mathrm{H} \\
 S _{2}^\mathrm{H} \\
 S _{3}^\mathrm{H}
\end{pmatrix},
\end{equation}
then, from Eq. (\ref{i6}) it is easy to see that $\vec{S}$ and
$\vec{S}^\mathrm{H}$ are related by a \emph{unitary} matrix $Q$,
\begin{equation}\label{i8}
\vec{S} = Q \vec{S}^\mathrm{H} ,
\end{equation}
where
\begin{equation}\label{i9}
Q = \begin{pmatrix}
  1 & 0 & 0 & 0 \\
  0 & 0 & 1 & 0 \\
  0 & 0 & 0 & -1 \\
  0 & -1 & 0 & 0
\end{pmatrix}.
\end{equation}
Now, let us consider a linear optical process described in the two
different notations as
\begin{equation}\label{i10}
\vec{S}_\mathrm{out} = M \vec{S}_\mathrm{in}, \qquad
\vec{S}_\mathrm{out}^\mathrm{H} = M^\mathrm{H}
\vec{S}_\mathrm{in}^\mathrm{H}.
\end{equation}
Then, it is easy to calculate
\begin{equation}\label{i11}
\begin{array}{rcl}
\vec{S}_\mathrm{out} = Q \vec{S}_\mathrm{out}^\mathrm{H} & = & Q
M^\mathrm{H}
\vec{S}_\mathrm{in}^\mathrm{H}\\
& = & Q M^\mathrm{H} Q^{-1} Q \vec{S}_\mathrm{in}^\mathrm{H}
\\
& = & \left[ Q M^\mathrm{H} Q^{-1} \right] \vec{S}_\mathrm{in}
\\
& \equiv & M \vec{S}_\mathrm{in},
\end{array}
\end{equation}
from which it follows
\begin{equation}\label{i12}
M = Q M^\mathrm{H} Q^{-1}.
\end{equation}
This is the sought relation between our definition of Mueller
matrix, and the optical one.

 These notes aim to be, from mathematical
point of view, as self-contained as possible; all formulae are
derived, the only omitted derivations are the ones which reduce to
an explicit calculation. For example the formula $\sigma_1
\sigma_2 = \im \sigma_3$ cannot be ``demonstrated'', it must be
checked by explicit calculation from the definition of the Pauli
matrices. However, all the omitted explicit calculations can be
easily done in few seconds with a computer program like
Mathematica.

As we already said, these notes focus on the \emph{mathematical}
aspects of the Mueller formalism, so no emphasis is given to any
physical process. For this reason in the first part of this script
we almost exclusively deal with the case of deterministic (or
Mueller-Jones) Mueller matrices which requires the knowledge of
the same amount of linear algebra results as the more general
case. However, all  formulae derived here can be straightforwardly
extended to the case of non-deterministic Mueller matrices.
\subsection{Notation}
A few words about the notation. We use three different kind of
indices: Latin, Greek and Calligraphic. Latin indices $i,j,k,
\ldots$ run from $0$ to $1$ and label the components of $2 \times
2 $ matrices and $2$-D vectors. Greek indices $\mu, \nu, \alpha,
\ldots$ run from $0$ to $3$ and label the components of $4 \times
4 $ matrices and $4$-D vectors. Finally, Calligraphic indices
$\mathcal{A}, \mathcal{B}, \mathcal{C} , \ldots$ run from $0$ to
$15$ and label the components of $16$-D vectors. In  these notes
the Einstein summation convention is used extensively, that is the
sum on repeated indices (Latin, Greek and Calligraphic) is
understood. For example
\begin{small}
\begin{equation}\label{i3}
a_\mu = \Lambda_{\mu \nu} b_\nu \Leftrightarrow a_\mu = \sum_{\nu
=0 }^3 \Lambda_{\mu \nu} b_\nu.
\end{equation}
\end{small}
We often use  the direct product of two matrices $A$ and $B$,
indicated with the symbol ``$\otimes$'':
\begin{small}
\begin{equation}\label{i4}
C = A \otimes B.
\end{equation}
\end{small}
For this kind of matrix product,  the standard convention for the
indices is the following:
\begin{small}
\begin{equation}\label{i5}
c_{ik,jl} = a_{ij} b_{kl}.
\end{equation}
\end{small}
It worths to note the order of the indices $j$ and $k$ in both
sides of this equation; it will play an important role in these
notes.
\section{Matrix bases}{}
In this section we study two different ways to represent  $2
\times 2$ matrices and the relations between different
representations.
\subsection{The Standard Basis}{}
Let $A \in \mathbb{C}^{2 \times 2}$ denotes a $2 \times 2$
complex-valued matrix defined in terms of its elements $[A]_{ij}
\equiv a_{ij}, \, (i,j=0,1)$ as
\begin{equation}\label{e1}
A = \begin{pmatrix}
  a_{00} & a_{01} \\
  a_{10} & a_{11}
\end{pmatrix}.
%
\end{equation}
Any $2 \times 2$ matrix can be put in one-to-one correspondence
with a complex $4$-vector $\vec{a} \in \mathbb{C}^4$  by writing
\begin{equation}\label{e2}
A = \begin{pmatrix}
  a_{00} & a_{01} \\
  a_{10} & a_{11}
\end{pmatrix} \equiv \begin{pmatrix}
  a_{0} & a_{1} \\
  a_{2} & a_{3}
\end{pmatrix},
%
\end{equation}
where
\begin{equation}\label{e2b}
\vec{a} =  \begin{pmatrix}
  a_{00} \\ a_{01} \\
  a_{10} \\ a_{11}
\end{pmatrix} \equiv  \begin{pmatrix}
  a_{0} \\ a_{1} \\
  a_{2} \\ a_{3}
\end{pmatrix}.
%
\end{equation}
  This rule is very simple and
can be easily extended to $n \times n$ matrices by defining
\begin{equation}\label{e3}
  a_{ij} \equiv   a_{n i + j},
%
\end{equation}
for $i,j = 0, \ldots, n -1$. This rule is so important that in the
remaining part of these notes we shall refer to  as the ``Rule''.
At this point it is important to notice that when we write a
vector $\vec{a}$ as in Eq. (\ref{e2b}), we are implicitly assuming
that its components $a_\mu, \; \mu = 0, \ldots,3$ are referred
  to the so-called \emph{standard basis} in
$\mathbb{R}^4$, that is
\begin{equation}\label{e2c}
\vec{a} =   a_{0} \begin{pmatrix}
  1 \\ 0 \\
  0 \\ 0
\end{pmatrix} +a_{1} \begin{pmatrix}
  0 \\ 1 \\
  0 \\ 0
\end{pmatrix}+a_{2} \begin{pmatrix}
  0 \\ 0 \\
  1 \\ 0
\end{pmatrix}+a_{3} \begin{pmatrix}
  0 \\ 0 \\
  0 \\ 1
\end{pmatrix}.
%
\end{equation}
Analogously Eq. (\ref{e2}) naturally suggests the possibility to
write
\begin{equation}\label{e4}
\begin{array}{ccl}
A & = & a_{0} \begin{pmatrix}
  1 & 0 \\
  0 & 0
\end{pmatrix}
+ a_{1} \begin{pmatrix}
  0 & 1 \\
  0 & 0
\end{pmatrix}
+a_{2} \begin{pmatrix}
  0 & 0 \\
  1 & 0
\end{pmatrix}
+a_{3} \begin{pmatrix}
  0 & 0 \\
  0 & 1
\end{pmatrix}\\\\
& \equiv & a_{\mu} \epsilon_{(\mu)},\qquad \qquad (\mu = 0,1,2,3),
\end{array}
%
\end{equation}
where summation on repeated indices is understood and the basis
matrices $ \epsilon_{(\mu)} \in \mathbb{R}^{2 \times 2}$ are
defined as
\begin{equation}\label{e5}
\epsilon_{(0)}  \equiv  \begin{pmatrix}
  1 & 0 \\
  0 & 0
\end{pmatrix}, \qquad
\epsilon_{(1)}  \equiv   \begin{pmatrix}
  0 & 1 \\
  0 & 0
\end{pmatrix}, \qquad
\epsilon_{(2)}  \equiv   \begin{pmatrix}
  0 & 0 \\
  1 & 0
\end{pmatrix},\qquad
\epsilon_{(3)}  \equiv   \begin{pmatrix}
  0 & 0 \\
  0 & 1
\end{pmatrix}.
%
\end{equation}
Then the numbers $\{a_\mu\}$, that we have found by using the Rule
Eq.(\ref{e3}), appear to be the components of the matrix $A$ with
respect to the basis $\{\epsilon_{(\mu)}\}$. In order to
demonstrate this, it is necessary to define a norm in the vector
space $\mathbb{C}^{2 \times 2}$ of the complex  $2 \times 2$
matrices.
%
It is  possible to introduce a norm in $\mathbb{C}^{2 \times 2}$
by defining the \emph{scalar product} $\{A,B\}$  between two
matrices $A$ and $B$ as
\begin{equation}\label{e6}
\begin{array}{ccl}
 \{A,B\} & = &   \displaystyle{\mathrm{Tr}\{A^\dagger B\} }\\
 & = & \displaystyle{a_{i j}^* b_{i j} }, \qquad (i,j = 0,1),
\end{array}
%
\end{equation}
where summation on repeated indices is again understood and
$A^\dagger$ denotes the Hermitian-conjugate  of $A$; that is
$A^\dagger = (A^T)^*= (A^*)^T$, where  $A^*$ and $A^T$ are the
complex conjugate and  the transpose  of $A$ respectively.
Moreover, since $\{A,B\}^* = (a_{i j}^* b_{i j})^* = a_{i j} b_{i
j}^*= \mathrm{Tr}\{B^\dagger A\}$, the result $\{A,B\}^* =
\{B,A\}$ follows.
 By
explicit calculation, one can see that the basis vectors
$\{\epsilon_{(\mu)}\}$ are orthonormal with respect to that norm:
\begin{equation}\label{e7}
\begin{array}{ccl}
\{\epsilon_{(\mu)},\epsilon_{(\nu)}\} & = &
\mathrm{Tr}\{\epsilon_{(\mu)}^T \epsilon_{(\nu)}\}\\
& = & \delta_{\mu \nu},
\end{array}
%
\end{equation}
where $\mu, \nu = 0,\dots,3$ and $\epsilon_{(\mu)}^\dagger =
{\epsilon}_{(\mu)}^T $ follows from Eq. (\ref{e5}). Now, having
introduced the norm Eq. (\ref{e6}), it is easy to calculate the
components of the matrix $A$ with respect to the basis
$\{\epsilon_{(\mu)} \}$ as
\begin{equation}\label{e8}
\begin{array}{ccl}
\{\epsilon_{(\mu)},A\} & = &
\{\epsilon_{(\mu)}, a_\nu\epsilon_{(\nu)}\}\\
& = & a_\nu \{\epsilon_{(\mu)}, \epsilon_{(\nu)}\}\\ & = &
a_{\mu},
\end{array}
%
\end{equation}
where $\mu, \nu = 0,1,2,3$.

Then we have shown that it is possible to associate with any
matrix $A \in \mathbb{C}^{2 \times 2}$ a vector $\vec{a} \in
\mathbb{C}^{4}$ and there are two different (but equivalent) ways
to calculate $\vec{a}$: we can either use the Rule given in Eq.
(\ref{e3})
\begin{equation}\label{e9}
a_{\mu = 2i + j}  = a_{ij}, \qquad (i,j= 0,1),
%
\end{equation}
or calculate explicitly
\begin{equation}\label{e10}
\begin{array}{ccl}
a_{\mu} & = & \{\epsilon_{(\mu)}, A\}\\
 & = & \mathrm{Tr}\{\epsilon_{(\mu)}^T A\}, \qquad (\mu= 0,1,2,3).
\end{array}
%
\end{equation}

Until now, $A$ was left  arbitrary, therefore Eq. (\ref{e10})
holds for \emph{any} $2 \times 2$ matrix.
If $A$ coincides with one of the basis matrices
$\epsilon_{(\alpha)}$, then Eq. (\ref{e10}) gives the components
$[\epsilon_{(\alpha)}]_\mu$ $(\mu = 0,\dots,3)$, of the matrix
$\epsilon_{(\alpha)}$ with respect to the basis
$\{\epsilon_{(\mu)}\}$:
\begin{equation}\label{e10b}
 \begin{array}{ccl}
[\epsilon_{(\alpha)}]_\mu & = & \{\epsilon_{(\mu)}, \epsilon_{(\alpha)}\}\\
 & = & \mathrm{Tr}\{\epsilon_{(\mu)}^T \epsilon_{(\alpha)}\}\\
 & = & \delta_{\mu \alpha },
\end{array}
%
\end{equation}
where $\alpha, \mu= 0,1,2,3$. Therefore we can build the basis
$4$-vectors $\vec{e}_{(\alpha)} \in \mathbb{R}^4$ associated to
the basis matrices $\epsilon_{(\alpha)}$ as
%
%
%
%
%
\begin{equation}\label{e10c}
\vec{e}_{(\alpha)} =
\begin{pmatrix}
[\epsilon_{(\alpha)}]_{00} \\
[\epsilon_{(\alpha)}]_{01} \\
[\epsilon_{(\alpha)}]_{10} \\
[\epsilon_{(\alpha)}]_{11}
\end{pmatrix} =
\begin{pmatrix}
[\epsilon_{(\alpha)}]_0 \\
[\epsilon_{(\alpha)}]_1 \\
[\epsilon_{(\alpha)}]_2 \\
[\epsilon_{(\alpha)}]_3
\end{pmatrix} =
\begin{pmatrix}
\delta_{0 \alpha } \\
\delta_{1 \alpha } \\
\delta_{2 \alpha } \\
\delta_{3 \alpha }
\end{pmatrix}, \qquad (\alpha = 0,1,2,3).
%
\end{equation}
 It is trivial to calculate from Eq. (\ref{e10c}) that
%
%
%
%
\begin{equation}\label{e11}
\vec{e}_{(0)} =
\begin{pmatrix}
1 \\
0 \\
0 \\
0
\end{pmatrix}, \qquad
\vec{e}_{(1)} =
\begin{pmatrix}
0 \\
1 \\
0 \\
0
\end{pmatrix}, \qquad
\vec{e}_{(2)} =
\begin{pmatrix}
0 \\
0 \\
1 \\
0
\end{pmatrix}, \qquad
\vec{e}_{(3)} =
\begin{pmatrix}
0 \\
0 \\
0 \\
1
\end{pmatrix};
%
\end{equation}
that is $\{\vec{e}_{(\mu)}\}$ is simply the standard basis in
$\mathbb{R}^4$. In summary, we have shown that there is a
one-to-one correspondence  between the standard basis
$\{\epsilon_{(\mu)} \in \mathbb{R}^{2 \times 2}\}$ and the
standard basis $\{\vec{e}_{(\mu)} \in \mathbb{R}^4\}$.
\subsection{The Pauli Basis}{}
Another basis commonly used in physics  is the so called Pauli
basis constituted by the $2 \times 2$ identity matrix and the
three Pauli matrices. Here we use a normalized version of the
Pauli matrices defined as
\begin{equation}\label{e12}
\begin{array}{cclcccl}
\sigma_{(0)}  &\equiv &
\displaystyle{\frac{1}{\sqrt{2}}\begin{pmatrix}
  1 & 0 \\
  0 & 1
\end{pmatrix}}, &&
\sigma_{(1)}  &\equiv &\displaystyle{ \frac{1}{\sqrt{2}}
\begin{pmatrix}
  0 & 1 \\
  1 & 0
\end{pmatrix}},
\\\\
\sigma_{(2)}  &\equiv &\displaystyle{ \frac{1}{\sqrt{2}}
\begin{pmatrix}
  0 & -\im \\
  \im & 0
\end{pmatrix}}, &&
\sigma_{(3)}  &\equiv &\displaystyle{  \frac{1}{\sqrt{2}}
\begin{pmatrix}
  1 & 0 \\
  0 & -1
\end{pmatrix}}.
\end{array}
%
\end{equation}
An explicit calculation shows that they satisfy the following
multiplication table:

\begin{center}
\begin{tabular}{c|ccccc}
  $\sqrt{2} \sigma_{(\mu)} \sigma_{(\nu)} $ & & $ \sigma_{(0)} $ & $ \sigma_{(1)} $ & $ \sigma_{(2)} $ & $ \sigma_{(3)}$ \\
  \hline  \
  $\sigma_{(0)} $ & &$   \sigma_{(0)} $ & $    \sigma_{(1)} $ & $         \sigma_{(2)} $ & $    \sigma_{(3)}$
  \\
  $\sigma_{(1)} $ & &$   \sigma_{(1)} $ & $    \sigma_{(0)} $ & $         \im \sigma_{(3)} $ & $    -\im \sigma_{(2)}$ \\
  $\sigma_{(2)} $ & &$   \sigma_{(2)} $ & $    -\im \sigma_{(3)} $ & $    \sigma_{(0)} $ & $    \im \sigma_{(1)}$ \\
  $\sigma_{(3)} $ & &$   \sigma_{(3)} $ & $    \im \sigma_{(2)} $ & $     -\im \sigma_{(1)} $ & $    \sigma_{(0)}$
  \\
\end{tabular}
\\
%
\end{center}
Moreover, again an explicit calculation shows that these matrices
are orthonormal with respect to the norm defined in Eq.
(\ref{e6}):
\begin{equation}\label{e13}
\begin{array}{ccl}
 \{\sigma_{(\mu)},\sigma_{(\nu)}\} & = &
\mathrm{Tr}\{\sigma_{(\mu)}^\dagger \sigma_{(\nu)}\}\\
& = & \delta_{\mu \nu},
\end{array}
%
\end{equation}
where $\mu, \nu = 0,1,2,3$. The Pauli basis is complete; in order
to show this we have to calculate the components
$[\sigma_{(\mu)}]_\alpha$  of the matrices $\sigma_{(\mu)}$ with
respect to the basis $\{ \epsilon_{(\mu)}\}$ in the way we learned
in the previous subsection (see Eq. (\ref{e10}) with $A =
\sigma_{(\mu)}$):
\begin{equation}\label{e16}
\begin{array}{ccl}
[\sigma_{(\mu)}]_\alpha & = &
\{\epsilon_{(\alpha)}, \sigma_{(\mu)}\}\\
 & = &
\mathrm{Tr}\{\epsilon_{(\alpha)}^T \sigma_{(\mu)}\}\\
&\equiv & \Lambda_{\alpha \mu},
\end{array},
%
\end{equation}
where $\mu, \alpha = 0,\ldots,3$, and in the last line we have
defined the $4 \times 4$ transformation matrix $\Lambda$ in terms
of its elements $[\Lambda]_{\alpha \mu} \equiv \Lambda_{\alpha
\mu} = \mathrm{Tr}\{\epsilon_{(\alpha)}^T \sigma_{(\mu)}\}$. An
explicit calculation shows that
\begin{equation}\label{e17}
\Lambda = \frac{1}{\sqrt{2}} \begin{pmatrix}
  1 & 0 & 0 & 1 \\
  0 & 1 &  -\im & 0 \\
  0 & 1 & \im & 0 \\
  1 & 0 & 0 & -1
\end{pmatrix},
%
\end{equation}
where Eqs. (\ref{e5},\ref{e12}) have been used.
In  the previous subsection we  shown how to build the basis
vectors $\{ \vec{e}_{(\alpha)} \}$ in $\mathbb{R}^4$ associated to
the
 basis matrices $\{\epsilon_{(\alpha)}\}$ in $\mathbb{R}^{2 \times 2}$. Analogously,
  we can now build the basis
vectors $\{ \vec{s}_{(\mu)} \} $ in $ \mathbb{C}^4$ associated to
basis matrices $\{ \sigma_{(\mu)}\} $ in $ \mathbb{C}^{2 \times
2}$. To this end, for a given $\mu$ we define the four components
$[\vec{s}_{(\mu)}]_\alpha, \; (\alpha = 0,\ldots,3) $ of the
vector $\vec{s}_{(\mu)}$, as $[\vec{s}_{(\mu)}]_\alpha \equiv
[\sigma_{(\mu)}]_\alpha$, that is
%
%
%
%
%
\begin{equation}\label{e18}
\vec{s}_{(\mu)} \equiv
\begin{pmatrix}
[\sigma_{(\mu)}]_0 \\
[\sigma_{(\mu)}]_1 \\
[\sigma_{(\mu)}]_2 \\
[\sigma_{(\mu)}]_3
\end{pmatrix} =
\begin{pmatrix}
\Lambda_{0 \mu } \\
\Lambda_{1 \mu } \\
\Lambda_{2 \mu } \\
\Lambda_{3 \mu }
\end{pmatrix},
%
\end{equation}
where $\mu = 0,\ldots,3$. From Eq. (\ref{e18}) is clear that the
$\mu$-th column of the matrix $\Lambda$ is made of the components
$[\vec{s}_{(\mu)}]_\alpha$  of the $4$-vector $\vec{s}_{(\mu)}$.
Alternatively we can find the vectors $\vec{s}_{(\mu)}$ by using
the Rule: $[\vec{s}_{(\mu)}]_{\alpha = 2i+j} =
[\sigma_{(\mu)}]_{ij} $. For example, for $\mu =2$ we have
\begin{equation}\label{e18b}
\sigma_{(2)} =  \frac{1}{\sqrt{2}}
\begin{pmatrix}
  0 & -\im \\
  \im & 0
\end{pmatrix} \equiv
\begin{pmatrix}
  [\vec{s}_{(2)}]_0 & [\vec{s}_{(2)}]_1 \\
  [\vec{s}_{(2)}]_2 & [\vec{s}_{(2)}]_3
\end{pmatrix},
%
\end{equation}
from which we deduce that
\begin{equation}\label{e18c}
\vec{s}_{(2)} = \begin{pmatrix}
[\vec{s}_{(2)}]_0\\
[\vec{s}_{(2)}]_1 \\
[\vec{s}_{(2)}]_2 \\
[\vec{s}_{(2)}]_3
\end{pmatrix} =  \frac{1}{\sqrt{2}}\begin{pmatrix}
0 \\
-\im \\
\im \\
0
\end{pmatrix}.
%
\end{equation}
From Eq. (\ref{e17}) it is easy to check by explicit calculation
that $\Lambda$ is unitary, that is $\Lambda^\dagger \Lambda =
\Lambda \Lambda^\dagger = \mathrm{I}_4$, where $\mathrm{I}_4$ is
the $4 \times 4$ identity matrix. In terms of the components with
respect to the basis $\{ \epsilon_{(\mu)}\}$ the relation
$\mathrm{I}_4 = \Lambda \Lambda^\dagger$ becomes
\begin{equation}\label{e19}
\begin{array}{ccl}
 \delta_{\alpha \beta} & = &  \displaystyle{\sum_{\mu = 0}^3 \Lambda_{\alpha \mu}
\Lambda_{\mu \beta }^\dagger}\\ & = &  \displaystyle{\sum_{\mu =
0}^3 \Lambda_{\alpha \mu} {\Lambda_{\beta \mu}^*}}\\
& = &  \displaystyle{\sum_{\mu = 0}^3 [\sigma_{(\mu)}]_\alpha
[\sigma_{(\mu)}^*]_\beta}.
\end{array}
%
\end{equation}
The first and the last line of Eq. (\ref{e19}) give us the
completeness relation (also called \emph{resolution of the
identity}) we are seeking:
\begin{equation}\label{e20}
 \displaystyle{\sum_{\mu = 0}^3 [\sigma_{(\mu)}]_\alpha
[\sigma_{(\mu)}^*]_\beta} =  \delta_{\alpha \beta}.
%
\end{equation}
By using Eq. (\ref{e16}) this relation can be written in less
involved form as
\begin{equation}\label{e21}
\begin{array}{ccl}
\displaystyle{\sum_{\mu = 0}^3 [\sigma_{(\mu)}]_\alpha
[\sigma_{(\mu)}^*]_\beta}& = &  \displaystyle{\sum_{\mu = 0}^3
\{\epsilon_{(\alpha)},\sigma_{(\mu)}\}\{\sigma_{(\mu)},\epsilon_{(\beta)}\}}\\
& = & \{\epsilon_{(\alpha)},\epsilon_{(\beta)}\}\\
& = &  \delta_{\alpha \beta},
\end{array}
%
\end{equation}
which is easier to understand.
It is useful for later purposes to write the completeness relation
in terms of the ``Latin'' matrix elements $[\sigma_{(\mu)}]_{ij},
(i,j=0,1)$. To this end we first associate the four indices
$i,j,k,l = 0, 1$ to the two indices $\alpha, \beta = 0, \ldots,3$,
by using the Rule:
\begin{equation}\label{e22}
\begin{array}{ccl}
  \alpha & = & 2i+j, \\
  \beta & = & 2k+l.
\end{array}
%
\end{equation}
Then, after noticing that $\delta_{\alpha \beta} = \delta_{2i+j,
 2k+l} =\delta_{i k}\delta_{j l}$, we  rewrite Eq. (\ref{e20})
 as
\begin{equation}\label{e23}
 \displaystyle{\sum_{\mu = 0}^3 [\sigma_{(\mu)}]_{ij}
[\sigma_{(\mu)}^*]_{kl}} = \delta_{i k}\delta_{j l}.
%
\end{equation}
From the definition in Eq. (\ref{e16}), it is obvious that
$\Lambda$ is the matrix that performs the change from the Pauli
basis $\{\sigma_{(\mu)} \}$ to the standard basis
$\{\epsilon_{(\mu)}\}$:
\begin{equation}\label{e14}
\begin{array}{ccl}
\sigma_{(\mu)} & = &  \epsilon_{(\nu)} \Lambda_{\nu \mu}\\
 \epsilon_{(\nu)} & = &  \sigma_{(\mu)}\Lambda_{\mu \nu}^\dagger,
\end{array}
%
\end{equation}
where $\mu, \nu = 0,\ldots,3 $ and summation on repeated indices
is understood. Previously we learned  that to any matrix
corresponds a vector, therefore the matrix $\Lambda$ also performs
the change from the basis $\{\vec{s}_{(\mu)} \}$ to the standard
basis $\{\vec{e}_{(\mu)}\}$. In fact, since by definition
$[\vec{s}_{(\mu)}]_\alpha = [\sigma_{(\mu)}]_\alpha =
\Lambda_{\alpha \mu}$, then
\begin{equation}\label{e14b}
\vec{s}_{(\mu)} = \vec{e}_{(\alpha)}[\vec{s}_{(\mu)}]_\alpha
=\vec{e}_{(\alpha)} \Lambda_{\alpha \mu},
%
\end{equation}
where $\alpha, \mu = 0,\ldots,3 $. This relation can be written in
fully matrix form by noticing that if $U$ denotes the unitary
transformation between the two basis: $\vec{s}_{(\mu)} = U
\vec{e}_{(\mu)}$, then it follows that $U_{\alpha \mu}  \equiv
(\vec{e}_{(\alpha)},U
\vec{e}_{(\mu)})=(\vec{e}_{(\alpha)},\vec{s}_{(\mu)}) =
\Lambda_{\alpha \mu}$, where the parentheses symbol
$(\vec{u},\vec{v})$ indicates the ordinary Euclidean scalar
product in $\mathbb{C}^n$
\begin{equation}\label{e14x}
(\vec{u},\vec{v}) = \sum_{\alpha = 0}^{n-1} {u_\alpha^*} v_\alpha
.
%
\end{equation}
So we have found that $U = \Lambda$ and, therefore,
\begin{equation}\label{e14c}
\vec{s}_{(\mu)} = \Lambda \vec{e}_{(\mu)}.
%
\end{equation}
\section{The Mueller formalism}
Let us consider a doublets of stochastic variables,
\begin{equation}\label{e14d}
\mathbf{E} = \begin{pmatrix}
  E_0 \\
  E_1
\end{pmatrix},
%
\end{equation}
which transform  under the action of a \emph{deterministic}
optical device, as
\begin{equation}\label{e14e}
\mathbf{E} \rightarrow \mathbf{E} '= T \mathbf{E},
%
\end{equation}
where the $ 2 \times 2$ complex-valued transformation matrix $T$
is known as the Jones matrix representing the optical device. We
do not make any hypothesis on the nature of the matrix $T$, it can
be arbitrary. The quantities $E_0, E_1$ in Eq. (\ref{e14d}) are
complex random variables described by a given  ensemble. Starting
from $E_0, E_1$ we can build the \emph{covariance matrix} (or
polarization matrix) $J \in \mathbb{C}^{2 \times 2}$ whose
elements are defined as
\begin{equation}\label{e14f}
J_{ij}= \la E_i E_j^* \ra \qquad (i,j = 0,1),
%
\end{equation}
where $\langle \cdot \rangle$ denotes the ensemble average. Note
that this average has nothing to do with any random medium, at
this stage we are just considering two components of the
electromagnetic field as two stochastic variables. By definition,
$J$ is Hermitian and nonnegative (or, positive semidefinite), that
is $(\mathbf{x},J \mathbf{x}) \geq 0$, $\forall \mathbf{x} \in
\mathbb{C}^2 $:
\begin{equation}\label{e14g}
\begin{array}{ccl}
(\mathbf{x},J \mathbf{x}) & = & x_i^* J_{ij} x_j\\
& = & x_i^* \la E_i E_j^* \ra x_j\\
& = & \la x_i^* E_i E_j^* x_j \ra\\
& = & \la( x_i^* E_i )(x_j^* E_j)^* \ra\\
& = & \la \left| x_i^* E_i  \right|^2 \ra\\
& = & \la \left| (\mathbf{x},\mathbf{E})  \right|^2 \ra\geq 0,
\end{array}
%
\end{equation}
where $i,j = 0,1$ and summation on repeated indices is understood.
Moreover, in the third line of Eq. (\ref{e14g}) we have exploited
the fact that, by hypothesis, the vector components $x_i$ are
deterministic variables and, therefore, are not affected by the
ensemble average.

As any other $2 \times 2$ matrix, $J$ can be written in the basis
$\{\sigma_{(\mu)}\}$ as
\begin{equation}\label{e14h}
J = S_\mu \sigma_{(\mu)} \qquad (\mu = 0, \ldots,3),
%
\end{equation}
where the components $S_\mu = \mathrm{Tr}\{\sigma_{(\mu)}J\}$ of
the $4$-vector $\vec{S}$ are known as the \emph{Stokes parameters}
of the field. Explicitly
\begin{equation}\label{e14hb}
J = \frac{1}{\sqrt{2}}  \begin{pmatrix}
  S_0 + S_3 & S_1 - \im S_2 \\
  S_1 + \im S_2 & S_0 - S_3
\end{pmatrix}.
%
\end{equation}
Form the formula above we see that
\begin{equation}\label{e14hc}
\mathrm{Tr} J =  \sqrt{2} S_0,
%
\end{equation}
while from the definition Eq. (\ref{e14f}) we have
\begin{equation}\label{e14hd}
\mathrm{Tr} J =  \la |E_0|^2\ra +  \la |E_1|^2\ra \equiv I,
%
\end{equation}
where with $I$ we denoted the total intensity of the beam. By
equating  Eq. (\ref{e14hc}) with Eq. (\ref{e14hd}) we obtain our
definiton of $S_0$:
\begin{equation}\label{e14he}
S_0 =  \frac{I}{ \sqrt{2}}.
%
\end{equation}

Under the transformation $T$, the polarization matrix $J$
transform as $J \rightarrow J'$ where, by definition,
\begin{equation}\label{e14i}
\begin{array}{ccl}
J_{ij}' & = & \la E_i' {E_j'}^{*} \ra\\
& = & T_{ik}\la E_k {E_l}^{*} \ra {T_{jl}^*}\\
& = & T_{ik} J_{kl} {T_{lj}^\dagger},
\end{array}
%
\end{equation}
or, in matrix form,
\begin{equation}\label{e14j}
J' = TJT^\dagger.
%
\end{equation}
From Eq. (\ref{e14j}) is clear that the transformed coherency
matrix $J'$ is still Hermitian and nonnegative. In correspondence
to the transformation $J \rightarrow J'$, the Stokes parameters
transform as $S_\mu \rightarrow S'_\mu$ where, by definition,
\begin{equation}\label{e14k}
\begin{array}{ccl}
S'_\mu & = & \mathrm{Tr}\{\sigma_{(\mu)}J'\}\\
& = & \mathrm{Tr}\{\sigma_{(\mu)}TJT^\dagger\}\\
& = & \mathrm{Tr}\{\sigma_{(\mu)}T S_\nu \sigma_{(\nu)}T^\dagger\}\\
& = & \mathrm{Tr}\{\sigma_{(\mu)}T \sigma_{(\nu)}T^\dagger\} S_\nu \\
& \equiv & M_{\mu \nu} S_\nu,
\end{array}
%
\end{equation}
where Eq. (\ref{e14h}) has been used in the third line and we have
defined the $4 \times 4$ Mueller matrix $M$ as
\begin{equation}\label{e14l}
\begin{array}{ccl}
M_{\mu \nu} & = & \mathrm{Tr} \left\{ \sigma_{(\mu)} T
\sigma_{(\nu)} T^\dagger \right\}\\
& = & \left\{ \sigma_{(\mu)}, T \sigma_{(\nu)} T^\dagger \right\},
\end{array}
%
\end{equation}
where $\mu, \nu = 0, \ldots,3$. It is easy to see that $M$ has
real elements:
\begin{equation}\label{e14m}
\begin{array}{ccl}
{M_{\mu \nu}^*} & = & \left\{ \sigma_{(\mu)}, T \sigma_{(\nu)} T^\dagger \right\}^*      \\
               & = & \left\{ T \sigma_{(\nu)} T^\dagger,  \sigma_{(\mu)}\right\}
                                \\
               & = &\mathrm{Tr}\left\{ T \sigma_{(\nu)}T^\dagger
                                 \sigma_{(\mu)} \right\}
                                \\
               & = &\mathrm{Tr}\left\{ \sigma_{(\mu)} T \sigma_{(\nu)}T^\dagger
                                 \right\}
                                \\
               & = & M_{\mu \nu},
\end{array}
%
\end{equation}
where the cyclic property  of the trace: $\mathrm{Tr}\{A B
\}=\mathrm{Tr}\{ B A \}$ has been used. En passant we may note
that if we write the Jones matrix $T$ in the Pauli basis as $T =
c_\alpha \sigma_{(\alpha)}$, where $c_\alpha = \mathrm{Tr} \{
\sigma_{(\alpha)} T \} $, then Eq. (\ref{e14l}) can be written as
\begin{equation}\label{e14l1}
\begin{array}{ccl}
M_{\mu \nu} & = & \mathrm{Tr} \left\{ \sigma_{(\mu)} T
\sigma_{(\nu)} T^\dagger \right\}\\
& = & c_\beta c_\alpha^* \mathrm{Tr}\left\{ \sigma_{(\mu)}
\sigma_{(\beta)} \sigma_{(\nu)} \sigma_{(\alpha)} \right\}
\\
& \equiv & C_{\beta \alpha} \mathrm{Tr}\left\{\sigma_{(\alpha)}
\sigma_{(\mu)} \sigma_{(\beta)} \sigma_{(\nu)}  \right\}
\\
& \equiv & C_{\beta \alpha} [\Gamma_{(\mu \nu)}]_{\alpha \beta}
\\
& = & \mathrm{Tr} \left\{C \Gamma_{(\mu \nu)} \right\},
\end{array}
%
\end{equation}
where the cyclic property of the trace has been used and we have
defined the \emph{coherency} matrix $C: C_{\beta \alpha} \equiv
c_\beta c_\alpha^*$ and the $16$ matrices $\{\Gamma_{(\mu \nu)} \}
:[\Gamma_{(\mu \nu)}]_{\alpha \beta} \equiv
\mathrm{Tr}\left\{\sigma_{(\alpha)} \sigma_{(\mu)}
\sigma_{(\beta)} \sigma_{(\nu)}  \right\}$. In the remaining part
of these notes we shall derive again the result in Eq.
(\ref{e14l1}) in two other different ways which are perhaps more
complex but also more physically clear.

Note that from Eq. (\ref{e14l}) it follows
\begin{equation}\label{e14l2}
M_{00} = \mathrm{Tr} \left\{ \sigma_{(0)} T \sigma_{(0)} T^\dagger
\right\} = \frac{1}{2} \mathrm{Tr} \left\{ T T^\dagger  \right\},
%
\end{equation}
therefore, when $T$ is unitary $ \mathrm{Tr} \left\{ T T^\dagger
\right\} =  \mathrm{Tr} \left\{ I_2  \right\} = 2$, which implies
\begin{equation}\label{e14l3}
M_{00} = 1.
%
\end{equation}
This is then the ``natural'' normalization of $M$.
\subsection{From the $M$ matrix to the $H$ matrix}
The Mueller matrix $M$ has not, in general, any particular
symmetry property. However it is possible to extract from it an
Hermitian matrix $H$ in the way we are going to show. Let us start
by writing $M$ in component form as
\begin{equation}\label{e24}
\begin{array}{ccl}
M_{\mu \nu} & = & \mathrm{Tr} \left\{ \sigma_{(\mu)} T
\sigma_{(\nu)} T^\dagger \right\}\\
& = &  [\sigma_{(\mu)}]_{mn} T_{np} [\sigma_{(\nu)}]_{pq}
T_{qm}^\dagger\\
& = &   T_{np} T_{mq}^* [\sigma_{(\mu)}]_{mn}
[\sigma_{(\nu)}]_{pq}\\
& = &   (T \otimes T^*)_{nm,pq} [\sigma_{(\mu)}]_{mn}
[\sigma_{(\nu)}]_{pq}\\
& \equiv &   F_{nm,pq} [\sigma_{(\mu)}]_{mn} [\sigma_{(\nu)}]_{pq}
\end{array}
%
\end{equation}
where we have defined the matrix $F \in \mathbb{C}^{4 \times 4}$
as
\begin{equation}\label{e24b}
F \equiv
 T \otimes T^*,
%
\end{equation}
which contains \emph{all} the information about the scattering
process. From Eq. (\ref{e24b}) it is clear that $F$ is not
Hermitian, however, we can extract out of it the Hermitian matrix
$H$ by doing a p\emph{artial} e\emph{xchange} of the r\emph{ows}
($\mathrm{Per}[\;]$) defined in the following way:
\begin{equation}\label{e25}
H = \mathrm{Per}[F] \quad \Leftrightarrow \quad  H_{np,m q} =
F_{nm,pq},
%
\end{equation}
%
%
%
%
where the indices $p$ and $m$ have been exchanged. This definition
clearly requires the matrices $H$ and $F$ to be written with four
indices, as if they were generated by a direct product of two $2
\times 2$ matrices; see, e.g., Eqs. (\ref{i4}-\ref{i5}). However,
this is unnecessary; actually after a careful examination of Eq.
(\ref{e25}) one can easily convince himself (or herself) that
 the effect of
the ``$\mathrm{Per}[\;]$'' operation on an arbitrary $4 \times 4$
matrix can be  written explicitly in matrix form as
\begin{equation}\label{e25b}
\mathrm{Per}\left[
\begin{pmatrix}
  a_{0} & b_{0} & c_{0} & d_{0} \\
  a_{1} & b_{1} & c_{1} & d_{1} \\
  a_{2} & b_{2} & c_{2} & d_{2} \\
  a_{3} & b_{3} & c_{3} & d_{3} \\
\end{pmatrix}
\right]
=
\begin{pmatrix}
  a_{0} & b_{0} & a_{1} & b_{1} \\
  c_{0} & d_{0} & c_{1} & d_{1} \\
  a_{2} & b_{2} & a_{3} & b_{3} \\
  c_{2} & d_{2} & c_{3} & d_{3} \\
\end{pmatrix}.
%
\end{equation}
This equation can be considered as a definition of the
$\mathrm{Per}[\;]$ operation alternative to the one given in Eq.
(\ref{e25}). The advantage of Eq. (\ref{e25b}) with respect to Eq.
(\ref{e25}) is that it does not require the $4 \times 4$ matrix to
be written as the direct product of two $2 \times 2$ sub-matrices,
but it is applicable to arbitrary matrices.

 The
matrix $H$ is Hermitian: this can be easily seen by first writing
explicitly $F$ in terms of the components $T_{ij}$ of $T$
\begin{equation}\label{e39}
F = \begin{pmatrix}
  T_{00}T_{00}^* & T_{00}T_{01}^* & T_{01}T_{00}^* & T_{01}T_{01}^* \\
  T_{00}T_{10}^* & T_{00}T_{11}^* & T_{01}T_{10}^* & T_{01}T_{11}^* \\
  T_{10}T_{00}^* & T_{10}T_{01}^* & T_{11}T_{00}^* & T_{11}T_{01}^* \\
  T_{10}T_{10}^* & T_{10}T_{11}^* & T_{11}T_{10}^* & T_{11}T_{11}^*
\end{pmatrix},
%
\end{equation}
and then by applying the $\mathrm{Per}[ \;]$ operation to $F$ to
obtain $H$:
\begin{equation}\label{e40}
\begin{array}{ccl}
H & = & \mathrm{Per}[F]
\\\\
 & = & \begin{pmatrix}
  T_{00}T_{00}^* & T_{00}T_{01}^* & T_{00}T_{10}^* & T_{00}T_{11}^* \\
  T_{01}T_{00}^* & T_{01}T_{01}^* & T_{01}T_{10}^* & T_{01}T_{11}^* \\
  T_{10}T_{00}^* & T_{10}T_{01}^* & T_{10}T_{10}^* & T_{10}T_{11}^* \\
  T_{11}T_{00}^* & T_{11}T_{01}^* & T_{11}T_{10}^* & T_{11}T_{11}^*
\end{pmatrix}
\\\\
& = & \begin{pmatrix}
  T_{00} \\
  T_{01} \\
  T_{10} \\
  T_{11}
\end{pmatrix} \begin{pmatrix}
  T^*_{00} & T^*_{01} & T^*_{10} & T^*_{11}
\end{pmatrix}\\\\
& = & \vec{h} \vec{h}^\dagger,
\end{array}
%
\end{equation}
where the diad $\vec{h} \vec{h}^\dagger$ is written in terms of
the $4$-vector $\vec{h}$ defined as
\begin{equation}\label{e40b}
\vec{h} = \begin{pmatrix}
  T_{00} \\
  T_{01} \\
  T_{10} \\
  T_{11}
\end{pmatrix},
%
\end{equation}
which is just the $4$-vector representing  $T$ in the basis
$\{\epsilon_{(\mu)}\}$:
\begin{equation}\label{e40c}
T = h_\mu \epsilon_{(\mu)}.
%
\end{equation}
Then, by using Eqs. (\ref{e40}-\ref{e40b}) we can write $H$ in
component form as
\begin{equation}\label{e40d}
H_{\mu \nu} = h_\mu {h_\nu^*},
%
\end{equation}
from which its Hermitian character is evident. Finally, by
combining Eq. (\ref{e24b}) and (\ref{e40c}) we get
\begin{equation}\label{e36}
\begin{array}{ccl}
F  & = & h_{\mu} \epsilon_{(\mu)} \otimes {h_\nu^*} \epsilon_{(\nu)}\\
   & = & h_{\mu} {h_\nu^*} \epsilon_{(\mu)} \otimes \epsilon_{(\nu)}\\
   & \equiv & H_{\mu \nu} \epsilon_{(\mu)} \otimes  \epsilon_{(\nu)} ,
\end{array}
%
\end{equation}
which shows that $H$ is just the representation of $F$ in the
basis $\{ \epsilon_{(\mu)} \otimes  \epsilon_{(\nu)} \}$.

 Now we
can continue the calculation of $M_{\mu \nu}$ by inserting Eq.
(\ref{e24b}) in Eq. (\ref{e24})  obtaining
\begin{equation}\label{e42}
\begin{array}{ccl}
M_{\mu \nu} & = & [\sigma_{(\mu)}]_{mn} (T \otimes T^*)_{nm,pq}
[\sigma_{(\nu)}]_{pq}\\
& = & [\sigma_{(\mu)}^*]_{nm} (T \otimes T^*)_{nm,pq}
[\sigma_{(\nu)}]_{pq}\\
& = & [\sigma_{(\mu)}^*]_{\alpha} (T \otimes T^*)_{\alpha \beta}
[\sigma_{(\nu)}]_{\beta},
\end{array}
%
\end{equation}
where in the second line we exploited the fact that the Pauli
matrices are Hermitian, so $[\sigma_{(\mu)}]_{mn} =
[\sigma_{(\mu)}^*]_{nm}$, and in the last line we used the Rule to
define $\alpha = 2n+m$ and $\beta = 2p+q$. But since $
[\sigma_{(\mu)}]_{\alpha} = \Lambda_{\alpha \mu}$, then
\begin{equation}\label{e43}
\begin{array}{ccl}
M_{\mu \nu} & = & \Lambda^*_{\alpha \mu}  (T \otimes T^*)_{\alpha
\beta}
\Lambda_{\beta \mu}\\
& = & \Lambda^\dagger_{ \mu \alpha}  (T \otimes T^*)_{\alpha
\beta}
\Lambda_{\beta \mu}\\
& = & [\Lambda^\dagger (T \otimes T^*) \Lambda]_{\mu \nu},
\end{array}
%
\end{equation}
or, in matrix form
\begin{equation}\label{e44}
M = \Lambda^\dagger (T \otimes T^*) \Lambda.
%
\end{equation}
This formula is particular relevant because it permits us to
define the matrix $F$ even when the Mueller matrix is
nondeterministic or, equivalently, when is not a Mueller-Jones
matrix. In fact, by rewriting Eq. (\ref{e44}) as
\begin{equation}\label{e44c}
M = \Lambda^\dagger F \Lambda,
%
\end{equation}
it is clear that we can invert it and define, in the general case
\begin{equation}\label{e44d}
F \equiv  \Lambda M \Lambda^\dagger.
%
\end{equation}
In the same spirit we can define $H$ in the general case by
starting from the last line of the Eq. (\ref{e24}) which can be
rewritten with the help of the Eq. (\ref{e25}) as
\begin{equation}\label{e44e}
\begin{array}{rcl}
M_{\mu \nu} & = & H_{np,m q}[\sigma_{(\mu)}]_{mn}
 [\sigma_{(\nu)}]_{pq}\\
 & = & H_{np,m q}[\sigma_{(\mu)}]_{mn}
 [\sigma_{(\nu)}^*]_{qp}\\
 & = & H_{np,m q}[\sigma_{(\mu)} \otimes \sigma_{(\nu)}^*]_{mq,np}\\
 & = & H_{\alpha\beta}[\sigma_{(\mu)} \otimes \sigma_{(\nu)}^*]_{\beta \alpha}\\
 & = & \mathrm{Tr} \left\{H (\sigma_{(\mu)} \otimes
 \sigma_{(\nu)}^*)\right\},
%
\end{array}
\end{equation}
where we used the Rule to define $\alpha = 2n + p$ and  $\beta = 2
m +q $. It is simple to invert this equation by using the
completeness relation Eq. (\ref{e23}) that here we rewrite
\begin{equation}\label{e23b}
 \displaystyle{\sum_{\mu = 0}^3 [\sigma_{(\mu)}]_{ij}
[\sigma_{(\mu)}^*]_{kl}} = \delta_{i k}\delta_{j l}.
%
\end{equation}
Then, by multiplying both sides of Eq. (\ref{e44e}) per
$[\sigma_{(\mu)}^*]_{ki}[\sigma_{(\nu)}^*]_{jl}$ and summing on
$\mu$ and $\nu$, we obtain
\begin{equation}\label{e26}
\begin{array}{ccl}
M_{\mu \nu}[\sigma_{(\mu)}^*]_{ki}[\sigma_{(\nu)}^*]_{jl} & = &
H_{np,m q}[\sigma_{(\mu)}]_{mn} [\sigma_{(\mu)}]_{ki}^*
[\sigma_{(\nu)}]_{pq}[\sigma_{(\nu)}]_{jl}^*
\\\\
& = & H_{np,m q}\delta_{mk}\delta_{ni} \delta_{pj}\delta_{ql}
\\\\
& = &  H_{ij,k l},
\end{array}
%
\end{equation}
which is the desired result. This equation can be put in matrix
form by noticing that
\begin{equation}\label{e27}
\begin{array}{ccl}
[\sigma_{(\mu)}^*]_{ki}[\sigma_{(\nu)}^*]_{jl} & = &
[\sigma_{(\mu)}]_{ik}[\sigma_{(\nu)}^*]_{jl}\\ & = &
(\sigma_{(\mu)} \otimes \sigma_{(\nu)}^*)_{ij , kl},
\end{array}
%
\end{equation}
which permits us to write
\begin{equation}\label{e28}
H_{i j,k l} = M_{\mu \nu} (\sigma_{(\mu)} \otimes
\sigma_{(\nu)}^*)_{ij , kl}.
%
\end{equation}
This formula is not very appealing because it contains both Latin
indices which run from $0$ to $1$, and Greek indices which run
from $0$ to $3$. This problem can be solved by using again the
Rule to define $\alpha = 2i+j$ and $\beta = 2k+l$. Finally, we can
rewrite Eq. (\ref{e28}) as
\begin{equation}\label{e29}
H_{\alpha \beta} = M_{\mu \nu} (\sigma_{(\mu)} \otimes
\sigma_{(\nu)}^*)_{\alpha \beta},
%
\end{equation}
or, in matrix form
\begin{equation}\label{e30}
H = \sum_{\mu, \nu}^{0,3} M_{\mu \nu} (\sigma_{(\mu)} \otimes
\sigma_{(\nu)}^*).
%
\end{equation}
We can consider this formula as the \emph{definition} of $H$ for
arbitrary $M$.
\subsection{The Coherency matrix $C$}
The relation between $H$ and $M$ is linear but quite involved, as
can be seen by writing explicitly $H$ in terms of the components
$M_{\mu \nu}$ of $M$:
\begin{equation}\label{e30b1}
\begin{array}{lcl}
H_{00} & = & \frac{1}{2} \left(  M_{00}  + M_{03} +      M_{30} +     M_{33}  \right),\\
H_{01} & = & \frac{1}{2} \left(  M_{01}  + M_{31} + \im  M_{02} + \im M_{32}  \right),\\
H_{02} & = & \frac{1}{2} \left(  M_{10}  + M_{13} - \im  M_{20} - \im M_{23}  \right),\\
H_{03} & = & \frac{1}{2} \left(  M_{11}  + M_{22} + \im  M_{12} - \im M_{21}  \right),
\end{array}
\end{equation}
\begin{equation}\label{e30b2}
\begin{array}{lcl}
H_{10} & = & \frac{1}{2} \left(     M_{01}  + M_{31} - \im M_{02}  - \im M_{32} \right),\\
H_{11} & = & \frac{1}{2} \left(     M_{00}  - M_{03} +     M_{30}  -     M_{33} \right),\\
H_{12} & = & \frac{1}{2} \left(     M_{11}  - M_{22} - \im M_{12}  - \im M_{21} \right),\\
H_{13} & = & \frac{1}{2} \left(     M_{10}  - M_{13} - \im M_{20}  + \im M_{23} \right),
\end{array}
\end{equation}
\begin{equation}\label{e30b3}
\begin{array}{lcl}
H_{20} & = & \frac{1}{2} \left( M_{10} + M_{13} + \im M_{20} + \im  M_{23}  \right),\\
H_{21} & = & \frac{1}{2} \left( M_{11} - M_{22} + \im M_{12} + \im  M_{21}  \right),\\
H_{22} & = & \frac{1}{2} \left( M_{00} + M_{03} -     M_{30} -      M_{33}  \right),\\
H_{23} & = & \frac{1}{2} \left( M_{01} - M_{31} + \im M_{02} - \im  M_{32}  \right),
\end{array}
\end{equation}
\begin{equation}\label{e30b4}
\begin{array}{lcl}
H_{30} & = & \frac{1}{2} \left( M_{11} + M_{22} - \im M_{12} + \im M_{21}  \right),\\
H_{31} & = & \frac{1}{2} \left( M_{10} - M_{13} + \im M_{20} - \im M_{23}  \right),\\
H_{32} & = & \frac{1}{2} \left( M_{01} - M_{31} - \im M_{02} + \im M_{32}  \right),\\
H_{33} & = & \frac{1}{2} \left( M_{00} - M_{03} -     M_{30} +     M_{33}  \right).
\end{array}
\end{equation}
From this formula we see that
\begin{equation}\label{e30c}
\mathrm{Tr}\{H\} = 2 M_{00}.
%
\end{equation}
If we choose the ``natural'' normalization  $M_{00} = 1$, it
follows $\mathrm{Tr}\{H\} = 2$. The matrix $H$ is not the only
Hermitian matrix we can extract from $M$, actually there are
infinitely many Hermitian matrices generated by $M$ which differ
from $H$ by a unitary transformation.  A particularly relevant
Hermitian matrix is the \emph{Coherency} matrix $C$ defined as the
representation of $F$ in the basis $\{\sigma_{(\mu)} \otimes
\sigma_{(\mu)}^*\}$. In order to find this representation, let us
first write the transformation matrix $T$ in both the bases
$\{\sigma_{(\mu)} \}$ and $\{\epsilon_{(\mu)} \}$ as
\begin{equation}\label{e33}
T =  c_{\mu} \sigma_{(\mu)} = h_{\mu} \epsilon_{(\mu)} , \qquad
(\mu = 0,1,2,3),
%
\end{equation}
and then let us calculate
\begin{equation}\label{e34}
\begin{array}{ccl}
F  & = & T \otimes T^*\\
   & = & c_{\mu} \sigma_{(\mu)} \otimes {c_{\nu}^*} \sigma_{(\nu)}^*\\
   & = & c_{\mu} {c_{\nu}^*} \sigma_{(\mu)} \otimes \sigma_{(\nu)}^*\\
   & \equiv & C_{\mu \nu} \sigma_{(\mu)} \otimes  \sigma_{(\nu)}^* ,
\end{array}
%
\end{equation}
where we have defined the coherency matrix elements as
\begin{equation}\label{e35}
 C_{\mu \nu} \equiv  c_{\mu} {c_{\nu}^*}, \qquad
(\mu, \nu = 0,1,2,3).
%
\end{equation}
By comparing Eq. (\ref{e36}) with Eq. (\ref{e34}) it appears
evident that $C$ and $H$ are different representations  of the
same matrix $F$, with respect to different bases. Therefore they
must be related by a unitary transformation: we want to find it.
To this end, let us recall that
%
%
%
%
if $A$ is an arbitrary $2 \times 2$ matrix in $\mathbb{C}^2$ which
can be represented in the two different bases
$\{\epsilon_{(\mu)}\}$ and $\{\sigma_{(\mu)}\}$  as
\begin{equation}\label{e31}
A = a_{\mu} \epsilon_{(\mu)} = b_{\mu} \sigma_{(\mu)}, \qquad (\mu
= 0,1,2,3),
%
\end{equation}
then the expansion coefficients $a_\mu$ and $b_\mu$ are related by
the change of basis matrix $\Lambda$ as
\begin{equation}\label{e32}
\begin{array}{ccl}
a_{\mu}  & = &
\{\epsilon_{(\mu)}, A\}\\
& = & \{\epsilon_{(\mu)}, b_{\nu} \sigma_{(\nu)}\}\\
& = & \{\epsilon_{(\mu)},\sigma_{(\nu)}\}  b_{\nu} \\
& = &  \Lambda_{\mu \nu} b_{\nu},
\end{array}
%
\end{equation}
or, in more compact form,
\begin{equation}\label{e32b}
\vec{a} = \Lambda \vec{b}.
%
\end{equation}
In our specific case we find, by using Eq. (\ref{e14}),
\begin{equation}\label{e37}
\begin{array}{ccl}
F  & = & C_{\mu \nu} \sigma_{(\mu)} \otimes  \sigma_{(\nu)}^*\\
   & = & C_{\mu \nu} \epsilon_{(\alpha)} \Lambda_{\alpha \mu}
            \otimes  \epsilon_{(\beta)} {\Lambda_{\beta \nu}^*}\\
   & = &  \Lambda_{\alpha \mu} C_{\mu \nu} \Lambda_{\nu \beta }^\dagger
                     \epsilon_{(\alpha)} \otimes
                     \epsilon_{(\beta)}\\
   & = &  [\Lambda C \Lambda^\dagger]_{\alpha \beta}
                     \epsilon_{(\alpha)} \otimes
                     \epsilon_{(\beta)}.
\end{array}
%
\end{equation}
The comparison of Eq. (\ref{e36}) with Eq. (\ref{e37}) reveals
that
\begin{equation}\label{e38}
H = \Lambda C \Lambda^\dagger.
%
\end{equation}
Finally, we can combine the results in Eq. (\ref{e30}) and
(\ref{e38}) to obtain the sought relation between $H$, $C$ and
$M$:
\begin{equation}\label{e44b}
\begin{array}{ccl}
C & = & \Lambda^\dagger  H
\Lambda\\
 & = & \displaystyle{\sum_{\mu, \nu}^{0,3} M_{\mu \nu}
 \Lambda^\dagger(\sigma_{(\mu)} \otimes
\sigma_{(\nu)}^*)\Lambda}\\
 & \equiv & \displaystyle{\sum_{\mu, \nu}^{0,3} M_{\mu \nu}
\Gamma_{(\mu \nu)}},
\end{array}
%
\end{equation}
where we have defined the $16$ Hermitian  matrices $\{\Gamma_{(\mu
\nu)}\}$ as
\begin{equation}\label{e44f}
\Gamma_{(\mu \nu)} \equiv \Lambda^\dagger(\sigma_{(\mu)} \otimes
\sigma_{(\nu)}^*)\Lambda.
%
\end{equation}
Note that since the Pauli basis is complete in $\mathbb{C}^{2
\times 2}$,  the direct products $\{\sigma_{(\mu)} \otimes
\sigma_{(\nu)}^*\}$ form a complete basis in $\mathbb{C}^{4 \times
4}$. Moreover, since $\Lambda$ is unitary, the matrices
$\sigma_{(\mu)} \otimes \sigma_{(\nu)}^*$ and $\Gamma_{(\mu \nu)}$
are equivalent, therefore the $16$ matrices $\{\Gamma_{(\mu
\nu)}\}$ form a complete basis in $\mathbb{C}^{4 \times 4}$. From
the matrix rule
\begin{equation}\label{e44fb}
\left( A \otimes B  \right)^\dagger =  A^\dagger \otimes
B^\dagger,
%
\end{equation}
and the definition Eq. (\ref{e44f}), it follows that the matrices
$\{\Gamma_{(\mu \nu)}\}$ are Hermitian.  Moreover, with the help
of the general matrix rules
\begin{equation}\label{e44g}
\mathrm{Tr}\{A\}\mathrm{Tr}\{B\} = \mathrm{Tr}\{A \otimes B\},
%
\end{equation}
and
\begin{equation}\label{e44h}
\left( A \otimes B \right)\left( C \otimes D \right) = AC \otimes
BD,
%
\end{equation}
we can show that the matrices $\Gamma_{(\mu \nu)}$ are also
orthonormal:
\begin{equation}\label{e44i}
\begin{array}{ccl}
\mathrm{Tr}\{\Gamma_{(\mu \nu)} \Gamma_{(\alpha \beta)} \} & = &
\mathrm{Tr}\{\Lambda^\dagger(\sigma_{(\mu)} \otimes
\sigma_{(\nu)}^*)\Lambda \Lambda^\dagger(\sigma_{(\alpha)} \otimes
\sigma_{(\beta)}^*)\Lambda\}\\
& = & \mathrm{Tr}\{\Lambda^\dagger(\sigma_{(\mu)} \otimes
\sigma_{(\nu)}^*)(\sigma_{(\alpha)} \otimes
\sigma_{(\beta)}^*)\Lambda\}\\
& = & \mathrm{Tr}\{\Lambda \Lambda^\dagger (\sigma_{(\mu)}
\sigma_{(\alpha)} \otimes \sigma_{(\nu)}^*\sigma_{(\beta)}^*)\}\\
& = & \mathrm{Tr}\{\sigma_{(\mu)}
\sigma_{(\alpha)} \otimes \sigma_{(\nu)}^*\sigma_{(\beta)}^*\}\\
& = & \mathrm{Tr}\{\sigma_{(\mu)}
\sigma_{(\alpha)}\} \mathrm{Tr}\{\sigma_{(\nu)}^*\sigma_{(\beta)}^*\}\\
& = & \delta_{\mu \alpha}\delta_{\nu \beta},
\end{array}
%
\end{equation}
where $\Lambda \Lambda^\dagger = I_4$ and the cyclic property of
the trace have been used.

 Now we use Eq. (\ref{e44i}) to invert
Eq. (\ref{e44b}) and express $M$ as function of $C$. By
multiplying both members of Eq. (\ref{e44b}) by $\Gamma_{(\alpha
\beta)}$ and by tacking the trace, we obtain
\begin{equation}\label{e44j}
\begin{array}{ccl}
\mathrm{Tr}\{  \Gamma_{(\alpha \beta)} C \}& = &
\displaystyle{\sum_{\mu, \nu}^{0,3} M_{\mu \nu}
\mathrm{Tr}\{\Gamma_{(\alpha \beta)} \Gamma_{(\mu \nu)}\} }\\& = &
\displaystyle{\sum_{\mu, \nu}^{0,3} M_{\mu \nu}\delta_{\mu
\alpha}\delta_{\nu \beta}}\\
& = & \displaystyle{M_{\alpha \beta}},
\end{array}
%
\end{equation}
that is
\begin{equation}\label{e44k}
M_{\mu \nu} = \mathrm{Tr}\{  \Gamma_{(\mu \nu)} C \}.
%
\end{equation}
 The Eqs. (\ref{e44b}) and (\ref{e44k}) can be put in a more compact form
  by using the Rule for $n=4$:
\begin{equation}\label{e44l}
(\mu \nu) \rightarrow 4 \mu + \nu \equiv \mathcal{A} \in
\{0,\ldots,15 \}.
%
\end{equation}
Then we can rewrite
\begin{equation}\label{e44m}
 C  =  \sum_{\mathcal{A} = 0}^{15} \Gamma_{(\mathcal{A})}
 m_\mathcal{A},
%
\end{equation}
where
\begin{equation}\label{e44n}
 m_\mathcal{A}  =  \mathrm{Tr}\{\Gamma_{(\mathcal{A})} C \},
%
\end{equation}
and where
\begin{equation}\label{e44o}
\vec{m} = \begin{pmatrix}
  M_{0} \\
  \vdots \\
  M_{15}
\end{pmatrix},
%
\end{equation}
is the $16$-vector associated to the matrix $M$ in the basis
$\{\Gamma_{(\mathcal{A})}\}$.

We conclude this section by writing explicitly the relation
between the matrix elements of $C$ and $M$. An explicit
calculation shows that if $C$ is written as
\begin{equation}\label{e44p}
C = \begin{pmatrix}
 a_0 + a  & c  - \im d  & h  + \im g  & i  - \im j  \\
 c  + \im d  & b_0 +  b  & e  + \im f  & k  - \im l  \\
 h  - \im g  & e  - \im f  & b_0 -  b  & m  + \im n  \\
 i  + \im j  & k  + \im l  & m  - \im n  & a_0 -  a
\end{pmatrix},
%
\end{equation}
then $M$ has the following form:
\begin{equation}\label{e44q}
M = \begin{pmatrix}
 a_0 + b_0 & c  +   n  & h  +   l  & i  +   f  \\
 c  - n  & a  +   b  & e  +   j  & k  +   g  \\
 h  - l  & e  -   j  & a  -   b  & m  +   d  \\
 i  - f  & k  -   g  & m  -   d    & a_0 -   b_0
\end{pmatrix}.
%
\end{equation}
\subsection{Alternative version}
In the literature can be found another method, more geometrical,
to find the matrices $\Gamma_{(\mu \nu)}$ and the result shown in
Eq. (\ref{e44k}). In this subsection we expose that method.

 Let $X, Y$ two matrices in $\mathbb{C}^{2
\times 2}$ and let us consider their product $Z = XY$. With
$\vec{x}, \vec{y}$ and $\vec{z}$ we denote the $4$-vectors
associated to $X,Y$ and $Z$ respectively,  with respect to the
Pauli basis:
\begin{equation}\label{e45}
\begin{array}{lclcccl}
  X & = & x_\mu \sigma_{(\mu)} & \Rightarrow & x_\mu & = & \mathrm{Tr} \{\sigma_{(\mu)} X \},  \\
  Y & = & y_\mu \sigma_{(\mu)} & \Rightarrow & y_\mu & = & \mathrm{Tr} \{\sigma_{(\mu)} Y \},  \\
  Z & = & z_\mu \sigma_{(\mu)} & \Rightarrow & z_\mu & = & \mathrm{Tr} \{\sigma_{(\mu)} Z \},  \\
\end{array}
%
\end{equation}
where $\mu = 0, \ldots, 3$ and summation on repeated indices is
understood. We want to find a formula which expresses $\vec{z}$ as
a function of $\vec{x}$ and $\vec{y}$. To this end let us write
\begin{equation}\label{e46}
\begin{array}{ccl}
  z_\mu & = & \mathrm{Tr} \{\sigma_{(\mu)} Z \}  \\
        & = & \mathrm{Tr} \{\sigma_{(\mu)} XY \} \\
        & = & x_\alpha y_\beta \mathrm{Tr} \{\sigma_{(\mu)}  \sigma_{(\alpha)}  \sigma_{(\beta)} \} \\
        & \equiv & x_\alpha y_\beta [\Upsilon_{(\mu)}]_{\alpha \beta},
\end{array}
%
\end{equation}
where we have defined the four matrices $\Upsilon_{(\mu)} \in
\mathbb{C}^{4 \times 4 }$ as
\begin{equation}\label{e47}
[\Upsilon_{(\mu)}]_{\alpha \beta} \equiv \mathrm{Tr}
\{\sigma_{(\mu)} \sigma_{(\alpha)}  \sigma_{(\beta)} \}.
%
\end{equation}
Then we can rewrite Eq. (\ref{e46}) in a compact form as
\begin{equation}\label{e48}
z_{\mu} = (\vec{x}^*, \Upsilon_{(\mu)} \vec{y}).
%
\end{equation}
Let us notice that because of the cyclic property of the trace
\begin{equation}\label{e49}
\mathrm{Tr} \{\sigma_{(\mu)} \sigma_{(\alpha)}  \sigma_{(\beta)}
\} = \mathrm{Tr} \{\sigma_{(\beta)} \sigma_{(\mu)}
\sigma_{(\alpha)} \} = \mathrm{Tr} \{\sigma_{(\alpha)}
\sigma_{(\beta)} \sigma_{(\mu)} \},
%
\end{equation}
we can write
\begin{equation}\label{e50}
[\Upsilon_{(\mu)}]_{\alpha \beta} =
[\Upsilon_{(\beta)}]_{\mu \alpha} =
[\Upsilon_{(\alpha)}]_{\beta \mu}.
%
\end{equation}
We shall exploit this property in a moment. Now, let us consider
the special case in which $Y = \sigma_{(\nu)} \Rightarrow y_\beta
= \delta_{\beta \nu} $. Then, from Eq. (\ref{e48}) follows that
\begin{equation}\label{e51}
\begin{array}{ccl}
z_\mu & = & x_\alpha [\Upsilon_{(\mu)}]_{\alpha \nu}\\
      & = &[\Upsilon_{(\nu)}]_{\mu \alpha} x_\alpha\\
      & = &[\Upsilon_{(\nu)} \vec{x}]_{\mu},
\end{array}
%
\end{equation}
where Eq. (\ref{e50}) has been used. Another special case is the
transposed one, that is when $X = \sigma_{(\gamma)} \Rightarrow
x_\alpha = \delta_{\alpha \gamma} $ and
\begin{equation}\label{e52}
\begin{array}{ccl}
z_\mu & = & [\Upsilon_{(\mu)}]_{\gamma \beta} y_\beta \\
      & = & [\Upsilon_{(\gamma)}]_{\beta \mu} y_\beta \\
      & = & [\Upsilon_{(\gamma)}^T \vec{y}]_{\mu}.
\end{array}
%
\end{equation}
The previous results can be summarized as follows:
\begin{equation}\label{e53}
\begin{array}{lclclcl}
Z & = & X \sigma_{(\nu)} & \Rightarrow & \vec{z} & = &
\Upsilon_{(\nu)}
\vec{x}\\
Z & = & \sigma_{(\mu)}Y & \Rightarrow & \vec{z} & = &
\Upsilon_{(\mu)}^T \vec{y}.
\end{array}
%
\end{equation}
Now we are equipped to consider the last, most complicated case $Z
= \sigma_{(\mu)} T \sigma_{(\nu)}$, where $T = c_\alpha
\sigma_{(\alpha)}$ is a given $2 \times 2$ matrix associated to
the vector $\vec{c}$, where
\begin{equation}\label{e53b}
\vec{c} = \begin{pmatrix}
  c_{0} \\ c_{1} \\
  c_{2} \\ c_{3}
\end{pmatrix}.
%
\end{equation}
 Then, by
putting $Y =T \sigma_{(\nu)} \Rightarrow \vec{y} =
\Upsilon_{(\nu)} \vec{c}$, it is easy to see that
\begin{equation}\label{e54}
\begin{array}{lcl}
\vec{z} & = & \Upsilon_{(\mu)}^T \vec{y}\\
        & = & \Upsilon_{(\mu)}^T \Upsilon_{(\nu)} \vec{c},
\end{array}
%
\end{equation}
where Eqs. (\ref{e53}) have been used. To summarize, we can write
\begin{equation}\label{e55}
\begin{array}{lcl}
\sigma_{(\mu)} T \sigma_{(\nu)} & \doteq & \Upsilon_{(\mu)}^T
\Upsilon_{(\nu)}
\vec{c}\\
& \equiv & \Gamma_{(\mu \nu)} \vec{c},
 \end{array}
%
\end{equation}
where we have defined the $16$  matrices $\Gamma_{(\mu \nu)} \in
\mathbb{C}^{4 \times 4}$ as
\begin{equation}\label{e55b}
\Gamma_{(\mu \nu)} \equiv \Upsilon_{(\mu)}^T \Upsilon_{(\nu)},
%
\end{equation}
 and the symbol ``$\doteq$'' stands for ``is represented by''.

Now we want to use these equations to calculate the matrix
elements $M_{\mu \nu}$ of the Mueller matrix, by using  Eq.
(\ref{e24}) that here we rewrite:
\begin{equation}\label{e56}
M_{\mu \nu}  =  \mathrm{Tr} \left\{ T^\dagger \sigma_{(\mu)} T
\sigma_{(\nu)} \right\}.
%
\end{equation}
Before doing that, notice that if $T = c_\alpha \sigma_{(\alpha)}
\doteq \vec{c}$ then $T^\dagger = c_\alpha^* \sigma_{(\alpha)}
\doteq \vec{c}^*$; and notice that if $A \doteq \vec{a}$ and $B
\doteq
\vec{b}$, then %
\begin{equation}\label{e57}
\begin{array}{lcl}
\{A, B\} & = & \mathrm{Tr} \left\{ A^\dagger B \right\}\\
         & = & a^*_{\mu}b_{\nu} \mathrm{Tr} \left\{ \sigma_{(\mu)} \sigma_{(\nu)} \right\}\\
         & = & a^*_{\mu}b_{\mu} \\
         & = & (\vec{a},\vec{}b).
\end{array}
%
\end{equation}
Finally, from Eqs. (\ref{e55}-\ref{e57}) it follows
straightforwardly that
\begin{equation}\label{e58}
\begin{array}{lcl}
M_{\mu \nu} & = & \mathrm{Tr} \left\{ T^\dagger \sigma_{(\mu)} T
\sigma_{(\nu)} \right\}\\
& = & (\vec{c},  \Gamma_{(\mu \nu)}
\vec{c})\\
& = &c_\beta c^*_\alpha  [ \Gamma_{(\mu \nu)}]_{\alpha \beta}
\\
& \equiv & C_{ \beta \alpha}  [ \Gamma_{(\mu \nu)}]_{\alpha \beta}
\\
& = & \mathrm{Tr} \left\{ C \Gamma_{(\mu \nu)}\right\},\\
\end{array}
%
\end{equation}
which coincides with the result found in Eq. (\ref{e44k}). To
complete the calculation we have to demonstrate that the $\{
\Gamma_{(\mu \nu)}\}$ matrices found in Eq. (\ref{e55b}) coincide
with the ones found in Eq. (\ref{e44f}). To this end we calculate
the matrix elements in both cases and then compare them. Let us
start from Eq. (\ref{e55b}) to write
\begin{equation}\label{e58b}
\begin{array}{lcl}
[\Gamma_{(\mu \nu)}]_{ \alpha \beta} & =
&\displaystyle{\sum_{\gamma = 0}^3 [\Upsilon_{(\mu)}^T]_{ \alpha
\gamma} [\Upsilon_{(\nu)}]_{\gamma \beta }
} \\
 & =
&\displaystyle{\sum_{\gamma = 0}^3 \mathrm{Tr}\{ \sigma_{(\mu)}
\sigma_{(\gamma)}  \sigma_{(\alpha)}\} \mathrm{Tr}\{
\sigma_{(\nu)} \sigma_{(\gamma)}  \sigma_{(\beta)}\}
} \\
 & =
&\displaystyle{\sum_{\gamma = 0}^3
 [\sigma_{(\mu)}]_{ij}
 [\sigma_{(\gamma)}]_{jk}
 [\sigma_{(\alpha)}]_{ki}
 [\sigma_{(\nu)}]_{lm}
 [\sigma_{(\gamma)}]_{mn}
 [\sigma_{(\beta)}]_{nl}
} \\
 & =
&\displaystyle{\left( \sum_{\gamma = 0}^3 [\sigma_{(\gamma)}]_{jk}
[\sigma_{(\gamma)}]_{mn}\right)
 [\sigma_{(\mu)}]_{ij}
 [\sigma_{(\alpha)}]_{ki}
 [\sigma_{(\nu)}]_{lm}
 [\sigma_{(\beta)}]_{nl}.
} \\
\end{array}
%
\end{equation}
From the completeness Eq. (\ref{e23}) we know that
\begin{equation}\label{e58c}
\begin{array}{lcl}
\displaystyle{\sum_{\gamma = 0}^3 [\sigma_{(\gamma)}]_{jk}
[\sigma_{(\gamma)}]_{mn} } & = & \displaystyle{ \sum_{\gamma =
0}^3 [\sigma_{(\gamma)}]_{jk}
[\sigma_{(\gamma)}^*]_{nm}}\\
& = & \displaystyle{ \delta_{jn} \delta_{km}},
\end{array}
%
\end{equation}
so that Eq. (\ref{e55b}) becomes
\begin{equation}\label{e58d}
\begin{array}{lcl}
[\Gamma_{(\mu \nu)}]_{ \alpha \beta} & = &
\displaystyle{\delta_{jn} \delta_{km}
 [\sigma_{(\mu)}]_{ij}
 [\sigma_{(\alpha)}]_{ki}
 [\sigma_{(\nu)}]_{lm}
 [\sigma_{(\beta)}]_{nl}
} \\
& = & \displaystyle{
 [\sigma_{(\mu)}]_{ij}
 [\sigma_{(\alpha)}]_{ki}
 [\sigma_{(\nu)}]_{lk}
 [\sigma_{(\beta)}]_{jl}
} \\
& = & \displaystyle{
 [\sigma_{(\alpha)}]_{ki} [\sigma_{(\mu)}]_{ij}
 [\sigma_{(\beta)}]_{jl}
 [\sigma_{(\nu)}]_{lk}
} \\
& = & \displaystyle{ \mathrm{Tr}\{
 \sigma_{(\alpha)} \sigma_{(\mu)}
 \sigma_{(\beta)}
 \sigma_{(\nu)}\}.
}
\end{array}
%
\end{equation}
The equality
\begin{equation}\label{e58e}
[\Upsilon_{(\mu)}^T \Upsilon_{(\nu)}]_{ \alpha \beta} =
\mathrm{Tr}\{
 \sigma_{(\alpha)} \sigma_{(\mu)}
 \sigma_{(\beta)}
 \sigma_{(\nu)}\},
%
\end{equation}
can be also easily checked by explicit calculation. Now we repeat
the calculation of $[\Gamma_{(\mu \nu)}]_{ \alpha \beta}$ starting
from  Eq. (\ref{e44f}):
\begin{equation}\label{e58f}
\begin{array}{lcl}
[\Gamma_{(\mu \nu)}]_{ \alpha \beta}
& = & \displaystyle{
 [ \Lambda^\dagger (\sigma_{(\mu)} \otimes
\sigma_{(\nu)}^*) \Lambda ]_{ \alpha \beta}
} \\
& = & \displaystyle{
 [ \Lambda^*]_{\gamma \alpha } [\sigma_{(\mu)} \otimes
\sigma_{(\nu)}^*]_{\gamma \varepsilon}  [\Lambda ]_{\varepsilon
\beta}
} \\
& = & \displaystyle{
 [ \sigma_{(\alpha )}^*]_{\gamma } [\sigma_{(\mu)} \otimes
\sigma_{(\nu)}^*]_{\gamma \varepsilon}  [\sigma_{(\beta)}
]_{\varepsilon }
},
\end{array}
%
\end{equation}
where we have used Eq. (\ref{e16}) in the last line. Now we use
the Rule to pass from the dummy $4$-D Greek indices to the dummy
$2$-D Latin indices and write
\begin{equation}\label{e58g}
\begin{array}{lcl}
[\Gamma_{(\mu \nu)}]_{ \alpha \beta}
& = & \displaystyle{
 [ \sigma_{(\alpha )}^*]_{ik} [\sigma_{(\mu)} \otimes
\sigma_{(\nu)}^*]_{ik ,jl}  [\sigma_{(\beta)} ]_{jl}
}\\
& = & \displaystyle{
 [ \sigma_{(\alpha )}^*]_{ik} [\sigma_{(\mu)}]_{ij}
[\sigma_{(\nu)}^*]_{kl}  [\sigma_{(\beta)} ]_{jl}
}\\
& = & \displaystyle{
 [ \sigma_{(\alpha )}]_{ki} [\sigma_{(\mu)}]_{ij}
 [\sigma_{(\beta)} ]_{jl}[\sigma_{(\nu)}]_{lk}
}\\
& = & \displaystyle{ \mathrm{Tr}\{
 \sigma_{(\alpha)} \sigma_{(\mu)}
 \sigma_{(\beta)}
 \sigma_{(\nu)}\}}.
\end{array}
%
\end{equation}
This complete our demonstration.

 Let us conclude this subsection by calculating
explicitly the $4$ matrices $\{ \Upsilon_{(\mu)} \}$. First of all
we notice that in the case in which one of the indices is zero,
then we have
\begin{equation}\label{e59}
\begin{array}{lclcl}
\, [\Upsilon_{(0)}]_{ \alpha \beta} & = &\displaystyle{
\frac{1}{\sqrt{2}}\mathrm{Tr} \left\{ \sigma_{(\alpha)}
\sigma_{(\beta)} \right\} }& = & \displaystyle{
\frac{1}{\sqrt{2}}\delta_{\alpha \beta}},\\
\, [\Upsilon_{(\mu)}]_{ 0 \beta} & = &\displaystyle{
\frac{1}{\sqrt{2}}\mathrm{Tr} \left\{ \sigma_{(\mu)}
\sigma_{(\beta)} \right\} }& = & \displaystyle{
\frac{1}{\sqrt{2}}\delta_{\mu \beta}},\\
 \, [
\Upsilon_{(\mu)}]_{\alpha 0} & = & \displaystyle{
\frac{1}{\sqrt{2}}\mathrm{Tr} \left\{
 \sigma_{(\mu)} \sigma_{(\alpha)} \right\}} & = &
 \displaystyle{ \frac{1}{\sqrt{2}}\delta_{\mu \alpha }}.
\end{array}
%
\end{equation}
In the case in which all indices are different from zero, we use
the following well known property of the Pauli matrices
\begin{equation}\label{e60}
\sigma_{(i)} \sigma_{(j)} = \frac{1}{\sqrt{2}}\left(\delta_{ij}
\sigma_{(0)} + \im \varepsilon_{ijl}\sigma_{(l)} \right),
%
\end{equation}
where $i,j,l = 1,2,3$ and $\varepsilon_{123} = -\varepsilon_{132}
=\varepsilon_{312} = -\varepsilon_{321} = \varepsilon_{231} =
-\varepsilon_{213}= 1 $ is the completely antisymmetric
Levi-Civita pseudo-tensor (all the unwritten components are zero);
to show that
\begin{equation}\label{e61}
\begin{array}{lcl}
[\Upsilon_{(i)}]_{ j k} & = & \mathrm{Tr} \left\{
\sigma_{i}\sigma_{j}\sigma_{k}\right\}\\\\
& = & \displaystyle{ \frac{1}{\sqrt{2}}\delta_{ij} \mathrm{Tr}
\left\{
 \sigma_{(0)} \sigma_{(k)} \right\}
 + \frac{\im}{\sqrt{2}}  \varepsilon_{ijl} \mathrm{Tr}
\left\{
 \sigma_{(l)} \sigma_{(k)}\right\}}\\\\
 & = & \displaystyle{\frac{\im}{\sqrt{2}}  \varepsilon_{ijk} }.
 \end{array}
%
\end{equation}
Finally, by collecting all these results, we can write explicitly:
\begin{equation}\label{e62}
\begin{array}{cclcccl}
  \Upsilon_{(0)} & = &
\displaystyle{ \frac{1}{\sqrt{2}} \begin{pmatrix}
    1 & 0 & 0 & 0 \\
    0 & 1 & 0 & 0 \\
    0 & 0 & 1 & 0 \\
    0 & 0 & 0 & 1 \
  \end{pmatrix} },& \quad & \Upsilon_{(1)} & = &
\displaystyle{ \frac{1}{\sqrt{2}}  \begin{pmatrix}
    0 & 1 & 0 & 0 \\
    1 & 0 & 0 & 0 \\
    0 & 0 & 0 & \im \\
    0 & 0 & -\im & 0 \
  \end{pmatrix} },  \\\\
  \Upsilon_{(2)} & = &
\displaystyle{ \frac{1}{\sqrt{2}}  \begin{pmatrix}
    0 & 0 & 1 & 0 \\
    0 & 0 & 0 & -\im \\
    1 & 0 & 0 & 0 \\
    0 & \im & 0 & 0 \
  \end{pmatrix} },& \quad & \Upsilon_{(3)} & = &
\displaystyle{ \frac{1}{\sqrt{2}}  \begin{pmatrix}
    0 & 0 & 0 & 1 \\
    0 & 0 & \im & 0 \\
    0 & -\im & 0 & 0 \\
    1 & 0 & 0 & 0 \
  \end{pmatrix}}.
\end{array}
%
\end{equation}
\section{The decomposition theorem}
In this section we show that a given Mueller matrix $M$ can be
written as a linear combination with positive coefficients of at
most $4$ Mueller-Jones matrices. Here we do \emph{not} adopt the
Einstein summation convention, therefore repeated indices must not
be summed. All sums will be written explicitly as in the right
side of Eq. (\ref{i3}).

In the Eq. (\ref{e35}) we have defined the Hermitian matrix $C$
and we have shown its relation with $H$ and $M$. Since $C$ is
Hermitian it can be diagonalized. Let $\vec{u}_{(\alpha)}, \,
(\alpha = 0,\ldots,3)$ the four eigenvectors of $C$ associated
with the four real eigenvalues $\lambda_{\alpha}$
\begin{equation}\label{e118}
C \vec{u}_{(\alpha)} = \lambda_{\alpha} \vec{u}_{(\alpha)},
\qquad(\alpha = 0,\ldots,3),
%
\end{equation}
where there is \emph{not} sum on repeated indices. The
eigenvectors of an Hermitian matrix can always be chosen
orthonormal, so we assume
\begin{equation}\label{e119}
(\vec{u}_{(\alpha)}, \vec{u}_{(\beta)}) = \delta_{\alpha \beta},
\qquad(\alpha, \beta = 0,\ldots,3).
%
\end{equation}
By tacking the scalar product of both sides of Eq. (\ref{e118})
with $\vec{u}_{(\beta)}$, we obtain
\begin{equation}\label{e120}
(\vec{u}_{(\beta)},C \vec{u}_{(\alpha)}) = \lambda_{\alpha}
(\vec{u}_{(\beta)},\vec{u}_{(\alpha)}) =
\lambda_{\alpha}\delta_{\beta  \alpha }, \qquad(\alpha, \beta =
0,\ldots,3),
%
\end{equation}
If we write explicitly the left side of this equation  we get
\begin{equation}\label{e121}
\begin{array}{ccl}
(\vec{u}_{(\beta)},C \vec{u}_{(\alpha)})
 & = & \displaystyle{ \sum_{\mu, \nu}^{0,3}
 [\vec{u}_{(\beta)}^*]_{\mu} C_{\mu \nu}
       [\vec{u}_{(\alpha)}]_{\nu} }\\
 & \equiv & \displaystyle{ \sum_{\mu, \nu}^{0,3}U^*_{\mu \beta} C_{\mu \nu} U_{\nu \alpha}}\\
 & = & \displaystyle{ \sum_{\mu, \nu}^{0,3}U^\dagger_{\beta \mu } C_{\mu \nu} U_{\nu \alpha}}\\
 & = & [ U^\dagger C U]_{\beta \alpha },
\end{array}
%
\end{equation}
where  we have defined the matrix $U$ as
\begin{equation}\label{e122}
U : \quad U_{\beta \alpha} \equiv  [\vec{u}_{(\alpha)}]_{\beta},
\qquad(\alpha, \beta = 0,\ldots,3).
%
\end{equation}
The matrix $U$ is unitary by definition:
\begin{equation}\label{e123}
\begin{array}{ccl}
[U^\dagger U]_{\alpha \beta}& = & \displaystyle{ \sum_{\mu =0}^3
U^*_{\mu
\alpha}U_{\mu \beta} }\\
& = & \displaystyle{ \sum_{\mu =0}^3 [\vec{u}_{(\alpha)}^*]_{\mu}
[\vec{u}_{(\beta)}]_{\mu} }\\
& = & \displaystyle{ (\vec{u}_{(\alpha)},\vec{u}_{(\beta)}) }\\
& = & \displaystyle{ \delta_{\alpha \beta} }.
\end{array}
\end{equation}
By comparing Eq. (\ref{e120}) with Eq. (\ref{e121}) we immediately
obtain
\begin{equation}\label{e124}
[ U^\dagger C U]_{\beta \alpha } = \lambda_{\alpha}\delta_{\beta
\alpha }, \qquad(\alpha, \beta = 0,\ldots,3),
%
\end{equation}
or, in matrix form
\begin{equation}\label{e125}
 U^\dagger C U = D,
%
\end{equation}
where $D = \mathrm{diag}\{\lambda_0, \lambda_1, \lambda_2,
\lambda_3\}$ or, explicitly
\begin{equation}\label{e126}
D = \begin{pmatrix}
 \lambda_{0} & 0& 0&0 \\
 0 & \lambda_{1} & 0 &0 \\
 0  & 0& \lambda_{2} & 0\\
 0  & 0& 0 & \lambda_{3}
\end{pmatrix}.
%
\end{equation}
Since $C$ is positive semidefinite, all its eigenvalues are
nonnegative: $\lambda _\mu \geq 0, \, (\mu = 0, \ldots,3)$.
Moreover, since from Eqs. (\ref{e30c},\ref{e38}) follow that
\begin{equation}\label{e126b}
\mathrm{Tr}\{C\} = \mathrm{Tr}\{\Lambda^\dagger H   \Lambda
\}=\mathrm{Tr}\{H\}=2,
%
\end{equation}
then $\lambda_0 + \lambda_1 + \lambda_2 + \lambda_3 =2$. Now we
want to write $C$ in terms of its eigenvalues; to this end we have
to invert Eq. (\ref{e125}) obtaining
\begin{equation}\label{e127}
 C  = U D U^\dagger,
%
\end{equation}
or, in components form
\begin{equation}\label{e128}
\begin{array}{ccl}
C_{ \alpha \beta}  & = & [U D U^\dagger]_{ \alpha \beta}\\
& = & \displaystyle{ \sum_{\mu, \nu}^{0,3}
U_{ \alpha \mu} [D]_{ \mu \nu} U^*_{ \beta \nu } } \\
 & = &\displaystyle{ \sum_{\mu, \nu}^{0,3}
  [\vec{u}_{(\mu)}]_{\alpha} \lambda_\mu \delta_{\mu \nu}
       [\vec{u}_{(\nu)}^*]_{\beta} }\\
 & = &\displaystyle{ \sum_{\mu=0}^{3} \lambda_\mu
  [\vec{u}_{(\mu)}]_{\alpha}
       [\vec{u}_{(\mu)}^*]_{\beta} }.\\
\end{array}
%
\end{equation}
If we indicate with $ \Omega_{(\mu)} \equiv
\vec{u}_{(\mu)}\vec{u}_{(\mu)}^\dagger$ the $4 \times 4$ Hermitian
diad whose elements are
\begin{equation}\label{e129}
\begin{array}{ccl}
[\Omega_{(\mu)}]_{\alpha \beta} & \equiv &
  [\vec{u}_{(\mu)}]_{\alpha}
       [\vec{u}_{(\mu)}^*]_{\beta}\\
       & = & U_{\alpha \mu} U_{\mu \beta}^\dagger,
\end{array}
%
\end{equation}
we can rewrite Eq. (\ref{e128}) in matrix form as
\begin{equation}\label{e131}
C = \sum_{\mu=0}^{3} \lambda_\mu \Omega_{(\mu)}.
%
\end{equation}
It is easy to see that the matrices $\{\Omega_{(\mu)}\}$ are
orthogonal:
\begin{equation}\label{e138}
\begin{array}{lcl}
\displaystyle{ \{ \Omega_{(\alpha)}, \Omega_{(\beta)}\}}
& = & \displaystyle{ \mathrm{Tr} \{
\Omega_{(\alpha)}^\dagger    \Omega_{(\beta)}\}}\\
& = & \displaystyle{ \sum_{\mu,\nu}^{0,3}
  [ \Omega_{(\alpha)}]_{\nu\mu}   [ \Omega_{(\beta)}]_{\mu \nu}}\\
& = & \displaystyle{ \sum_{\mu,\nu}^{0,3}
   U_{\nu \alpha} U_{\alpha \mu}^\dagger   U_{\mu \beta} U_{\beta \nu}^\dagger}\\
& = & \displaystyle{ \sum_{\nu=0}^{3}
    U_{\beta \nu}^\dagger U_{\nu \alpha} \sum_{\mu=0}^{3}U_{\alpha \mu}^\dagger   U_{\mu \beta}}\\
& = & \delta_{\beta \alpha}  \delta_{\alpha \beta}\\
& = & \delta_{\alpha \beta}
\end{array}
%
\end{equation}
where in the second line we have exploited the fact that the $\{
\Omega_{(\alpha)} \}$ are Hermitian.
 We are now close to our goal; let us notice that from Eqs.
(\ref{e44c}-\ref{e34}-\ref{e44f}) follow that
\begin{equation}\label{e132}
\begin{array}{ccl}
 M & = &\displaystyle{\Lambda^\dagger F \Lambda  }\\
 & = &\displaystyle{ \sum_{\mu,\nu}^{0,3} C_{\mu\nu}
  \Lambda^\dagger \left( \sigma_{(\mu)}  \otimes \sigma_{(\nu)}^* \right)\Lambda
  }\\
   & = &\displaystyle{ \sum_{\mu,\nu}^{0,3} C_{\mu\nu}
   \Gamma_{(\mu\nu)}
  }.
\end{array}
%
\end{equation}
Now we insert Eq. (\ref{e131}) in Eq. (\ref{e132}) to obtain
\begin{equation}\label{e133}
\begin{array}{ccl}
 M
  & = &\displaystyle{ \sum_{\alpha=0}^{3} \lambda_\alpha
  \sum_{\mu,\nu}^{0,3}
  [ \Omega_{(\alpha)}]_{\mu\nu}
   \Gamma_{(\mu\nu)}
  }\\
  & \equiv &\displaystyle{ \sum_{\alpha=0}^{3} \lambda_\alpha
  \Phi_{(\alpha)}
  },
\end{array}
%
\end{equation}
where we have defined the four Mueller-Jones matrices
$\Phi_{(\alpha)}, \, (\alpha = 0,\ldots,3) $ as
\begin{equation}\label{e134}
\Phi_{(\alpha)} \equiv   \sum_{\mu,\nu}^{0,3}
  [ \Omega_{(\alpha)}]_{\mu\nu}
   \Gamma_{(\mu\nu)}.
%
\end{equation}
These matrices are real, in fact
\begin{equation}\label{e135}
\begin{array}{lcl}
\Phi_{(\alpha)}^* & =  & \displaystyle{\sum_{\mu,\nu}^{0,3}
  [ \Omega_{(\alpha)}^*]_{\mu\nu}
   \Gamma_{(\mu\nu)}^*}\\
   & =& \displaystyle{\sum_{\mu,\nu}^{0,3}
  [ \Omega_{(\alpha)}]_{\nu\mu}
   \Gamma_{(\nu\mu)}}\\
   & = & \displaystyle{\Phi_{(\alpha)}},
\end{array}
%
\end{equation}
since both $\Omega_{(\alpha)}$ and $\Gamma_{(\nu\mu)}$ are
Hermitian matrices.
 Actually we have still to
demonstrate that the $\{\Phi_{(\alpha)}\}$ are Mueller-Jones
matrices. To do that we need two simple partial results. The first
comes from Eq. (\ref{e58g}) which shows that
\begin{equation}\label{e136}
\begin{array}{lcl}
[\Gamma_{(\mu \nu)}]_{ 00}
& = & \displaystyle{ \mathrm{Tr}\{
 \sigma_{(0)} \sigma_{(\mu)}
 \sigma_{(0)}
 \sigma_{(\nu)}\}}\\
 & = & \displaystyle{\mathrm{Tr}\{
 \sigma_{(\mu)}

 \sigma_{(\nu)}\}/2}\\
  & = & \displaystyle{\delta_{\mu \nu}/2}.
\end{array}
%
\end{equation}
The second result we need is the orthonormality of the
$\{\Phi_{(\alpha)}\}$:
\begin{equation}\label{e137}
\begin{array}{lcl}
\{\Phi_{(\alpha)}, \Phi_{(\beta)}\}
& = & \displaystyle{ \mathrm{Tr}\{
\Phi_{(\alpha)}^\dagger \Phi_{(\beta)}\}}\\
& = & \displaystyle{ \sum_{\mu,\nu}^{0,3} \sum_{\gamma,\tau}^{0,3}
  [ \Omega_{(\alpha)}^*]_{\mu\nu}   [ \Omega_{(\beta)}]_{\gamma \tau}
   \mathrm{Tr}\{\Gamma_{(\mu \nu)} \Gamma_{(\gamma \tau)}  \}}\\
& = & \displaystyle{ \sum_{\mu,\nu}^{0,3} \sum_{\gamma,\tau}^{0,3}
  [ \Omega_{(\alpha)}^*]_{\mu\nu}   [ \Omega_{(\beta)}]_{\gamma \tau}
  \delta_{\mu \gamma}  \delta_{\nu \tau}}\\
& = & \displaystyle{ \sum_{\mu,\nu}^{0,3}
  [ \Omega_{(\alpha)}]_{\nu\mu}   [ \Omega_{(\beta)}]_{\mu \nu}}\\
%
%
%
%
%
%
& = & \displaystyle{ \{ \Omega_{(\alpha)}, \Omega_{(\beta)}\}}\\
& = & \displaystyle{ \delta_{\alpha \beta}}.
\end{array}
%
\end{equation}
It is now straightforward to calculate
 from Eqs. (\ref{e134},\ref{e136})
\begin{equation}\label{e139}
\begin{array}{lcl}
\displaystyle{[\Phi_{(\alpha)}]_{00}}
& = & \displaystyle{ \sum_{\mu,\nu}^{0,3}
  [ \Omega_{(\alpha)}]_{\mu\nu}
   [\Gamma_{(\mu\nu)}]_{00}
 }\\
 & = & \displaystyle{ \sum_{\mu=0}^{3}
  [ \Omega_{(\alpha)}]_{\mu \mu}/2
 }\\
  & = & \displaystyle{ \sum_{\mu=0}^{3}
  U_{\mu \alpha} U_{\alpha \mu}^\dagger/2
 }\\
   & = & 1/2,
\end{array}
%
\end{equation}
while from Eqs. (\ref{e137}) we get $\mathrm{Tr} \{
\Phi_{(\alpha)}^T \Phi_{(\alpha)} \} =1$. A necessary and
sufficient condition for a Mueller matrix $M$ to be a
Mueller-Jones matrix is $\mathrm{Tr} \{M^T M \} = (2M_{00})^2$. In
our case we have
\begin{equation}\label{e140}
\frac{\mathrm{Tr} \{ \Phi_{(\alpha)}^T \Phi_{(\alpha)} \}}{
(2[\Phi_{(\alpha)}]_{00})^2} =1, \qquad(\alpha = 0, \ldots,3),
%
\end{equation}
therefore the $\{ \Phi_{(\alpha)}\}$ are genuine Mueller-Jones
matrices. This step complete the demonstration of the
decomposition theorem. In the next subsection we shall derive this
result once more by explicit construction of the matrices  $\{
\Phi_{(\alpha)}\}$.
\subsection{A step backward: from $M$ to $T$}
Now that we learned how to decompose $M$, we want to make a step
backward in order to see if it is possible to find such a kind of
decomposition for the $2 \times 2$ matrix $J'$ introduced in Eq.
(\ref{e14j}). To this end we seek a different form for the
matrices $\{ \Phi_{\alpha}  \}$. We start by rewriting  Eq.
(\ref{e134}) with the help Eq. (\ref{e44f}) of as
\begin{equation}\label{e141}
\begin{array}{lcl}
\Phi_{(\alpha)} & =  &
 \displaystyle{
 \sum_{\mu,\nu}^{0,3}
 [ \Omega_{(\alpha)}]_{\mu\nu}
\Gamma_{(\mu\nu)}
   }\\
& =  &
 \displaystyle{
 \sum_{\mu,\nu}^{0,3}
 [ \vec{u}_{(\alpha)}]_{\mu}
 [ \vec{u}_{(\alpha)}^*]_{\nu}
\Lambda^\dagger \left( \sigma_{(\mu)} \otimes \sigma_{(\nu)}^*
\right) \Lambda
   }\\
& =  &
 \displaystyle{
\Lambda^\dagger \left( \sum_{\mu=0}^{3} [
\vec{u}_{(\alpha)}]_{\mu} \sigma_{(\mu)} \otimes
 \sum_{\nu=0}^{3} [ \vec{u}_{(\alpha)}^*]_{\nu} \sigma_{(\nu)}^* \right) \Lambda
   }\\
& = &  \displaystyle{ \Lambda^\dagger \left( T_{(\alpha)} \otimes
T_{(\alpha)}^* \right) \Lambda
   },
\end{array}
%
\end{equation}
where we have defined the four $2 \times 2$ Jones matrices $\{
T_{(\alpha)} \}$ as
\begin{equation}\label{e142}
T_{(\alpha)} \equiv \sum_{\mu=0}^{3} [ \vec{u}_{(\alpha)}]_{\mu}
\sigma_{(\mu)}.
%
\end{equation}
The result in Eq. (\ref{e141}) shows once again that the
$\Phi_{(\alpha)}$ are genuine Mueller-Jones matrices. At this
point we can rewrite Eq. (\ref{e133}) as
\begin{equation}\label{e143}
M = \sum_{\alpha=0}^{3} \lambda_{\alpha}\Lambda^\dagger \left(
T_{(\alpha)} \otimes T_{(\alpha)}^* \right) \Lambda,
%
\end{equation}
and compare it with Eq. (\ref{e44}). Then it appears that in the
general case comprising also nondeterministic Mueller matrices the
single Jones matrix $T$ must be substituted by the set of the four
Jones matrices  $\{ T_{(\alpha)} \}$ following the recipe given
above. In the same way, if we \emph{assume a priori} that in the
general case Eq. (\ref{e14j}) must be substituted by
\begin{equation}\label{e144}
J' = \sum_{\alpha=0}^{3} \lambda_{\alpha} T_{(\alpha)} J
T_{(\alpha)}^\dagger ,
%
\end{equation}
and rewrite Eqs. (\ref{e24},\ref{e42}-\ref{e44}), we obtain again
Eq. (\ref{e143}). Then the decomposition of $J'$ we were looking
for has been found. Note that since $\lambda_\alpha \geq 0$, we
can always rewrite Eq. (\ref{e144}) as
\begin{equation}\label{e144b}
\begin{array}{rcl}
J' & = & \displaystyle{ \sum_{\alpha=0}^{3} \lambda_{\alpha}
T_{(\alpha)} J
T_{(\alpha)}^\dagger}\\
 & = & \displaystyle{ \sum_{\alpha=0}^{3} \left( \sqrt{ \lambda_{\alpha}} T_{(\alpha)} \right) J
\left( \sqrt{ \lambda_{\alpha}} T_{(\alpha)}^\dagger \right) }\\
 & \equiv & \displaystyle{ \sum_{\alpha=0}^{3}  A_{(\alpha)}  J
 A_{(\alpha)}^\dagger  }
\end{array}
%
\end{equation}
where we have defined $ A_{(\alpha)} = \sqrt{ \lambda_{\alpha}}
T_{(\alpha)}$. In quantum optics and quantum information  Eq.
(\ref{e144b}) is known as ``Kraus decomposition''.

At this point there are two things to be noted. The first is about
the normalization of the Jones matrices $T_{(\alpha)}$. In fact,
it is easy to see
\begin{equation}\label{e144c}
\begin{array}{rcl}
\mathrm{Tr} \{ T_{(\alpha)}^\dagger T_{(\alpha)} \} & = &
\displaystyle{ \mathrm{Tr} \{ \sum_{\mu=0}^{3} [
\vec{u}_{(\alpha)}]_{\mu}^* \sigma_{(\mu)} \sum_{\nu=0}^{3} [
\vec{u}_{(\alpha)}]_{\nu} \sigma_{(\nu)}  \}
 \}
}\\
& = & \displaystyle{ \sum_{\mu,\nu=0}^{3} U_{\mu \alpha}^* U_{\nu
\alpha} \mathrm{Tr} \{\sigma_{(\mu)}\sigma_{(\nu)} \}
}\\
& = & \displaystyle{ \sum_{\mu=0}^{3} U_{\alpha \mu }^\dagger
U_{\mu \alpha}
}\\
& = & \displaystyle{  [U^\dagger U]_{\alpha \alpha} =1.
}\\
\end{array}
%
\end{equation}
This result may seem surprising because if the $T_{(\alpha)}$ were
unitary, then the result would have been $\mathrm{Tr} \{
T_{(\alpha)}^\dagger T_{(\alpha)} \} = 2$. However, surprising or
not, this result is correct and consistent with the normalization
we adopted. The second thing is about trace-preserving processes.
A Kraus decomposition maintains the trace of the coherency matrix
$J$, if and only if
\begin{equation}\label{e144d}
 \sum_{\alpha=0}^{3}  A_{(\alpha)}^\dagger
 A_{(\alpha)} = I_2,
%
\end{equation}
where $I_2$ is the $2 \times 2$ identity matrix. Let us see
whether this is true or not in our case:
\begin{equation}\label{e144e}
\begin{array}{rcl}
\displaystyle{ \sum_{\alpha=0}^{3}  A_{(\alpha)}^\dagger
 A_{(\alpha)} } & = &
\displaystyle{
 \sum_{\alpha=0}^{3} \lambda_\alpha T_{(\alpha)}^\dagger
 T_{(\alpha)}
}\\
& = & \displaystyle{ \sum_{\alpha=0}^{3} \lambda_\alpha
\sum_{\mu,\nu}^{0,3} [\vec{u}_{(\alpha)}]_\mu^* \sigma_{(\mu)}
[\vec{u}_{(\alpha)}]_\nu \sigma_{(\nu)}
}\\
& = & \displaystyle{ \sum_{\mu,\nu}^{0,3} \sigma_{(\mu)}
\sigma_{(\nu)} \sum_{\alpha=0}^{3} \lambda_\alpha
[\vec{u}_{(\alpha)}]_\nu [\vec{u}_{(\alpha)}]_\mu^*
}\\
& = & \displaystyle{ \sum_{\mu,\nu}^{0,3} \sigma_{(\nu)} C_{\nu
\mu} \sigma_{(\mu)}},
\end{array}
%
\end{equation}
where Eq. (\ref{e128}) has been used. From the definition  Eq.
(\ref{e35}), we can write $C_{\nu \mu} = \la c_\nu c^*_\mu \ra$,
where the brackets indicate the average with respect to an
ensemble that represent a generic medium. Then, Eq. (\ref{e144e})
can be rewritten as
\begin{equation}\label{e144f}
\begin{array}{rcl}
\displaystyle{ \sum_{\alpha=0}^{3}  A_{(\alpha)}^\dagger
 A_{(\alpha)} } & = &
\displaystyle{ \sum_{\mu,\nu}^{0,3} \sigma_{(\nu)} C_{\nu \mu}
\sigma_{(\mu)}}\\
& = & \displaystyle{ \sum_{\mu,\nu}^{0,3} \sigma_{(\nu)} \la c_\nu
c^*_\mu \ra
\sigma_{(\mu)}}\\
& = & \displaystyle{ \left\la  \sum_{\mu,\nu}^{0,3} \sigma_{(\nu)}
c_\nu c^*_\mu
\sigma_{(\mu)} \right\ra}\\
& = & \displaystyle{ \left\la  \sum_{\nu=0}^{3} c_\nu
\sigma_{(\nu)} \sum_{\mu=0}^{3}c^*_\mu
\sigma_{(\mu)} \right\ra}\\
& = & \displaystyle{ \left\la  T T^\dagger \right\ra},
\end{array}
%
\end{equation}
where Eq. (\ref{e33}) has been used. The it is clear that for a
\emph{non-depolarizing} medium $ T T^\dagger = I_2$ which implies
\begin{equation}\label{e144g}
 \sum_{\alpha=0}^{3}  A_{(\alpha)}^\dagger
 A_{(\alpha)}= I_2.
%
\end{equation}

In summary, for \emph{any} Mueller $M$ we can calculate the
associate Hermitian matrix $C$. Then, by diagonalizing $C$ we find
its eigenvectors $\{ \vec{u}_{(\alpha)} \}$ whose components
constitutes the Jones matrices $\{ T_{(\alpha)} \}$. Finally we
can find the transformation rule for the covariance matrix $J$ as
in Eq. (\ref{e144}).

A small comment is in order. Until now we have used the matrix $C$
instead of $H$ because it is expressed in terms of measurable
quantities. However, from computational point of view the use of
the matrix $H$ reveals to be more advantageous. This can be seen
in the following manner: let us multiply both sides of  Eq.
(\ref{e118}) by $\Lambda$ and exploit the fact that $\Lambda$ is
unitary:
\begin{equation}\label{e145}
\left( \Lambda C \Lambda^\dagger \right)  \Lambda
\vec{u}_{(\alpha)} = \lambda_{\alpha} \Lambda \vec{u}_{(\alpha)}
\Leftrightarrow H \vec{v}_{(\alpha)} = \lambda_{\alpha}
\vec{v}_{(\alpha)},
\qquad(\alpha = 0,\ldots,3),
%
\end{equation}
where Eq. (\ref{e38}) has been used and we have written the
eigenvectors $\vec{v}_{(\alpha)}$ of $H$ as
\begin{equation}\label{e146}
\vec{v}_{(\alpha)} =  \Lambda \vec{u}_{(\alpha)},
\qquad(\alpha = 0,\ldots,3),
%
\end{equation}
At this point we can jump directly to Eq. (\ref{e142}) to write
$T_{(\alpha)}$ in terms of $\vec{v}_{(\alpha)}$ as
\begin{equation}\label{e147}
\begin{array}{lcl}
T_{(\alpha)} & =  &
 \displaystyle{
\sum_{\mu=0}^{3} [ \vec{u}_{(\alpha)}]_{\mu} \sigma_{(\mu)}
   }\\
& =  &
 \displaystyle{
\sum_{\mu=0}^{3} [ \Lambda^\dagger \vec{v}_{(\alpha)}]_{\mu}
\sigma_{(\mu)}
   }\\
& =  &
 \displaystyle{
\sum_{\mu, \nu}^{0,3} \sigma_{(\mu)} \Lambda^\dagger_{\mu \nu} [
\vec{v}_{(\alpha)}]_{\nu}
   }\\
& =  &
 \displaystyle{
\sum_{\nu=0}^{3} \epsilon_{(\nu)}  [ \vec{v}_{(\alpha)}]_{\nu}
   },
\end{array}
%
\end{equation}
where Eq. (\ref{e14}) has been used. It is clear then, that the
representation of $T_{(\alpha)}$ in the basis $\{\epsilon_{(\mu)}
\}$ is very simple, being
\begin{equation}\label{e148}
T_{(\alpha)} = \begin{pmatrix}
  [ \vec{v}_{(\alpha)}]_{0} & [ \vec{v}_{(\alpha)}]_{1} \\
  [ \vec{v}_{(\alpha)}]_{2} & [ \vec{v}_{(\alpha)}]_{3}
\end{pmatrix},
\qquad(\alpha = 0,\ldots,3),
%
\end{equation}
which is very advantageous from computational point  of view.
\section{Mueller matrix in the standard basis}
In this Section we introduce a new Mueller matrix $\mathcal{M}$
defined with respect to the standard basis.
Let $J$ and $J'$ be the covariance matrices that describe the
input and output light beams entering and leaving a given optical
system, respectively.
 We assume the system to be a
\emph{linear}, \emph{passive} optical element described by the
linear map $\mathcal{M}$:
\begin{equation}\label{e150}
\mathcal{M}:J \rightarrow J' = \mathcal{M}[J].
%
\end{equation}
The above \emph{linear} relation can be explicitly written in
terms of cartesian components as
\begin{equation}\label{e160}
 J_{ij}' = \mathcal{M}_{ij,kl}J_{kl}, \qquad (i,j,k,l \in \{ 0,1\}),
%
\end{equation}
or, by using the Rule
\begin{equation}\label{e170}
 J_{\mu}' = \mathcal{M}_{\mu \nu}J_{\nu}, \qquad (\mu, \nu \in \{0, 1,2, 3 \}),
%
\end{equation}
where $\mu = 2i+j$, $\nu  = 2k+l$, and $ J_{\alpha} =
\{\epsilon_{(\alpha)}, J\} = \mathrm{Tr} \{\epsilon_{(\alpha)}^T
J\}$ are the components of the covariance matrix $J$ with respect
to the standard basis $\{\epsilon_{(\alpha)}\} , \, (\alpha = 0,
\ldots,3)$. Equation (\ref{e170}) is analogous to Eq.
(\ref{e14k}), the difference being that the former is written with
respect to the standard basis, while the latter with respect to
the Pauli basis. Then, it is clear that $\mathcal{M}_{\mu \nu}$ is
just the Mueller matrix written in the standard basis. This
statement can be easily proved by calculating
\begin{equation}\label{e180}
 \begin{array}{ccl}
J_{\mu} & = & \{\epsilon_{(\mu)}, J \} \\
& = & \mathrm{Tr}\{\epsilon_{(\mu)}^\dagger J \}\\
& = & \Lambda_{\mu \nu}\mathrm{Tr}\{\sigma_{(\nu)} J \}\\
& \equiv & \Lambda_{\mu \nu} S_\nu,
\end{array}
%
\end{equation}
where  Eq. (\ref{e14}) was used in the third line (in fact, we
have just rewritten Eq. (\ref{e32})). Now, if we insert Eq.
(\ref{e180}) for both $J_\nu$ and $J_\mu'$ into Eq. (\ref{e170})
we obtain
\begin{equation}\label{e190}
\Lambda_{\mu \alpha} S_\alpha'  =  \mathcal{M}_{\mu \nu}
\Lambda_{\nu \beta} S_\beta ,
%
\end{equation}
which reads, in vectorial form
\begin{equation}\label{e200}
\Lambda \vec{S}'  =  \mathcal{M} \Lambda \vec{S}\Rightarrow
\vec{S}'  = \Lambda^\dagger \mathcal{M} \Lambda \vec{S}.
%
\end{equation}
Since we know that $\vec{S}'  = M \vec{S}$, then from Eq.
(\ref{e200}) it straightforwardly follows the desired relation
between $\mathcal{M}$ and $M$:
\begin{equation}\label{e210}
M  = \Lambda^\dagger \mathcal{M} \Lambda.
%
\end{equation}
Finally, from Eqs. (\ref{e25},\ref{e44c}) it follows that
\begin{equation}\label{e215}
 \mathcal{M} = F, \qquad H = \mathrm{Per}[\mathcal{M}].
%
\end{equation}
It is possible to write $\mathcal{M}$ directly in terms of the
matrix elements of $H$. To this end, let us indicate with $\{
E_{(\mu \nu)} \}$ the standard basis in $\mathbb{R}^{4 \times 4}$
defined as
\begin{equation}\label{e215b}
[E_{(\mu \nu)}]_{\alpha \beta} = \delta_{\mu \alpha} \delta_{\nu
\beta}.
%
\end{equation}
An explicit calculation shows that
\begin{equation}\label{e215c}
\mathrm{Per}[E_{(\mu \nu)}] = \epsilon_{(\mu)} \otimes
\epsilon_{(\nu)}.
%
\end{equation}
However, this equality can also be easily proved in the following
way: Let us write
\begin{equation}\label{e215c1}
 \begin{array}{ccl}
[E_{(\mu \nu)}]_{\alpha \beta} & = & \delta_{\mu \alpha}
\delta_{\nu
\beta}\\
 & = & [\epsilon_{(\mu)}]_{\alpha}
[\epsilon_{(\nu)}]_{\beta}
\end{array}
%
\end{equation}
where Eq. (\ref{e10b}) has been used. Now we can use the Rule to
write $\alpha = 2i +j$ and $\beta = 2k +l$ and rewrite Eq.
(\ref{e215c1}) as
\begin{equation}\label{e215c2}
 \begin{array}{ccl}
[E_{(\mu \nu)}]_{\alpha \beta} & = & [E_{(\mu \nu)}]_{ij,kl}\\
 & = & [\epsilon_{(\mu)}]_{ij}
[\epsilon_{(\nu)}]_{kl}\\
 & = & [\epsilon_{(\mu)} \otimes
 \epsilon_{(\nu)}]_{ik, jl}\\
 & = & \left[ \mathrm{Per}[\epsilon_{(\mu)} \otimes
 \epsilon_{(\nu)}] \right]_{ij, kl}\\
 & = & \left[ \mathrm{Per}[\epsilon_{(\mu)} \otimes
 \epsilon_{(\nu)}] \right]_{\alpha \beta}\\
\end{array}
%
%
\end{equation}
where Eq. (\ref{e25}) has been used. By comparing the first and
the last row of Eq. (\ref{e215c2}) we obtain
\begin{equation}\label{e215c3}
 E_{(\mu \nu)}=  \mathrm{Per}[\epsilon_{(\mu)} \otimes
 \epsilon_{(\nu)}].
 %
\end{equation}
Since for an arbitrary matrix $A \in \mathbb{C}^{4 \times 4}$ the
following relations hold
\begin{equation}\label{e215d}
\mathrm{Per}[\mathrm{Per}[A]] = A, \qquad A = \sum_{\alpha,
\beta}^{0,3} A_{\alpha \beta} E_{(\alpha  \beta)},
%
\end{equation}
then Eq. (\ref{e215c}) follows and, moreover, we can write
\begin{equation}\label{e215e}
 \begin{array}{ccl}
\mathcal{M} & = & \displaystyle{\mathrm{Per}[H]} \\
            & = & \displaystyle{\sum_{\alpha,
\beta}^{0,3} H_{\alpha \beta} \mathrm{Per}[ E_{(\alpha  \beta)}] }\\
            & = &\displaystyle{ \sum_{\alpha,
\beta}^{0,3} H_{\alpha \beta} \left( \epsilon_{(\alpha)} \otimes
\epsilon_{(\beta)} \right)},
\end{array}
%
\end{equation}
which is just the sought relation.
\subsection{$\mathcal{M}$ as a positive map}
In this subsection, we assume that the linear map $\mathcal{M}$ is
a completely positive (CP) map. In this case we can write the
transformation law of $J$ as a Kraus decomposition:
\begin{equation}\label{e220}
J' = \sum_{\alpha=0}^3 A_{(\alpha)} J  A_{(\alpha)}^\dagger.
%
\end{equation}
In the standard basis
\begin{equation}\label{e230}
 A_{(\alpha)} = \sum_{\beta=0}^3 \epsilon_{(\beta)} \mathcal{A}_{\beta
 \alpha},
%
\end{equation}
where
\begin{equation}\label{e240}
\mathcal{A}_{\beta \alpha} \equiv [A_{(\alpha)}]_\beta =
\mathrm{Tr} \{ \epsilon_{(\beta)}^\dagger A_{(\alpha)} \}.
%
\end{equation}
If we substitute Eq. (\ref{e230}) into   Eq. (\ref{e220}) we
obtain
\begin{equation}\label{e250}
 \begin{array}{ccl}
J' & = & \displaystyle{\sum_{\alpha=0}^3 A_{(\alpha)} J  A_{(\alpha)}^\dagger }\\
& = & \displaystyle{ \sum_{\alpha=0}^3 \left( \sum_{\beta=0}^3
\epsilon_{(\beta)} \mathcal{A}_{\beta
 \alpha} \right)  J  \left( \sum_{\gamma=0}^3 \epsilon_{(\gamma)}^\dagger \mathcal{A}_{\gamma
 \alpha}^* \right)  }\\
& = & \displaystyle{  \sum_{\alpha,\beta,\gamma}
\mathcal{A}_{\beta \alpha} \mathcal{A}_{
 \alpha \gamma}^\dagger \left( \epsilon_{(\beta)}   J  \epsilon_{(\gamma)}^\dagger \right)}\\
& = & \displaystyle{\sum_{\beta,\gamma} \left( \mathcal{A}
\mathcal{A}^\dagger\right)_{\beta \gamma} \left(
\epsilon_{(\beta)} J \epsilon_{(\gamma)}^\dagger \right) }\\
&\equiv & \displaystyle{\sum_{\beta,\gamma} \chi_{\beta \gamma}
\left( \epsilon_{(\beta)} J \epsilon_{(\gamma)}^\dagger \right) },
\end{array}
%
\end{equation}
where we have defined the  Hermitian, positive semidefinite  $4
\times 4$ matrix $\chi$ as:
\begin{equation}\label{e260}
\chi \equiv  \mathcal{A} \mathcal{A}^\dagger.
%
\end{equation}
Now, in order to compare   Eq. (\ref{e250}) with Eq. (\ref{e160})
we have to write the latter in terms of cartesian components as
\begin{equation}\label{e270}
 \begin{array}{ccl}
J_{ij}' & = &  \displaystyle{\sum_{\beta,\gamma} \chi_{\beta
\gamma} \left( \epsilon_{(\beta)} J \epsilon_{(\gamma)}^\dagger
\right)_{ij} }\\
& = &  \displaystyle{\sum_{\beta,\gamma} \chi_{\beta \gamma} [
\epsilon_{(\beta)}]_{ik} J_{kl} [\epsilon_{(\gamma)}^\dagger]_{lj}
 }\\
 & = &  \displaystyle{\left\{ \sum_{\beta,\gamma} \chi_{\beta \gamma} [
\epsilon_{(\beta)}]_{ik} [\epsilon_{(\gamma)}]_{jl} \right\}
J_{kl}
 }\\
  & \equiv &  \displaystyle{\mathcal{M}_{ij,kl}
J_{kl}
 },
\end{array}
%
\end{equation}
where
\begin{equation}\label{e280}
 \begin{array}{ccl}
\displaystyle{\mathcal{M}_{ij,kl}} & = &
 \displaystyle{ \sum_{\beta,\gamma} \chi_{\beta \gamma} [
\epsilon_{(\beta)}]_{ik} [\epsilon_{(\gamma)}]_{jl}
 }\\
 & = &
 \displaystyle{ \sum_{\beta,\gamma} \chi_{\beta \gamma} [
\epsilon_{(\beta)} \otimes \epsilon_{(\gamma)}]_{ij,kl}
 },
\end{array}
%
\end{equation}
or, in matrix form
\begin{equation}\label{e290}
\mathcal{M} =
 \sum_{\beta,\gamma} \chi_{\beta \gamma} \left(
\epsilon_{(\beta)} \otimes \epsilon_{(\gamma)}\right).
%
\end{equation}
This Equation should be compared with Eq. (\ref{e215e}) to write
the identity
\begin{equation}\label{e295}
H = \chi.
%
\end{equation}
Therefore, we conclude that when $\mathcal{M}$ is a completely
positive map, its associated $H$ matrix is positive semidefinite.

At this point it may be instructive to write explicitly the
relation between $\mathcal{M}$ and $\chi$ (or $H$) in terms of
their elements.  Since
\begin{equation}\label{e300}
[\epsilon_{(\mu)}]_{ij} = \delta_{\mu,2i+j},
%
\end{equation}
then
\begin{equation}\label{e310}
 \begin{array}{ccl}
\displaystyle{\mathcal{M}_{ij,kl}} & = &
 \displaystyle{ \sum_{\beta,\gamma} \chi_{\beta \gamma} [
\epsilon_{(\beta)}]_{ik} [\epsilon_{(\gamma)}]_{jl}
 }\\
 & = &
 \displaystyle{ \sum_{\beta,\gamma} \chi_{\beta \gamma}
 \delta_{\beta,2i+k} \delta_{\gamma,2j+l}
 }\\
 & = &
 \displaystyle{ \chi_{2i+k,2j+l}},
\end{array}
%
\end{equation}
or, in matrix form
\begin{equation}\label{e320}
\chi = H = \begin{pmatrix}
  \mathcal{M}_{00,00} & \mathcal{M}_{00,01} & \mathcal{M}_{01,00} & \mathcal{M}_{01,01} \\
  \mathcal{M}_{00,10} & \mathcal{M}_{00,11} & \mathcal{M}_{01,10} & \mathcal{M}_{01,11} \\
  \mathcal{M}_{10,00} & \mathcal{M}_{10,01} & \mathcal{M}_{11,00} & \mathcal{M}_{11,01} \\
  \mathcal{M}_{10,10} & \mathcal{M}_{10,11} & \mathcal{M}_{11,10} & \mathcal{M}_{11,11}
\end{pmatrix}.
%
\end{equation}
As expected, we found again the relation $H = \mathrm{Per}
[\mathcal{M}]$, as it is clear from a visual inspection of Eq.
(\ref{e320}).
\section{Classical Mueller matrices and quantum entangled states \textit{or}:
 Quantum measurement of a classical Mueller matrix}
In this section we deal with the problem of determining the $4
\times 4$ density matrix representing a two-photon state, when the
photon pair is scattered by a ``medium'' classically describable
by a Mueller matrix. Here, with the word ``medium'' we denote any
linear optical device, either deterministic or random, which
scatters the photons. We consider two possible configurations: In
the first one, a single scatterer interacts with only one of the
two photons. In the second configuration there are two spatially
separated media, each of them interacting with a single photon
belonging to the photon pair. The relevant literature for the
problem under consideration is listed below:
\begin{small}
\begin{description}
  \item[[ 7]] A. Peres and D. R. Terno,
  \textit{J. Mod. Opt.} {\bf 50}, 1165 (2003).
  \item[[ 8]] N. H. Lindner, A. Peres, and D. R. Terno, \textit{J. Phys. A}
  {\bf 36}, L449 (2003).
    \item[[ 9]] A. Peres and D. R. Terno,
  \textit{Rev. Mod. Phys.} {\bf 76}, 93 (2004).
  \item[[ 10]] A. Aiello and J. P. Woerdman, \textit{Phys. Rev. A} {\bf 70}, 023808 (2004).
  \item[[ 11]] N. H. Lindner and D. R. Terno, \textit{J. Mod. Opt.}
  {\bf 52}, 1177 (2005).
\end{description}
\end{small}

As it is in the style of these notes, we shall follow a didactic
approach, so all the main formulas will be explicitly calculated
step by step.
\subsection{Rewriting the decomposition theorem}
Let us begin by rewriting Eq. (\ref{e144}) as:
\begin{equation}\label{q10}
J \rightarrow J' = \sum_{\alpha = 0}^3 p_\alpha S_{(\alpha)} J
S_{(\alpha)}^\dagger,
%
\end{equation}
where we have defined
\begin{equation}\label{q20}
 p_\alpha \equiv \frac{\lambda_\alpha}{2 M_{00}}, \qquad
 S_{(\alpha)} \equiv \sqrt{2 M_{00}} T_{(\alpha)},
%
\end{equation}
in such a way that
\begin{equation}\label{q30}
\sum_{\alpha = 0}^3 p_\alpha =1, \qquad \mathrm{Tr} \{
S_{(\alpha)}^\dagger  S_{(\alpha)} \} = 2 M_{00}, \quad (\alpha =
0,1,2,3),
%
\end{equation}
where Eq. (\ref{e144c}) has been used. Now, we exploit the
isomorphism between the classical covariance matrix $J$  and the
quantum density matrix $\rho$ and make the ansatz that  a single
photon initially prepared in the quantum state $\rho$, after the
interaction with a medium classically described by Eq. (\ref{q10})
can be described by the density matrix $\rho'$ defined as:
\begin{equation}\label{q40}
\rho \rightarrow \rho' = \sum_{\alpha = 0}^3 p_\alpha S_{(\alpha)}
\rho S_{(\alpha)}^\dagger.
%
\end{equation}
\subsection{Single- and two-photon quantum states}
Let us denote with
\begin{equation}\label{q50}
\{ | i \ra \} = \{|0 \ra, |1 \ra \}, \qquad (i = 0,1),
%
\end{equation}
the basis kets representing two orthogonal linear polarization
states of a photon. These states are often indicated as horizontal
$|H\ra$ and vertical $|V\ra$, respectively. Here we follow the
convention
\begin{equation}\label{q60}
 | 0 \ra = |H \ra, \quad |1\ra = |V\ra.
%
\end{equation}
By definition these states form an orthonormal and complete basis:
\begin{equation}\label{q65}
\la i | j \ra = \delta_{ij}, \quad (i,j \in \{0,1\}), \qquad
\sum_{i=0}^1 |i \ra \la i | = \hat{1}.
%
\end{equation}
As usual, we put them in correspondence with the standard basis in
$\mathbb{R}^2$ $\{ \vec{f}_{(i)} \}$:
\begin{equation}\label{q70}
 | 0 \ra \doteq \vec{f}_{(0)}=\left(%
\begin{array}{c}
  1 \\
  0 \\
\end{array}%
\right) , \qquad  | 1 \ra \doteq \vec{f}_{(1)} =\left(%
\begin{array}{c}
  0 \\
  1 \\
\end{array}%
\right).
%
\end{equation}
In a similar manner, the dual basis $\{ \la i | \}, \; (i=0,1)$ is
associated with $\{ \vec{f}_{(i)}^{\, \dagger} \}$:
\begin{equation}\label{q80}
 \la 0 | \doteq \vec{f}_{(0)}^{\, \dagger} = (%
\begin{array}{cc}
  1 & 0 \\
\end{array}%
) , \qquad  \la 1 | \doteq \vec{f}_{(1)}^{\, \dagger} = (%
\begin{array}{cc}
  0 & 1 \\
\end{array}%
).
%
\end{equation}
The two-photon polarization \emph{standard} basis can be built by
tacking the direct product between single photon states, as
follows:
\begin{equation}\label{q90}
| \alpha = 2i +j \ra =|i \ra \otimes |j\ra \equiv  |i j \ra, \quad
(i,j \in \{0,1\}, \; \alpha \in \{0, \ldots 3 \}),
%
\end{equation}
where the Rule has been used to write $\alpha = 2i +j$. It is
straightforward to show that
\begin{equation}\label{q100}
\begin{array}{lcl}
  \la \alpha | \beta \ra & = & (\la i | \otimes \la j|)(| k \ra \otimes |l \ra) \\
   & = & \la i |k \ra  \la j |l \ra \\
   & = & \delta_{ik} \delta_{jl} \\
   & = & \delta_{2i + j,2k+l}  \\
   & = & \delta_{\alpha \beta}.  \\
\end{array}
%
\end{equation}
In the literature it is often used the so-called Bell basis $\{
|b_{(\alpha)} \ra \}$ defined as
\begin{equation}\label{q110}
|b_{(\alpha)} \ra  = \hat{B} | \alpha \ra, \qquad (\alpha = 0,
\ldots,3),
%
\end{equation}
where the unitary operator $\hat{B}$ is represented with respect
to the standard basis $\{ |\alpha \ra \}$ by the unitary matrix
$B$
\begin{equation}\label{q120}
B = \frac{1}{\sqrt{2}} \left(%
\begin{array}{cccc}
  1 & 0 & 0 & 1 \\
  1 & 0 & 0 & -1 \\
  0 & 1 & 1 & 0 \\
  0 & 1 & -1 & 0 \\
\end{array}%
\right).
%
\end{equation}
In explicit form we have
\begin{equation}\label{q130}
\begin{array}{lclcl}
  |\psi^+ \ra & = &  |b_{(0)} \ra & = & \frac{1}{\sqrt{2}} \left( |00\ra + |11\ra \right),\\
  |\psi^- \ra & = &  |b_{(1)} \ra & = & \frac{1}{\sqrt{2}} \left( |00\ra - |11\ra \right),\\
  |\phi^+ \ra & = &  |b_{(2)} \ra & = & \frac{1}{\sqrt{2}} \left( |01\ra + |10\ra \right),\\
  |\phi^- \ra & = &  |b_{(3)} \ra & = & \frac{1}{\sqrt{2}} \left( |01\ra - |10\ra \right),\\
\end{array}%
%
\end{equation}
where the first column displays the most common notation for the
Bell states.

Four single-photon operators $\{ \hat{\epsilon}_{(\alpha)}, \;
(\alpha = 0,\ldots, 3) \}$ may be formed by tacking the direct
product between a single-photon bra and a single-photon ket as
follows:
\begin{equation}\label{q140}
\hat{\epsilon}_{(\alpha)} \equiv |i \ra \la j|, \qquad (\alpha = 2
i +j; \; \; i,j \in \{0,1\}).
%
\end{equation}
These operators can be straightforwardly put in a one-to-one
correspondence with the elements of the standard basis
$\{\epsilon_{(\alpha)} \}$ in $\mathbb{R}^{2 \times 2}$:
\begin{equation}\label{q150}
\hat{\epsilon}_{(\alpha)} \doteq \epsilon_{(\alpha)} =
\vec{f}_{(i)} \otimes \vec{f}_{(j)}^{\, \dagger}, \qquad (\alpha =
2 i +j; \; \; i,j \in \{0,1\}),
%
\end{equation}
where
\begin{equation}\label{q160}
\begin{array}{lcl}
[\epsilon_{(\alpha)}]_{kl} & = &  [\vec{f}_{(i)} \otimes \vec{f}_{(j)}^{\, \dagger}]_{kl}\\
   & = &  [\vec{f}_{(i)}]_k [\vec{f}_{(j)}^*]_{l} \\
   & = &  \delta_{ik} \delta_{jl} \\
   & = &  \delta_{2i+j, 2k +l} \\
   & = &  \delta_{\alpha \beta},
\end{array}%
%
\end{equation}
where $\beta = 2k + l$, in agreement with Eq. (\ref{e10b}).
\subsection{Two-photon density matrix and scattering processes}
 An arbitrary two-photon state can be described by a
density operator $\hat{\rho}$ as
\begin{equation}\label{q170}
\begin{array}{lcl}
 \displaystyle{ \hat{\rho}} & = &
  \displaystyle{ \sum_{\alpha, \beta}^{0,3} \mathcal{D}_{\alpha \beta} | \alpha \ra \la \beta | }\\
   & = &  \displaystyle{ \sum_{i,j,k,l}^{0,1} \mathcal{D}_{ij,kl} | ij \ra \la kl | } \\
   & = &  \displaystyle{ \sum_{i,j,k,l}^{0,1} \mathcal{D}_{ij,kl} | i \ra \la k | \otimes |j \ra \la l | }  \\
   & = &  \displaystyle{ \sum_{i,j,k,l}^{0,1} { \widetilde{\mathcal{D}}}_{ik,jl} | i \ra \la k | \otimes |j \ra \la l | } \\
   & = &
  \displaystyle{ \sum_{\mu, \nu}^{0,3} { \widetilde{ \mathcal{D}}}_{\mu \nu} \hat{\epsilon}_{(\mu)}\otimes \hat{\epsilon}_{(\nu)}},\\
\end{array}%
%
\end{equation}
where $\mu = 2i+k, \; \nu = 2j + l$ and
\begin{equation}\label{q180}
\widetilde{\mathcal{D}}_{ik,jl} = \mathcal{D}_{ij,kl} \quad
\Leftrightarrow \quad \widetilde{\mathcal{D}} =
\mathrm{Per}[\mathcal{D}].
%
\end{equation}
At this point we can work directly with the matrix representation
of the operators and deal with the density matrix $\rho$
corresponding to the operator $\hat{\rho}$:
\begin{equation}\label{q190}
\hat{\rho} \doteq \rho = \sum_{\mu, \nu}^{0,3}
{\widetilde{\mathcal{D}}}_{\mu \nu} {\epsilon}_{(\mu)}\otimes
{\epsilon}_{(\nu)},
%
\end{equation}
where  Eq. (\ref{q150}) has been used. Before going ahead, we need
to derive two intermediate results. The first one is a simple
calculation: Because of the completeness of the Pauli basis we can
always write:
\begin{equation}\label{q200}
\sigma_{(\alpha)} \sigma_{(\mu)} \sigma_{(\beta)} = \sum_{\nu=0}^3
K_{\alpha \mu \beta \nu} \sigma_{(\nu)},
%
\end{equation}
where, by definition
\begin{equation}\label{q210}
K_{\alpha \mu \beta \nu} = \mathrm{Tr} \{ \sigma_{(\alpha)}
\sigma_{(\mu)} \sigma_{(\beta)} \sigma_{(\nu)} \} = [\Gamma_{(\mu
\nu)}]_{\alpha \beta},
%
\end{equation}
where Eq. (\ref{e58g}) has been used. Moreover, we note that from
the definition (\ref{q210}) it immediately follows that
\begin{equation}\label{q215}
\begin{array}{lcl}
 \displaystyle{ [\Gamma_{(\mu
\nu)}]_{\alpha \beta}} & = &
  \displaystyle{\mathrm{Tr} \{ \sigma_{(\alpha)}
\sigma_{(\mu)} \sigma_{(\beta)} \sigma_{(\nu)} \} }\\
& = &
  \displaystyle{\mathrm{Tr} \{ \sigma_{(\beta)} \sigma_{(\nu)} \sigma_{(\alpha)}
\sigma_{(\mu)}   \} }\\
& = &
 \displaystyle{ [\Gamma_{(\nu
\mu)}]_{ \beta \alpha}}.
\end{array}%
%
\end{equation}
The second result we need is also a simple calculation: First,
from Eq. (\ref{q20}) we write
\begin{equation}\label{q220}
S_{(\alpha)} = \sum_{\beta =0}^3 S_{\alpha \beta}
\sigma_{(\beta)},
%
\end{equation}
where, by definition,
\begin{equation}\label{q230}
\begin{array}{lcl}
 \displaystyle{ S_{\alpha \beta}} & = &
  \displaystyle{\mathrm{Tr} \{  \sigma_{(\beta)}  S_{(\alpha)} \}}\\
   & = &    \displaystyle{ \sqrt{2 M_{00} } \, \sum_{\mu = 0}^3[\vec{u}_{(\alpha)}]_\mu \mathrm{Tr}
    \{  \sigma_{(\beta)}  \sigma_{(\mu)} \}} \\
   & = &   \displaystyle{ \sqrt{2 M_{00} } \, [\vec{u}_{(\alpha)}]_\beta }  \\
   & = &  \displaystyle{ \sqrt{2 M_{00} } \, U_{\beta \alpha} }, \\
\end{array}%
%
\end{equation}
and Eqs. (\ref{q20},\ref{e142},\ref{e122}) have been used. Then,
by using Equations (\ref{e14},\ref{q210},\ref{q230}) we can
calculate the following quantity that will be used later:
\begin{equation}\label{q240}
\begin{array}{lcl}
 \displaystyle{ \sum_{\gamma = 0}^3 p_\gamma S_{(\gamma)} \epsilon_{(\eta)}S_{(\gamma)}^\dagger} & = &
 \displaystyle{ \sum_{\gamma,\alpha,\beta}^{0,3} p_\gamma S_{\gamma \alpha}  S_{\gamma \beta}^*
 \sigma_{(\alpha)}\epsilon_{(\eta)} \sigma_{(\beta)}} \\
   & = &    \displaystyle{ \sum_{\gamma,\alpha,\beta,\mu}^{0,3} \Lambda_{\mu \eta}^\dagger
    p_\gamma S_{\gamma \alpha}  S_{\gamma \beta}^*
 \sigma_{(\alpha)}\sigma_{(\mu)} \sigma_{(\beta)}} \\
   & = &    \displaystyle{ \sum_{\gamma,\alpha,\beta,\mu, \nu}^{0,3} \Lambda_{\mu \eta}^\dagger
    p_\gamma S_{\gamma \alpha}  S_{\gamma \beta}^*
[\Gamma_{(\mu
\nu)}]_{\alpha \beta} \sigma_{(\nu)}} \\
   & = &    \displaystyle{ \sum_{\alpha,\beta,\mu, \nu, \tau}^{0,3} \Lambda_{\mu \eta}^\dagger
   \left( \sum_{\gamma=0}^{3} p_\gamma S_{\gamma \alpha}  S_{\gamma
   \beta}^*\right)
[\Gamma_{(\mu \nu)}]_{\alpha \beta} \Lambda_{\tau \nu}
\epsilon_{(\tau)}} .
\end{array}%
%
\end{equation}
Moreover, from Eqs. (\ref{e124},\ref{e125},\ref{e127}) it follows
that
\begin{equation}\label{q250}
\begin{array}{lcl}
 \displaystyle{ \sum_{\gamma=0}^{3}
    p_\gamma S_{\gamma \alpha}  S_{\gamma \beta}^*} & = &
  \displaystyle{ \sum_{\gamma=0}^{3} S_{\gamma \alpha}
    \frac{\lambda_\gamma}{2 M_{00}} S_{\gamma \beta}^*} \\
 & = &
  \displaystyle{ \sum_{\gamma=0}^{3} \sqrt{2 M_{00}}U_{\alpha \gamma }
    \frac{\lambda_\gamma}{2 M_{00}} \sqrt{2 M_{00}} U_{ \beta \gamma}^* } \\
 & = &
  \displaystyle{ \sum_{\gamma=0}^{3}U_{\alpha \gamma }
    {\lambda_\gamma}U_{ \gamma \beta}^\dagger } \\
 & = &
  \displaystyle{ \sum_{\gamma, \varsigma}^{0,3}U_{\alpha \gamma }
    {\lambda_\gamma} \delta_{\gamma \varsigma}U_{ \varsigma \beta}^\dagger } \\
 & = &
  \displaystyle{ [U D U^\dagger]_{\alpha \beta}} \\
 & = &
  \displaystyle{ C_{\alpha \beta}} ,
\end{array}%
%
\end{equation}
therefore we can rewrite Eq. (\ref{q240}) as
\begin{equation}\label{q260}
\begin{array}{lcl}
 \displaystyle{ \sum_{\gamma = 0}^3 p_\gamma S_{(\gamma)} \epsilon_{(\eta)}S_{(\gamma)}^\dagger}
 & = &
   \displaystyle{ \sum_{\alpha,\beta,\mu, \nu, \tau}^{0,3} \Lambda_{\mu \eta}^\dagger
   C_{\alpha \beta}
[\Gamma_{(\mu \nu)}]_{\alpha \beta} \Lambda_{\tau \nu}
\epsilon_{(\tau)}}\\
 & = &
   \displaystyle{ \sum_{\alpha,\beta,\mu, \nu, \tau}^{0,3} \Lambda_{\mu \eta}^\dagger
   C_{\alpha \beta}
[\Gamma_{(\nu \mu)}]_{ \beta \alpha} \Lambda_{\tau \nu}
\epsilon_{(\tau)}}\\
 & = &
   \displaystyle{ \sum_{\mu, \nu, \tau}^{0,3} \Lambda_{\mu \eta}^\dagger
  \mathrm{Tr} \{ C
\Gamma_{(\nu \mu)} \} \Lambda_{\tau \nu}
\epsilon_{(\tau)}}\\
 & = &
   \displaystyle{ \sum_{\mu, \nu, \tau}^{0,3} \Lambda_{\tau \nu} M_{\nu \mu}\Lambda_{\mu \eta}^\dagger
\epsilon_{(\tau)}}\\
 & = &
   \displaystyle{ \sum_{\tau=0}^{3} [ \Lambda M
   \Lambda^\dagger]_{\tau \eta}
\epsilon_{(\tau)}}\\
 & = &
   \displaystyle{ \sum_{\tau=0}^{3}  \mathcal{M}_{\tau \eta}
\epsilon_{(\tau)}},
\end{array}%
%
\end{equation}
where Eqs. (\ref{e44k},\ref{e210},\ref{q250}) have been used.

At this point we have collected all the results necessary to
calculate explicitly  the transformation law of the density
matrix:
\begin{equation}\label{q270}
\begin{array}{lcl}
 \displaystyle{ \rho'}
 & = &
   \displaystyle{ \sum_{\gamma=0}^{3} p_\gamma \left( S_{(\gamma)} \otimes I_2 \right)
    \rho \left( S_{(\gamma)}^\dagger \otimes I_2 \right) }\\
 & = &
   \displaystyle{ \sum_{\gamma, \eta, \zeta }^{0,3} \widetilde{\mathcal{D}}_{\eta \zeta} p_\gamma
    \left( S_{(\gamma)} \otimes I_2 \right)
      \left( \epsilon_{(\eta)} \otimes \epsilon_{(\zeta)} \right) \left( S_{(\gamma)}^\dagger \otimes I_2 \right) }\\
 & = &
   \displaystyle{ \sum_{\eta, \zeta }^{0,3} \widetilde{\mathcal{D}}_{\! \eta \zeta} \left( \sum_{\gamma=0 }^{3}  p_\gamma
    S_{(\gamma)}  \epsilon_{(\eta)} S_{(\gamma)}^\dagger  \right) \otimes \epsilon_{(\zeta)}  }\\
 & = &
   \displaystyle{ \sum_{\eta, \zeta ,\tau}^{0,3}  \mathcal{M}_{\tau \eta}\widetilde{\mathcal{D}}_{\! \eta \zeta}
\epsilon_{(\tau)} \otimes \epsilon_{(\zeta)}  }\\
 & = &
   \displaystyle{ \sum_{ \zeta ,\tau}^{0,3} [ \mathcal{M} \widetilde{\mathcal{D}}]_{\tau \zeta}
\epsilon_{(\tau)} \otimes \epsilon_{(\zeta)}  }\\
 & = &
   \displaystyle{ \sum_{ \zeta ,\tau}^{0,3} [ \mathcal{M} \widetilde{\mathcal{D}}]_{\tau \zeta}
\mathrm{Per}[E_{(\tau \zeta)}] }\\
 & = &
   \displaystyle{\mathrm{Per}\left[ \sum_{ \zeta ,\tau}^{0,3} [ \mathcal{M} \widetilde{\mathcal{D}}]_{\tau \zeta}
E_{(\tau \zeta)}\right] }\\
 & = &
   \displaystyle{\mathrm{Per} [  \mathcal{M} \widetilde{\mathcal{D}} ] },
\end{array}%
%
\end{equation}
where Eq. (\ref{e215c3}) has been used. Let us note that, by
definition, from  Eq. (\ref{q190}) it trivially follows that
\begin{equation}\label{q280}
\begin{array}{lcl}
 \displaystyle{ \rho'}
 & = &
   \displaystyle{ \sum_{\alpha, \beta}^{0,3} \widetilde{\mathcal{D}}_{\alpha \beta}'
   \epsilon_{(\alpha)} \otimes \epsilon_{(\beta)}  }\\
 & = &
   \displaystyle{ \sum_{\alpha, \beta}^{0,3} \widetilde{\mathcal{D}}_{\alpha \beta}'
\mathrm{Per}[E_{(\alpha \beta)}] }\\
 & = &
   \displaystyle{ \mathrm{Per}[\widetilde{\mathcal{D}}'] }.
\end{array}%
%
\end{equation}
Finally, by equating Eq.  (\ref{q270}) with Eq.  (\ref{q280}), we
obtain
\begin{equation}\label{q290}
\begin{array}{lcl}
\widetilde{\mathcal{D}}' & = & \mathcal{M} \widetilde{\mathcal{D}}\\
 &  =  & \Lambda M \Lambda^\dagger  \widetilde{\mathcal{D}},
\end{array}%
%
\end{equation}
which, when $\det\{ \widetilde{\mathcal{D}} \} \neq 0$, can be
inverted to give:
\begin{equation}\label{q300}
M = \Lambda^\dagger \widetilde{\mathcal{D}}'
(\widetilde{\mathcal{D}})^{-1} \Lambda.
%
\end{equation}
This result shows that the knowledge given by a \emph{single}
input quantum state (represented in this case by $\mathcal{D}$) is
sufficient to uniquely determine the classical Mueller matrix
representing the scatterer.

Equation (\ref{q290}) relates the \emph{Cartesian} coordinates in
the standard basis of the input and output density matrices $\rho$
and $\rho'$ respectively. However, in the classical Mueller-Stokes
formalism the observables are referred to the Pauli basis rather
than to the standard one. To illustrate this point let us consider
the density matrices $\rho^A$ and $\rho^B$ of two independent
photons
\begin{equation}\label{q310}
\rho^F = \sum_{\alpha = 0}^3 S^F_\alpha \sigma_{(\alpha)}, \qquad
(F = A,B),
%
\end{equation}
and let build the corresponding two-photon density matrix
$\rho^{AB}$ in the usual way:
\begin{equation}\label{q320}
\begin{array}{lcl}
\rho^{AB} & = & \rho^{A} \otimes \rho^{B}\\
 &  =  & \displaystyle{\sum_{\alpha, \beta}^{0,1}  S^A_\alpha S^B_\beta \sigma_{(\alpha)} \otimes
 \sigma_{(\beta)}}\\
 &  \equiv  & \displaystyle{\sum_{\alpha, \beta}^{0,1} D^{AB}_{\alpha \beta} \sigma_{(\alpha)} \otimes
 \sigma_{(\beta)}},
\end{array}%
%
\end{equation}
where we have defined the $16$ two-photon Stokes parameters as:
\begin{equation}\label{q322}
 D^{AB}_{\alpha \beta} \equiv S^A_\alpha S^B_\beta.
%
\end{equation}
 From now on we suppress the superscript $AB$
and we seek the relation between the two $4 \times 4$ matrices
$\mathcal{\widetilde{D}}$ and $D$ defined by the following
relations:
\begin{equation}\label{q330}
\begin{array}{lcl}
\rho & = & \displaystyle{\sum_{\alpha, \beta}^{0,1} D_{\alpha
\beta} \sigma_{(\alpha)} \otimes
 \sigma_{(\beta)}} \\
 & =  & \displaystyle{\sum_{\alpha, \beta}^{0,1} \widetilde{\mathcal{D}}_{\alpha \beta} \epsilon_{(\alpha)} \otimes
 \epsilon_{(\beta)}}.
\end{array}%
%
\end{equation}
By using Eq. (\ref{e14}) it trivially follows
\begin{equation}\label{q340}
\begin{array}{lcl}
\rho & = & \displaystyle{\sum_{\alpha, \beta}^{0,1} D_{\alpha
\beta} \sigma_{(\alpha)} \otimes
 \sigma_{(\beta)}} \\
 &  =  & \displaystyle{\sum_{\alpha, \beta, \mu, \nu}^{0,1}
 \Lambda_{\mu \alpha} D_{\alpha \beta}  \Lambda_{\nu \beta} \epsilon_{(\mu)} \otimes
 \epsilon_{(\nu)}}\\
  &  =  & \displaystyle{\sum_{\mu, \nu}^{0,1}
 [ \Lambda D  \Lambda^T]_{\mu \nu} \epsilon_{(\mu)} \otimes
 \epsilon_{(\nu)}}\\
  &  =  & \displaystyle{\sum_{\mu, \nu}^{0,1}
\widetilde{\mathcal{D}}_{\mu \nu} \epsilon_{(\mu)} \otimes
 \epsilon_{(\nu)}}.
\end{array}%
%
\end{equation}
So, we found
\begin{equation}\label{q350}
\widetilde{\mathcal{D}} = \Lambda D \Lambda^T \quad
\Leftrightarrow \quad D = \Lambda^\dagger  \widetilde{\mathcal{D}} \Lambda^*,%
\end{equation}
where we have used the fact that $ \Lambda^T \Lambda^* = I_4$.
Finally, by multiplying both sides of Eq. (\ref{q290}) from left
by $\Lambda^\dagger$ and from right by $\Lambda^*$ we obtain
\begin{equation}\label{q360}
D' = M D.
\end{equation}
This relation is the ``quantum-equivalent'' to the classical one
relating input and output Stokes vectors. Then, in the Pauli basis
the expression for the Mueller matrix becomes very simple:
\begin{equation}\label{q370}
 M = D' D^{-1}.
\end{equation}

We can use alternatively Eq. (\ref{q300}) or Eq. (\ref{q370}) to
determine what classical Mueller matrix is necessary to achieve a
certain quantum state. For example, suppose that we seek a
scatterer that produces a Maximally Entangled Mixed State (MEMS)
when interacting with an individual photon belonging to an
entangled pair  prepared in the ``singlet state'', namely $|
b_{(3)} \ra$ as given in the last row of Eq. (\ref{q130}). The
output MEMS is characterized by the density matrix \footnotemark
\footnotetext{W. J. Munro, D. F. V. James, A. G. White, and P. G.
Kwiat, \textit{Maximizing the entanglement of two mixed qubits},
 Phys. Rev. A \textbf{64}  R030302 (2001).} in the standard basis
\begin{equation}\label{q380}
{\mathcal{D}} ' = \left(%
\begin{array}{cccc}
  g(\gamma) & 0 & 0 & \gamma/2 \\
  0 & 1-2g(\gamma) & 0 & 0 \\
  0 & 0 & 0 & 0 \\
  \gamma/2 & 0 & 0 & g(\gamma) \\
\end{array}%
\right),
\end{equation}
where
\begin{equation}\label{q390}
g(\gamma) = \left\{
\begin{array}{ll}
  \gamma/2, & \gamma \geq 2/3, \\
  1/3 & \gamma < 2/3, \\
\end{array}
\right.
\end{equation}
while the input singlet state is described by
\begin{equation}\label{q400}
{\mathcal{D}}  = \left(%
\begin{array}{cccc}
  0 & 0 & 0 & 0 \\
  0 & 1/2 & -1/2 & 0 \\
  0 & -1/2 & 1/2 & 0 \\
  0 & 0 & 0 & 0 \\
\end{array}%
\right).
\end{equation}
If we substitute Eq. (\ref{q380}) and Eq. (\ref{q400}) into Eq.
(\ref{q300}) we obtain straightforwardly
\begin{equation}\label{q410}
M  = \left(%
\begin{array}{cccc}
  1 & 0 & 0 & 1-2g(\gamma) \\
  0 & -\gamma & 0 & 0 \\
  0 & 0 & \gamma & 0 \\
  1-2g(\gamma) & 0 & 0 & 1-4g(\gamma) \\
\end{array}%
\right).
\end{equation}

As a last example, we consider the case of an output Werner state
represented by
\begin{equation}\label{q420}
{\mathcal{D}} ' = \left(%
\begin{array}{cccc}
  (1 - p)/4 & 0 & 0 & 0 \\
  0 & (1 + p)/4 & -p/2 & 0 \\
  0 & -p/2 & (1 + p)/4 & 0 \\
  0 & 0 & 0 & (1 - p)/4 \\
\end{array}%
\right),
\end{equation}
and again a singlet input state. In this case it is easy to see
that the required Mueller matrix can be written as
\begin{equation}\label{q430}
M  = \left(%
\begin{array}{cccc}
  1 & 0 & 0 & 0 \\
  0 & p & 0 & 0 \\
  0 & 0 & p & 0 \\
  0 & 0 & 0 & p \\
\end{array}%
\right).
\end{equation}
\subsection{Multi-mode states}
Until now we dealt with two- and four-dimensional Hilbert spaces,
since we considered only polarization degrees of freedom of
photons. However, photons also posses other degrees of freedom
that, although apparently irrelevant, may play an important role.
In this subsection we consider photons as physical systems with
many degrees of freedom, including the polarization ones that will
be regarded as the \emph{relevant} ones.

Let us consider a finite-dimensional ``bare bones'' version of the
electromagnetic field. It consists of $2N$ independent
one-dimensional harmonic oscillators each of them characterized by
two quantum numbers: the ``mode'' number $n \in \{0,1,\ldots,
N-1\}$ and the ``polarization'' number $\alpha \in \{0,1 \}$. For
a given $n$ the two oscillators labelled by the pairs $\{n,\alpha
=0 \}$ and $\{ n, \alpha =1\}$ ``oscillate'' along two mutually
orthogonal directions fixed by the two (possibly complex) unit
vectors $\vepsilon_{n0}$ and $ \vepsilon_{n1}$, respectively:
\begin{equation}\label{q465}
\left( \vepsilon_{n\alpha},\vepsilon_{n\beta} \right) =
\delta_{\alpha \beta}, \qquad (\alpha, \beta \in \{ 0,1\}).
%
\end{equation}
 A third unit
vector $ \vepsilon_{n3} $ orthogonal to the other two remains
automatically fixed by the relation
\begin{equation}\label{q470}
\vepsilon_{n2} = \vepsilon_{n0} \times \vepsilon_{n1}.
%
\end{equation}
It is important to note that in the theory  there is \emph{not} a
third harmonic oscillator labelled by $\{n,\alpha =2 \}$ that
oscillates along  $ \vepsilon_{n2} $. However, from a geometrical
point of view the introduction of $ \vepsilon_{n2} $ is necessary
to write the resolution of the identity in a $3$-dimensional space
as
\begin{equation}\label{q480}
\sum_{\ii = 0}^2 \vepsilon_{n\ii} \vepsilon_{n\ii}^\dagger = I_3,
%
\end{equation}
where $I_3$ is the $3 \times 3$ identity matrix. A set of $N$
projection matrices $\{ \mathcal{P}_n \}$ (and the complementary
ones $\{ \mathcal{Q}_n \}$) projecting onto the \emph{physical}
directions of oscillation of the system, can be easily build as
\begin{equation}\label{q490}
\begin{array}{lcl}
\mathcal{P}_n & = & \displaystyle{\sum_{\alpha=0}^1
\vepsilon_{n\alpha}
\vepsilon_{n\alpha}^\dagger}, \\
\mathcal{Q}_n& = & \displaystyle{ \vepsilon_{n 2} \vepsilon_{n
2}^\dagger},
\end{array}%
%
\end{equation}
and $\mathcal{P}_n + \mathcal{Q}_n = I_3$. Each harmonic
oscillator is characterized by its annihilation and creation
operators $\hat{a}_{n \alpha}$ and $\hat{a}_{n \alpha}^\dagger$
respectively, that satisfy the canonical commutation rules:
\begin{equation}\label{q440}
\left[ \hat{a}_{n \alpha},\hat{a}_{m \beta}^\dagger \right] =
\delta_{nm} \delta_{\alpha \beta}.
\end{equation}
The Hamiltonian of the system is just the sum of the Hamiltonians
of the $2N$ harmonic oscillators:
\begin{equation}\label{q445}
\hat{H} = \frac{1}{2} \sum_{n=0}^{N-1} \sum_{\alpha =0}^1 \omega_n
\left( \hat{a}_{n \alpha}^\dagger  \hat{a}_{n \alpha} + \hat{a}_{n
\alpha} \hat{a}_{n \alpha}^\dagger  \right),
\end{equation}
where $\hbar =1$ and $\omega_n \geq 0$.
The single-particle states $\{ | n \alpha \ra \}$ are built from
the vacuum state $| 0 \ra$ in the usual way
\begin{equation}\label{q450}
| n \alpha \ra = \hat{a}_{n \alpha}^\dagger |0 \ra.
\end{equation}
Finally, the resolution of the identity can  be written as
\begin{equation}\label{q460}
\begin{array}{lcl}
\mathbb{I} & = & \displaystyle{\mathbb{I}_0 + \mathbb{I}_1 + \ldots} \\
& = & \displaystyle{| 0 \ra \la 0 |  + \sum_{n=0}^{N-1}
\sum_{\alpha =0}^1 |n \alpha \ra \la n \alpha |  + \sum \{
\mathrm{multiparticle \; \;  states}\}}.
\end{array}%
%
\end{equation}

Now that our system is well defined, we try to build a Positive
Operator Valued Measure (POVM) in order to determine the
\emph{relevant} density matrix pertaining to the \emph{relevant}
polarization degrees of freedom. Let $\{ \vf_{\, \ii} \}$ denotes
an orthonormal and complete basis in $\mathbb{C}^3$:
\begin{equation}\label{q500}
\left( \vf_\ii, \vf_\jj \right) = \delta_{\ii \jj}, \qquad
\sum_{\ii=0}^2 \vf_\ii \vf_\ii^\dagger = I_3, \qquad (\ii, \jj \in
\{ 0,1,2\}).
%
\end{equation}
By using Eq. (\ref{q480}) for each mode $n$ we can write
\begin{equation}\label{q510}
\begin{array}{lcl}
\displaystyle{\vf_{ \, \ii}} & = & \displaystyle{I_3 \cdot\vf_{ \, \ii} } \\
& = & \displaystyle{\sum_{\jj = 0}^2 \vepsilon_{n\jj} \vepsilon_{n\jj}^\dagger \cdot\vf_{ \, \ii} } \\
& = & \displaystyle{\sum_{\jj = 0}^2 \vepsilon_{n\jj} \left( \vepsilon_{n\jj}, \vf_{ \, \ii} \right)} \\
& \equiv & \displaystyle{\sum_{\jj = 0}^2 \vepsilon_{n\jj} F_{n\jj
\ii}},
\end{array}%
%
\end{equation}
where $F_{n\jj \ii} \equiv \left( \vepsilon_{n\jj}, \vf_{ \, \ii}
\right)$. Then, we define the \emph{physical} vectors $\bff_{n
\ii} $ associated to the mode $n$ as
\begin{equation}\label{q520}
\begin{array}{lcl}
\displaystyle{\bff_{n \ii}} & = & \displaystyle{\mathcal{P}_n \vf_\ii} \\
& = & \displaystyle{\sum_{\alpha = 0}^1 \vepsilon_{n\alpha} \vepsilon_{n\alpha}^\dagger \cdot\vf_{ \ii} } \\
& = & \displaystyle{\sum_{\alpha = 0}^1 \vepsilon_{n\alpha} \left( \vepsilon_{n\alpha}, \vf_{ \ii} \right)} \\
& \equiv & \displaystyle{\sum_{\alpha = 0}^1 \vepsilon_{n\alpha}
F_{n\alpha \ii}},
\end{array}%
%
\end{equation}
These vectors are not of unit length nor mutually orthogonal:
\begin{equation}\label{q530}
\begin{array}{lcl}
\displaystyle{\left( \bff_{n \ii}, \bff_{n \jj} \right)}
& = & \displaystyle{\left( \mathcal{P}_n \vf_\ii, \mathcal{P}_n \vf_\jj \right)} \\
& = & \displaystyle{\left( \vf_\ii, \mathcal{P}_n \vf_\jj \right)} \\
& = & \displaystyle{ \mathcal{P}_{n \ii \jj } },
\end{array}%
%
\end{equation}
where, by definition, $\mathcal{P}_{n \ii \ii } \geq 0$.

Now we are ready to write the \emph{single-mode} operator
$\hat{\mathbf{F}}_{n \ii}$ acting on the physical states of the
system as
\begin{equation}\label{q540}
\displaystyle{\hat{\mathbf{F}}_{n \ii}} = \left\{
\begin{array}{cll}
\displaystyle{ \sum_{\alpha =0}^1 \frac{\bff_{n \ii}}{\sqrt{\left(
\bff_{n \ii}, \bff_{n \ii} \right)}} \left( \bff_{n \ii}
 , \vepsilon_{n \alpha}   \right) \hat{a}_{n \alpha}, } & & \left(
\bff_{n \ii}, \bff_{n \ii} \right) \neq 0,\\
0 & & \left( \bff_{n \ii}, \bff_{n \ii} \right) = 0.
\end{array}%
\right.
%
\end{equation}
Then, we can use this operator to build the \emph{multi-mode}
Hermitian positive semidefinite ``intensity'' operator
$\hat{F}_\ii$ as
\begin{equation}\label{q550}
\begin{array}{lcl}
\displaystyle{\hat{F}_\ii}
& = & \displaystyle{\sum_{n=0}^{N-1} \hat{\mathbf{F}}_{n \ii}^\dagger \cdot \hat{\mathbf{F}}_{n \ii}}\\
& = & \displaystyle{\sum_{n=0}^{N-1} \sum_{\alpha ,\beta}^{0,1}
\left[ \frac{\bff_{n \ii}^\dagger }{\sqrt{\left( \bff_{n \ii},
\bff_{n \ii} \right)}} \left( \bff_{n \ii}, \vepsilon_{n \alpha}
\right)^* \hat{a}_{n \alpha}^\dagger \right] \cdot \left[
\frac{\bff_{n \ii} }{\sqrt{\left( \bff_{n \ii}, \bff_{n \ii}
\right)}} \left(  \bff_{n \ii}, \vepsilon_{n \beta} \right)
\hat{a}_{n \beta} \right]
}\\
& = & \displaystyle{\sum_{n=0}^{N-1} \sum_{\alpha ,\beta}^{0,1}
 \left( \bff_{n \ii},
\vepsilon_{n \alpha} \right)^* \left(  \bff_{n \ii}, \vepsilon_{n
\beta} \right)\hat{a}_{n
\alpha}^\dagger  \hat{a}_{n \beta}}\\
& = & \displaystyle{\sum_{n=0}^{N-1}\sum_{\alpha ,\beta}^{0,1}
\left( \vepsilon_{n \alpha},  \bff_{n \ii} \right) \left(  \bff_{n
\ii}, \vepsilon_{n \beta} \right)\hat{a}_{n
\alpha}^\dagger  \hat{a}_{n \beta}}\\
& = & \displaystyle{\sum_{n=0}^{N-1}\sum_{\alpha ,\beta}^{0,1}
\left( \vepsilon_{n \alpha},  \vf_{n \ii} \right) \left(  \vf_{n
\ii}, \vepsilon_{n \beta} \right)\hat{a}_{n \alpha}^\dagger
\hat{a}_{n \beta}},
\end{array}%
%
\end{equation}
where the last step trivially follows from the fact that
$\mathcal{P}_n \vepsilon_{n \alpha} =  \vepsilon_{n \alpha}$ and,
therefore,
\begin{equation}\label{q560}
\begin{array}{lcl}
\displaystyle{\left( \vepsilon_{n \alpha},  \vf_{n \ii} \right)}
& = &
\displaystyle{\left( \mathcal{P}_n \vepsilon_{n \alpha},  \vf_{n \ii} \right)}\\
& = &
\displaystyle{\left(  \vepsilon_{n \alpha},  \mathcal{P}_n \vf_{n \ii} \right)}\\
& = & \displaystyle{\left(  \vepsilon_{n \alpha}, \bff_{n \ii}
\right)}.
\end{array}%
%
\end{equation}
At this point it is easy to see that the three operators
$\{\hat{F}_0,\hat{F}_1,\hat{F}_2 \}$ form a POVM in the
one-particle space:
\begin{equation}\label{q570}
\begin{array}{lcl}
\displaystyle{\hat{F}}
& = & \displaystyle{\sum_{\ii=0}^{2} \hat{F}_{\ii}}\\
& = & \displaystyle{\sum_{n=0}^{N-1}\sum_{\alpha ,\beta}^{0,1}
\sum_{\ii=0}^{2} \left( \vepsilon_{n \alpha},  \vf_{n \ii} \right)
\left(  \vf_{n \ii}, \vepsilon_{n \beta} \right)\hat{a}_{n
\alpha}^\dagger
\hat{a}_{n \beta}}\\
& = & \displaystyle{\sum_{n=0}^{N-1}\sum_{\alpha ,\beta}^{0,1} (
\vepsilon_{n \alpha},  \sum_{\ii=0}^{2} \vf_{n \ii}  \vf_{n
\ii}^\dagger \cdot \vepsilon_{n \beta} )\hat{a}_{n \alpha}^\dagger
\hat{a}_{n \beta}}\\
& = & \displaystyle{\sum_{n=0}^{N-1}\sum_{\alpha ,\beta}^{0,1}
\left( \vepsilon_{n \alpha}, \vepsilon_{n \beta} \right)\hat{a}_{n
\alpha}^\dagger
\hat{a}_{n \beta}}\\
& = & \displaystyle{\sum_{n=0}^{N-1}\sum_{\alpha =0}^{1}
\hat{a}_{n \alpha}^\dagger
\hat{a}_{n \alpha}}\\
& = & \displaystyle{\hat{N}},
\end{array}%
%
\end{equation}
where $\hat{N}$ is the particle-number operator and Eqs.
(\ref{q465}) and (\ref{q500}) have been used.
\subsection{Reconstruction of the density matrix}
Let $\mathcal{R} = \{\vx, \vy, \vz \}$ be an orthonormal Cartesian
coordinate system in $\mathbb{R}^3$ and let $\mathcal{U}$,
$\mathcal{V}$ and $\mathcal{W}$ three mutually unbiased bases for
$\mathbb{C}^3$ defined as
\begin{equation}\label{q580}
\begin{array}{lcl}
\displaystyle{\mathcal{U} = \{\vu_0, \vu_1, \vu_2 \}}
& = & \displaystyle{\{\vx, \vy, \vz \}},\\
\displaystyle{\mathcal{V} = \{\vv_0, \vv_1, \vv_2 \}}& = &
\displaystyle{\left\{\frac{\vx + \vy}{\sqrt{2}},
\frac{\vx - \vy}{\sqrt{2}}, \vz \right\}},\\
\displaystyle{\mathcal{W} = \{\vw_0, \vw_1, \vw_2 \}}& = &
\displaystyle{\left\{\frac{\vx + \imath \vy}{\sqrt{2}}, \frac{\vx
- \imath \vy}{\sqrt{2}}, \vz \right\}}.
\end{array}%
%
\end{equation}
From a physical point of view, these three bases correspond to the
three pairs of mutually orthogonal polarization directions
($\mathcal{U}, \mathcal{V},\mathcal{W}$: linear
horizontal-vertical, linear $45^\circ$-$135^\circ$, and circular
right-left, respectively), selected by a polarizer whose planar
surface is orthogonal to $\vz$. We want to calculate the Stokes
parameter of a beam of light (either classical or quantum). To
this end, let us imagine to repeat the construction of the POVM
outlined in the previous section for each of the basis set
$\mathcal{U}$, $\mathcal{V}$ and $\mathcal{W}$, thus obtaining
three different POVMs denoted with $\hat{U}_\ii$, $\hat{V}_\ii$
and $\hat{W}_\ii$, respectively. For example, if in Eq.
(\ref{q550}) we substitute $\vf_{n \ii}$ with $\vu_{n \ii}$, we
obtain
\begin{equation}\label{q585}
\displaystyle{\hat{U}_\ii}  =
\displaystyle{\sum_{n=0}^{N-1}\sum_{\alpha ,\beta}^{0,1} \left(
\vepsilon_{n \alpha},  \vu_{n \ii} \right) \left(  \vu_{n \ii},
\vepsilon_{n \beta} \right)\hat{a}_{n \alpha}^\dagger \hat{a}_{n
\beta}, \qquad (\ii = 0,1,2)}.
%
%
\end{equation}
In exactly the same manner we may obtain $\hat{V}_\ii$ and
$\hat{W}_\ii$. As a subsequent step we introduce, in analogy with
classical optics, four Hermitian ``Stokes'' operators defined as
follows:
\begin{equation}\label{q590}
\begin{array}{lcl}
\displaystyle{\hat{S}_{(0)}}
& = & \displaystyle{\frac{1}{\sqrt{2}} \left( \hat{U}_0 + \hat{U}_1 \right)},\\
\displaystyle{\hat{S}_{(1)}}
& = & \displaystyle{\frac{1}{\sqrt{2}} \left( \hat{V}_0 - \hat{V}_1 \right)},\\
\displaystyle{\hat{S}_{(2)}}
& = & \displaystyle{\frac{1}{\sqrt{2}} \left( \hat{W}_0 - \hat{W}_1 \right)},\\
\displaystyle{\hat{S}_{(3)}}
& = & \displaystyle{\frac{1}{\sqrt{2}} \left( \hat{U}_0 - \hat{U}_1 \right)}.\\
\end{array}%
%
\end{equation}
For sake of clarity, we introduce the six operators $\{ \hat{E}_X
\}, \; (X = 0, \ldots, 5)$ defined as
\begin{equation}\label{q600}
\left[%
\begin{array}{c}
  \hat{E}_0 \\
  \hat{E}_1 \\
  \hat{E}_2 \\
  \hat{E}_3 \\
  \hat{E}_4 \\
  \hat{E}_5 \\
\end{array}%
\right] \equiv
\left[%
\begin{array}{c}
  \hat{U}_0 \\
  \hat{U}_1 \\
  \hat{V}_0 \\
  \hat{V}_1 \\
  \hat{W}_0 \\
  \hat{W}_1 \\
\end{array}%
\right],
%
\end{equation}
in such a way that we can rewrite Eq. (\ref{q590}) in a compact
form as
\begin{equation}\label{q610}
\hat{S}_{(\mathcal{A})} = \sum_{X=0}^5 P_{\mathcal{A} X}
\hat{E}_X, \qquad (\mathcal{A} \in \{0, \ldots,3 \}),
%
\end{equation}
where we have defined the $4 \times 6$ matrix $P$ as
\begin{equation}\label{q620}
P \equiv \frac{1}{\sqrt{2}}\left(%
\begin{array}{cccccc}
  1 & 1 & 0 & 0 & 0 & 0 \\
  0 & 0 & 1 & -1 & 0 & 0 \\
  0 & 0 & 0 & 0 & 1 & -1 \\
  1 & -1 & 0 & 0 & 0 & 0 \\
\end{array}%
\right),
%
\end{equation}
and $P P^\dagger = I_4$. It is instructive to write explicitly the
operators $\{ \hat{S}_{(\mathcal{A})}\}$:
\begin{equation}\label{q630}
\begin{array}{lcl}
\displaystyle{\hat{S}_{(\mathcal{A})}} & = &
\displaystyle{\sum_{X=0}^5 P_{\mathcal{A} X}
\hat{E}_X}\\
 & = &
\displaystyle{\sum_{X=0}^5 P_{\mathcal{A} X}
\sum_{n=0}^{N-1}\sum_{\alpha ,\beta}^{0,1} \left( \vepsilon_{n
\alpha},  \vxi_{n x} \right) \left(  \vxi_{n x}, \vepsilon_{n
\beta} \right)\hat{a}_{n \alpha}^\dagger \hat{a}_{n \beta},}
\end{array}%
%
\end{equation}
where $\vxi = \vxi(X) \in \{\vu,\vv,\vw \}$ and $x = x(X) \in \{
0,1\}$. Then, we can rewrite Eq. (\ref{q630}) as
\begin{equation}\label{q640}
\displaystyle{\hat{S}_{(\mathcal{A})}}  =  \displaystyle{
\sum_{n=0}^{N-1}\sum_{\alpha ,\beta}^{0,1} ( \vepsilon_{n \alpha},
\left[ \sum_{X=0}^5 P_{\mathcal{A} X} \vxi_{n x}
 \vxi_{n x}^\dagger \right] \cdot  \vepsilon_{n \beta}
)\hat{a}_{n \alpha}^\dagger \hat{a}_{n \beta},}
%
\end{equation}
where an explicit calculation  shows that
\begin{equation}\label{q650}
\begin{array}{lcl}
\displaystyle{\sum_{X=0}^5 P_{\mathcal{A} X} \vxi_{n x}
 \vxi_{n x}^\dagger} & = & \displaystyle{ \left(%
\begin{array}{ccc}
  \left[ \sigma_{( \mathcal{A}) } \right]_{00} &  \left[ \sigma_{( \mathcal{A})} \right]_{01} & 0 \\
  \left[ \sigma_{( \mathcal{A}) } \right]_{10} &  \left[ \sigma_{( \mathcal{A})} \right]_{11} & 0 \\
  0                             &                              0 & 0 \\
\end{array}%
\right)}\\
& \equiv & \displaystyle{\Omega_{\mathcal{(A)}}},
\end{array}
%
\end{equation}
where $\{ \sigma_{( \mathcal{A})} \}, \; (\mathcal{A} \in \{0,
\ldots,3 \})$ are the $2 \times 2$ Pauli matrices, and Eqs.
(\ref{q580}) and (\ref{q620}) have been used. Finally we can write
\begin{equation}\label{q660}
\begin{array}{lcl}
\displaystyle{\hat{S}_{(\mathcal{A})}} & = & \displaystyle{
\sum_{n=0}^{N-1}\sum_{\alpha ,\beta}^{0,1} ( \vepsilon_{n \alpha},
\Omega_{\mathcal{(A)}}   \vepsilon_{n \beta} )\hat{a}_{n
\alpha}^\dagger
\hat{a}_{n \beta}}\\
& = & \displaystyle{ \sum_{n=0}^{N-1}\sum_{\alpha ,\beta}^{0,1} (
\vvarepsilon_{n \alpha}, \sigma_{\mathcal{(A)}}   \vvarepsilon_{n
\beta} )\hat{a}_{n \alpha}^\dagger \hat{a}_{n \beta},}
\end{array}%
%
\end{equation}
where with $\vvarepsilon_{n \beta} \in \mathbb{C}^{\,2}$ we have
denoted the restriction of $\vepsilon_{n \beta}$ to a
two-dimensional subspace:
\begin{equation}\label{q670}
\vvarepsilon_{n \beta} \equiv \left(
\begin{array}{c}
  \left[ \vepsilon_{n \beta} \right]_0 \\
  \left[ \vepsilon_{n \beta} \right]_1 \\
\end{array}
\right).
%
\end{equation}
Of course, the two-dimensional vectors $\{ \vvarepsilon_{n
\alpha}\}$ are not unit length nor mutually orthogonal. Now we can
use this result to calculate
\begin{equation}\label{q680}
\begin{array}{lcl}
\displaystyle{\la n \nu |\hat{S}_{(\mathcal{A})}| m \mu \ra} & = &
\displaystyle{ \sum_{p=0}^{N-1}\sum_{\alpha ,\beta}^{0,1} (
\vvarepsilon_{p \alpha}, \sigma_{\mathcal{(A)}}   \vvarepsilon_{p
\beta} )\la n \nu | \hat{a}_{p \alpha}^\dagger
\hat{a}_{p \beta}| m \mu \ra}\\
 & = &
\displaystyle{ \sum_{p=0}^{N-1}\sum_{\alpha ,\beta}^{0,1} (
\vvarepsilon_{p \alpha}, \sigma_{\mathcal{(A)}}   \vvarepsilon_{p
\beta} )\la 0 | \hat{a}_{n \nu} \hat{a}_{p \alpha}^\dagger
\hat{a}_{p \beta}\hat{a}_{m \mu}^\dagger| 0 \ra}\\
 & = &
\displaystyle{ \delta_{nm} ( \vvarepsilon_{n \nu},
\sigma_{\mathcal{(A)}} \vvarepsilon_{n \mu} )}.
\end{array}%
%
\end{equation}
At this point we have all the ingredients necessary to calculate
the expectation value $\bigl\la \hat{S}_{(\mathcal{A})} \bigr\ra$
with respect to the generic state described by $\hat{\rho}$:
\begin{equation}\label{q690}
\hat{\rho}  = \sum_{m,n}^{0,N-1} \sum_{\mu, \nu}^{0,1} \rho_{m
\mu,n \nu} | m \mu \ra \la n \nu |.
%
\end{equation}
Then
\begin{equation}\label{q700}
\begin{array}{lcl}
\displaystyle{\bigl\la \hat{S}_{(\mathcal{A})} \bigr\ra} & = &
\displaystyle{ \mathrm{Tr} \bigl\{ \hat{\rho} \hat{S}_{(\mathcal{A})} \bigr\}}\\
& = & \displaystyle{ \sum_{m,n}^{0,N-1} \sum_{\mu, \nu}^{0,1}
\rho_{m
\mu,n \nu} \mathrm{Tr} \bigl\{ | m \mu \ra \la n \nu | \hat{S}_{(\mathcal{A})} \bigr\}}\\
& = & \displaystyle{ \sum_{m,n}^{0,N-1} \sum_{\mu, \nu}^{0,1}
\rho_{m
\mu,n \nu}   \la n \nu | \hat{S}_{(\mathcal{A})} | m \mu \ra}\\
& = & \displaystyle{ \sum_{n=0}^{N-1} \sum_{\mu, \nu}^{0,1}
\rho_{n \mu,n \nu} ( \vvarepsilon_{n \nu},
\sigma_{\mathcal{(A)}} \vvarepsilon_{n \mu} )}\\
& \equiv & \displaystyle{ \sum_{n=0}^{N-1} \mathrm{Tr} \bigl\{
\mathcal{D}_n \sigma_{n \mathcal{(A)}}\bigr\}},
\end{array}%
%
\end{equation}
where we have defined the $2 \times 2$ single-mode matrices
$\mathcal{D}_n$ and $\sigma_{n \mathcal{(A)}}$ as:
\begin{equation}\label{q710}
\begin{array}{lcl}
\displaystyle{\bigl[ \mathcal{D}_n \bigr]_{\alpha \beta}} & = &
\displaystyle{ \rho_{n \alpha,n \beta},}\\
\displaystyle{\bigl[ \sigma_{n \mathcal{(A)}} \bigr]_{\alpha
\beta}} & = & \displaystyle{( \vvarepsilon_{n \alpha},
\sigma_{\mathcal{(A)}} \vvarepsilon_{n \beta} )},
\end{array}%
%
\end{equation}
and $\alpha, \beta \in \{0,1 \}$.

In a \emph{paraxial regime} of propagation there is a ``dominant''
mode of the field, say $n = n_0$, and one can assume
\begin{equation}\label{q712}
( \vvarepsilon_{n \alpha}, \sigma_{\mathcal{(A)}} \vvarepsilon_{n
\beta} ) \cong ( \vvarepsilon_{n_0 \alpha}, \sigma_{\mathcal{(A)}}
\vvarepsilon_{n_0 \beta} ), \qquad \forall n \in \{0, \ldots, N-1
\}.
%
\end{equation}
Since we always have the freedom to choose our reference frame, in
this case it is convenient to choose the two polarization vectors
$\{ \vepsilon_{n_0 \alpha} \}$ associated to the mode $n_0$ in
such a way that:
\begin{equation}\label{q750}
\vepsilon_{n_0 0}   =
 \vx = \left(%
\begin{array}{c}
  1 \\
  0 \\
  0 \\
\end{array}%
\right), \qquad
\vepsilon_{n_0 1}   =
 \vy = \left(%
\begin{array}{c}
  0 \\
  1 \\
  0 \\
\end{array}%
\right).
%
\end{equation}
From the definition (\ref{q720}) it trivially follows
\begin{equation}\label{q760}
\vvarepsilon_{n_0 0}   =
\left(%
\begin{array}{c}
  1 \\
  0 \\
\end{array}%
\right), \qquad
\vvarepsilon_{n_0 1}   =
 \left(%
\begin{array}{c}
  0 \\
  1 \\
\end{array}%
\right),
%
\end{equation}
which implies
\begin{equation}\label{q770}
( \vvarepsilon_{n_0 \alpha}, \sigma_{\mathcal{(A)}}
\vvarepsilon_{n_0 \beta} ) = \left[ \sigma_{\mathcal{(A)}}
\right]_{\alpha \beta}.
%
\end{equation}
Then, from Eqs. (\ref{q700},\ref{q712},\ref{q770}) it follows that
\begin{equation}\label{q714}
\begin{array}{lcl}
\displaystyle{\bigl\la \hat{S}_{(\mathcal{A})} \bigr\ra} & = &
\displaystyle{ \sum_{n=0}^{N-1} \mathrm{Tr} \bigl\{
\mathcal{D}_n \sigma_{n \mathcal{(A)}}\bigr\}}\\
& \cong & \displaystyle{ \sum_{n=0}^{N-1} \mathrm{Tr} \bigl\{
\mathcal{D}_n \sigma_{ \mathcal{(A)}}\bigr\}}\\
& \equiv & \displaystyle{ \mathrm{Tr} \bigl\{ \mathcal{D} \sigma_{
\mathcal{(A)}}\bigr\}},
\end{array}%
%
\end{equation}
where we have defined the $2 \times 2$ matrix $\mathcal{D}$
 as
\begin{equation}\label{q716}
\mathcal{D} \equiv  \displaystyle{ \sum_{n=0}^{N-1}
\mathcal{D}_n}, \qquad \mathrm{or,} \qquad [\mathcal{D} ]_{\alpha
\beta} = \sum_{n=0}^{N-1} \rho_{n \alpha, n \beta},
%
\end{equation}
which coincides with the naive definition of reduced density
matrix.
\subsection{The relevant density matrix}
 At this point we are ready to calculate the
\emph{relevant} density operator $\hat{\rho}^R$ for the
polarization degrees of freedom. However, before doing so, let us
apply the formulas written above to the simple case in which a
single mode of the field, say again $n_0$, is excited. In this
case, by definition
\begin{equation}\label{q720}
\rho_{m \mu, n \nu} = \delta_{m n_0} \delta_{n n_0} \rho_{n_0
\mu,n_0,\nu} \equiv \delta_{m n_0} \delta_{n n_0} \rho_{ \mu
\nu}^0,
%
\end{equation}
and if we substitute Eq. (\ref{q720}) into Eq. (\ref{q700}) we
obtain
\begin{equation}\label{q730}
\hat{\rho}  =  \sum_{\mu, \nu}^{0,1} \rho_{ \mu \nu}^0 | n_0 \mu
\ra \la n_0 \nu |.
%
\end{equation}
From Eq. (\ref{q700}) we can easily calculate
\begin{equation}\label{q740}
\begin{array}{lcl}
\displaystyle{\bigl\la \hat{S}_{(\mathcal{A})} \bigr\ra} & = &
\displaystyle{ \mathrm{Tr} \bigl\{ \hat{\rho} \hat{S}_{(\mathcal{A})} \bigr\}}\\
& = & \displaystyle{  \sum_{\mu, \nu}^{0,1} \rho_{ \mu \nu}^0
\mathrm{Tr}
\bigl\{ | n_0 \mu \ra \la n_0 \nu | \hat{S}_{(\mathcal{A})} \bigr\}}\\
& = & \displaystyle{ \sum_{\mu, \nu}^{0,1} \rho_{
\mu \nu}^0   \la n_0 \nu | \hat{S}_{(\mathcal{A})} | n_0 \mu \ra}\\
& = & \displaystyle{  \sum_{\mu, \nu}^{0,1} \rho_{\mu \nu}^0 (
\vvarepsilon_{n_0 \nu}, \sigma_{\mathcal{(A)}} \vvarepsilon_{n_0
\mu} )}.
\end{array}%
%
\end{equation}
%
By substituting Eq. (\ref{q770}) into Eq. (\ref{q740}) we
immediately obtain
\begin{equation}\label{q780}
\begin{array}{lcl}
\displaystyle{\bigl\la \hat{S}_{(\mathcal{A})} \bigr\ra} & = &
\displaystyle{ \sum_{\mu, \nu}^{0,1}\rho_{ \mu \nu}^0  \left[ \sigma_{\mathcal{(A)}}  \right]_{\nu \mu}}\\
& = & \displaystyle{ \mathrm{Tr} \bigl\{\rho^0
\sigma_{\mathcal{(A)}}\bigr\}}.
\end{array}%
%
\end{equation}
since, by definition, $\mathrm{Tr} \{ \rho^0 \} = \mathrm{Tr}
\{\hat{\rho} \}=1 $, then $\bigl\la \hat{S}_{({0})} \bigr\ra = 1/
\sqrt{2}$ and, after a simple calculation,  we obtain explicitly
\begin{equation}\label{q790}
\hat{\rho} \doteq \rho^0 = \frac{1}{\sqrt{2}} \left(%
\begin{array}{cc}
 \bigl\la \hat{S}_{( {0})} \bigr\ra +
 \bigl\la \hat{S}_{( {3})} \bigr\ra &
 \bigl\la \hat{S}_{( {1})} \bigr\ra - \imath
 \bigl\la \hat{S}_{( {2})} \bigr\ra \\
 \bigl\la \hat{S}_{( {1})} \bigr\ra + \imath
 \bigl\la \hat{S}_{( {2})} \bigr\ra  &
 \bigl\la \hat{S}_{( {0})} \bigr\ra -
 \bigl\la \hat{S}_{( {3})} \bigr\ra \\
\end{array}%
\right).
%
\end{equation}
for later purposes it is useful to define the four scaled real
parameters $\{s_{\mathcal{A}} \}$ as
\begin{equation}\label{q800}
 \quad s_\mathcal{A} = \frac{1}{\sqrt{2}}
\frac{ \bigl\la \hat{S}_{( \mathcal{A})}\bigr\ra}{ \bigl\la
\hat{S}_{( {0})}\bigr\ra}, \qquad (\mathcal{A} \in \{0,\ldots,3
\}),
%
\end{equation}
and rewrite Eq. (\ref{q790}) as
\begin{equation}\label{q810}
\rho^0 = \frac{1}{\sqrt{2}} \left(%
\begin{array}{cc}
 s_0 + s_3 &
 s_1 - \imath s_2 \\
 s_1 + \imath s_2 &
 s_0 - s_3 \\
\end{array}%
\right).
%
\end{equation}

Now we go back to the general multi-mode case. Let us suppose that
we have measured or calculated the four values
$\{\bigl\la\hat{S}_{(\mathcal{A})} \bigr\ra \}$. Two cases are
possible: Either
\begin{equation}\label{q820}
\bigl\la\hat{S}_{(0)} \bigr\ra^2 < \sum_{\mathcal{B} = 1}^3
\bigl\la\hat{S}_{(\mathcal{B})} \bigr\ra^2, \quad \mathrm{or}
\quad \bigl\la\hat{S}_{(0)} \bigr\ra^2 \geq \sum_{\mathcal{B} =
1}^3 \bigl\la\hat{S}_{(\mathcal{B})} \bigr\ra^2.
%
\end{equation}
If the first case occurs, then it is \emph{not} possible to
calculate a relevant density matrix for the system (this may
happen because of unwanted experimental errors). Vice versa, if
the second case occur, then $\hat{\rho}^R$ can be obtained
straightforwardly.
 This result
can be achieved in three steps: First, by using Eq. (\ref{q800})
we calculated the four scaled parameters $\{ s_\mathcal{A} \}$;
second, we introduce the four relevant observables $\{
\hat{\sigma}_{(\mathcal{A})}^R \}$ such that
\begin{equation}\label{q830}
 \mathrm{Tr} \bigl\{ \hat{\rho}^R
\hat{\sigma}_{(\mathcal{A})}^R \bigr\} =  s_\mathcal{A},
%
\end{equation}
where, inspired by Eq. (\ref{q660}), we make the ansatz:
\begin{equation}\label{q840}
\hat{\sigma}_{(\mathcal{A})}^R \doteq \sigma_{(\mathcal{A})},
\qquad (\mathcal{A} \in \{0, \ldots 3 \}).
%
\end{equation}
Finally, we calculate the less biased $\hat{\rho}^R$ by using the
maximum entropy criterion which leads to \footnotemark
\footnotetext{R. Balian, \textit{Incomplete descriptions and
relevant entropies}, Am. J. Phys. \textbf{67} (12), 1078 (1999).}
\begin{equation}\label{q850}
\hat{\rho}^R \doteq \rho^R = \exp \left[ - \Psi -
\sum_{\mathcal{B}=1}^3 \gamma_{\mathcal{B}}
{\sigma}_{(\mathcal{B})}\right],
%
\end{equation}
where the normalization term $\Psi$ ensure the condition
$\mathrm{Tr} \hat{\rho}^R = 1, $ and each of the three lagrange
multipliers $\gamma_{\mathcal{B}}, \; (\mathcal{B} =1,2,3)$ is
associated with each constraint Eq. (\ref{q830}). After a
straightforward calculation it follows that
\begin{equation}\label{q860}
\hat{\rho}^R \doteq \mathcal{D}^R = \sum_{\mathcal{A}=0}^3
s_\mathcal{A} \sigma_{(\mathcal{A})}.
%
\end{equation}
This is our result, now we seek a relation between input and
output relevant density operator in a multi-mode scattering
process.
\subsection{Input and output relations}
Let us consider now a generic \emph{linear} scattering process
that transform the input single-photon density operator
$\hat{\rho}^\mathrm{in}$ into  the output single-photon density
operator $\hat{\rho}^\mathrm{out}$:
\begin{equation}\label{q870}
\begin{array}{lcl}
\displaystyle{\hat{\rho}^\mathrm{out}} & = &
\displaystyle{ \mathcal{L}[\hat{\rho}^\mathrm{in}]}\\
& = & \displaystyle{ \sum_i \hat{A}_i \hat{\rho}^\mathrm{in}
\hat{A}_i^\dagger},
\end{array}%
%
\end{equation}
where
\begin{equation}\label{q880}
\sum_i \hat{A}_i  \hat{A}_i^\dagger = \hat{I}.
%
\end{equation}
The relevant quantities to calculate are the transition amplitudes
\begin{equation}\label{q890}
\la m \mu | \hat{A}_i | n \nu \ra \equiv A_{i,\mu \nu} (m,n),
%
\end{equation}
where with $ A_{i} (m,n) $ we have denoted the  $2 \times 2$
matrix whose elements are $ A_{i,\mu \nu} (m,n)$. Then, we can
rewrite Eq. (\ref{q870}) as
\begin{equation}\label{q900}
\rho_{a \alpha, b \beta}^\mathrm{out} =  \sum_{m,n}^{0,
N-1}\sum_{\mu ,\nu}^{0,1} \sum_i A_{i,\alpha \mu} (a,m) \rho_{m
\mu, n \nu}^\mathrm{in} A_{i,\nu \beta}^\dagger (n,b).
%
\end{equation}
From the algebra of the Pauli matrices it is easy to see that for
any given pair of modes $\{m, n\}$ one can always write
\begin{equation}\label{q910}
\rho_{m \mu, n \nu} = \sum_{\mathcal{A} =0}^3
\mathcal{S}_\mathcal{A}(m,n) [\sigma_{(\mathcal{A})}]_{\mu \nu},
%
\end{equation}
where we have defined
\begin{equation}\label{q920}
\mathcal{S}_\mathcal{A}(m,n) = \sum_{\mu, \nu}^{0,1} \rho_{m \mu,
n \nu} [\sigma_{(\mathcal{A})}]_{\nu \mu}  , \qquad (\mathcal{A}
\in \{ 0, \ldots, 3 \}).
%
\end{equation}
If we use Eq. (\ref{q910}) in both sides of  Eq. (\ref{q900}) we
obtain
\begin{equation}\label{q930}
\begin{array}{lcl}
\displaystyle{\mathcal{S}_\mathcal{A}^\mathrm{out}(a,b)} & = &
\displaystyle{\sum_{\alpha, \beta}^{0,1} \rho_{a \alpha, b
\beta}^\mathrm{out}
[\sigma_{(\mathcal{A})}]_{\beta \alpha}}\\
& = & \displaystyle{\sum_{\alpha, \beta}^{0,1} \sum_{m,n}^{0,
N-1}\sum_{\mu ,\nu}^{0,1} \sum_i A_{i,\alpha \mu} (a,m) \rho_{m
\mu, n \nu}^\mathrm{in} A_{i,\nu \beta}^\dagger (n,b)
[\sigma_{(\mathcal{A})}]_{\beta \alpha}}\\
& = & \displaystyle{\sum_{\alpha, \beta}^{0,1} \sum_{m,n}^{0,
N-1}\sum_{\mu ,\nu}^{0,1} \sum_i A_{i,\alpha \mu} (a,m)
\sum_{\mathcal{B} =0}^3 \mathcal{S}_\mathcal{B}^\mathrm{in}(m,n)
[\sigma_{(\mathcal{B})}]_{\mu \nu}
A_{i,\nu \beta}^\dagger (n,b)
[\sigma_{(\mathcal{A})}]_{\beta \alpha}}\\
& = & \displaystyle{\sum_{\mathcal{B} =0}^3 \sum_{m,n}^{0, N-1}
 \mathcal{S}_\mathcal{B}^\mathrm{in}(m,n)   \sum_i \sum_{\alpha, \beta, \mu ,\nu}^{0,1} A_{i,\alpha \mu} (a,m)
[\sigma_{(\mathcal{B})}]_{\mu \nu}
A_{i,\nu \beta}^\dagger (n,b)
[\sigma_{(\mathcal{A})}]_{\beta \alpha}}\\
& = & \displaystyle{\sum_{\mathcal{B} =0}^3 \sum_{m,n}^{0, N-1}
 \mathcal{S}_\mathcal{B}^\mathrm{in}(m,n)   \sum_i \mathrm{Tr} \left\{ \sigma_{(\mathcal{A})} A_{i} (a,m)
\sigma_{(\mathcal{B})}
A_{i}^\dagger (n,b)
 \right\}}\\
& \equiv & \displaystyle{\sum_{\mathcal{B} =0}^3 \sum_{m,n}^{0,
N-1} M_{\mathcal{A} \mathcal{B} }(a,b,m,n)
\mathcal{S}_\mathcal{B}^\mathrm{in}(m,n)   },
\end{array}%
%
%
\end{equation}
where, by analogy with the definition of a classical Mueller
matrix, we have defined
\begin{equation}\label{q940}
M_{\mathcal{A} \mathcal{B} }(a,b,m,n) \equiv \sum_i \mathrm{Tr}
\left\{ \sigma_{(\mathcal{A})} A_{i} (a,m)
\sigma_{(\mathcal{B})}
A_{i}^\dagger (n,b)
 \right\}.
%
\end{equation}
From Eqs. (\ref{q700})  and (\ref{q910}) it follows that
\begin{equation}\label{q950}
\begin{array}{lcl}
\displaystyle{\bigl\la \hat{S}_{(\mathcal{A})} \bigr\ra}
& = &
\displaystyle{ \sum_{n=0}^{N-1} \sum_{\mu, \nu}^{0,1}
\rho_{n \mu,n \nu} ( \vvarepsilon_{n \nu},
\sigma_{\mathcal{(A)}} \vvarepsilon_{n \mu} )}\\
& = & \displaystyle{\sum_{\mathcal{B} =0}^3 \sum_{n=0}^{N-1}
\sum_{\mu, \nu}^{0,1}
 \mathcal{S}_\mathcal{B}(n,n)
[\sigma_{(\mathcal{B})}]_{\mu \nu} [\sigma_{n(\mathcal{A})}]_{\nu \mu}}\\
& = & \displaystyle{\sum_{\mathcal{B} =0}^3 \sum_{n=0}^{N-1}
\mathcal{S}_\mathcal{B}(n,n) \mathrm{Tr} \left\{
\sigma_{(\mathcal{B})} \sigma_{n(\mathcal{A})} \right\}}\\
& \equiv & \displaystyle{\sum_{\mathcal{B} =0}^3 \sum_{n=0}^{N-1}
\Delta_{\mathcal{A} \mathcal{B}} (n) \mathcal{S}_\mathcal{B}(n,n)
},
\end{array}%
%
\end{equation}
where
\begin{equation}\label{q960}
\Delta_{\mathcal{A} \mathcal{B}} (n) \equiv \mathrm{Tr} \left\{
 \sigma_{n(\mathcal{A})} \sigma_{(\mathcal{B})} \right\}.
%
\end{equation}
Note that in the paraxial approximation (see Eq. (\ref{q712}))
$\Delta_{\mathcal{A} \mathcal{B}} (n) \cong \delta_{\mathcal{A}
\mathcal{B}} $.
By using the results in Eqs. (\ref{q930}) and (\ref{q950}) we can
write
\begin{equation}\label{q970}
\begin{array}{lcl}
\displaystyle{\bigl\la \hat{S}_{(\mathcal{A})}
\bigr\ra^\mathrm{out}}
& = &  \displaystyle{\sum_{\mathcal{B} =0}^3 \sum_{n=0}^{N-1}
\Delta_{\mathcal{A} \mathcal{B}} (n)
\mathcal{S}_\mathcal{B}^\mathrm{out}(n,n) }\\
& = &  \displaystyle{\sum_{\mathcal{B} =0}^3 \sum_{n=0}^{N-1}
\Delta_{\mathcal{A} \mathcal{B}} (n)
 \sum_{\mathcal{C} =0}^3
\sum_{p,q}^{0, N-1} M_{\mathcal{B} \mathcal{C} }(n,n,p,q)
\mathcal{S}_\mathcal{C}^\mathrm{in}(p,q) }\\
& = &  \displaystyle{\sum_{\mathcal{B},\mathcal{C}
}^{0,3}\sum_{p,q,n}^{0, N-1}  \Delta_{\mathcal{A} \mathcal{B}} (n)
 M_{\mathcal{B} \mathcal{C} }(n,n,p,q)
\mathcal{S}_\mathcal{C}^\mathrm{in}(p,q) }.
\end{array}%
%
\end{equation}
In a realistic experimental configuration, the input beam of light
has a well defined polarization irrespective of the spatial and
temporal coherency properties of the beam itself. This means that
it is possible to write
\begin{equation}\label{q980}
\rho_{m \mu, n \nu} = \widetilde{\rho}_{mn, \mu \nu} \cong R_{mn}
r _{\mu \nu},
%
\end{equation}
namely
\begin{equation}\label{q990}
\rho \cong R \otimes r,
%
\end{equation}
where $R$ and $r$ are a $N \times N$ and a $2 \times 2$ matrices,
respectively, and $\mathrm{Tr} \{ R \} = 1 = \mathrm{Tr} \{ r \}
$. Note that this factorization has been made upon the matrix
$\rho$ representing the operator $\hat{\rho}$ and not on the
operator itself, where it would have been meaningless. With this
assumption we can write
\begin{equation}\label{q1000}
\begin{array}{lcl}
\displaystyle{\mathcal{S}_\mathcal{C}^\mathrm{in}(p,q) }
& = &  \displaystyle{ \sum_{\alpha, \beta}^{0,1} \rho_{p \alpha, q
\beta}^{\, \mathrm{in}}
[\sigma_{(\mathcal{C})}]_{\beta \alpha} }\\
& \cong &  \displaystyle{ R_{pq}^{\, \mathrm{in}} \sum_{\alpha,
\beta}^{0,1}  r_{\alpha \beta}^{\, \mathrm{in}}
[\sigma_{(\mathcal{C})}]_{\beta \alpha} }\\
& \equiv &  \displaystyle{ R_{pq}^{\, \mathrm{in}}
S_{\mathcal{C}}^{\, \mathrm{in}}},

\end{array}%
%
\end{equation}
where $S_{\mathcal{C}}^{\, \mathrm{in}} \equiv \mathrm{Tr} \{
  r^{\, \mathrm{in}} \sigma_{(\mathcal{C})} \} $, and we use this result
in Eq. (\ref{q970}) to obtain
\begin{equation}\label{q1100}
\begin{array}{lcl}
\displaystyle{\bigl\la \hat{S}_{(\mathcal{A})}
\bigr\ra^\mathrm{out}}
& = &  \displaystyle{\sum_{\mathcal{B},\mathcal{C}
}^{0,3}\sum_{n,p,q}^{0, N-1}  \Delta_{\mathcal{A} \mathcal{B}} (n)
 M_{\mathcal{B} \mathcal{C} }(n,n,p,q)
\mathcal{S}_\mathcal{C}^\mathrm{in}(p,q) }\\
& = &  \displaystyle{\sum_{\mathcal{B},\mathcal{C}
}^{0,3}\sum_{n,p,q}^{0, N-1}  \Delta_{\mathcal{A} \mathcal{B}} (n)
 M_{\mathcal{B} \mathcal{C} }(n,n,p,q)
R_{pq}^{\, \mathrm{in}}
S_{\mathcal{C}}^{\, \mathrm{in}} }\\
& = &  \displaystyle{\sum_{\mathcal{C} = 0}^{3}
\left[\sum_{n,p,q}^{0, N-1} \sum_{\mathcal{B}=0 }^{3}
\Delta_{\mathcal{A} \mathcal{B}} (n)
 M_{\mathcal{B} \mathcal{C} }(n,n,p,q)
R_{pq}^{\, \mathrm{in}} \right]
S_{\mathcal{C}}^{\, \mathrm{in}} }\\
& \equiv &  \displaystyle{\sum_{\mathcal{C} = 0}^{3}
 M_{\mathcal{A} \mathcal{C} } S_{\mathcal{C}}^{\, \mathrm{in}}
},
\end{array}%
%
\end{equation}
where we have defined the effective $4 \times 4$ Mueller matrix
$M$ as
\begin{equation}\label{q1110}
 M_{\mathcal{A} \mathcal{C} } \equiv \sum_{p,q,n}^{0, N-1} \sum_{\mathcal{B}=0 }^{3}
\Delta_{\mathcal{A} \mathcal{B}} (n)
 M_{\mathcal{B} \mathcal{C} }(n,n,p,q)
R_{pq}^{\, \mathrm{in}}.
%
\end{equation}
From Eqs. (\ref{q950},\ref{q980},\ref{q1000}), it follows that for
a paraxial input beam
\begin{equation}\label{q1120}
\begin{array}{lcl}
\displaystyle{\bigl\la \hat{S}_{(\mathcal{A})}}
\bigr\ra^{\mathrm{in}}
& = & \displaystyle{\sum_{\mathcal{B} =0}^3 \sum_{n=0}^{N-1}
\Delta_{\mathcal{A} \mathcal{B}} (n)
\mathcal{S}_\mathcal{B}^{\mathrm{in}}(n,n) }\\
& \cong & \displaystyle{\sum_{\mathcal{B} =0}^3 \sum_{n=0}^{N-1}
\delta_{\mathcal{A} \mathcal{B}}  R^{\, \mathrm{in}}_{nn}
S_\mathcal{B}^{\mathrm{in}} }\\
& = & \displaystyle{S_\mathcal{A}^{\mathrm{in}} \sum_{n=0}^{N-1}
  R^{\, \mathrm{in}}_{nn}
 }\\
& = & \displaystyle{S_\mathcal{A}^{\mathrm{in}} \mathrm{Tr} \{
R^{\, \mathrm{in}} \}
 }\\
& = & \displaystyle{S_\mathcal{A}^{\mathrm{in}} }.
\end{array}%
%
\end{equation}
Finally, by comparing Eq. (\ref{q1100}) with Eq. (\ref{q1120}) we
found the sought relation between $\bigl\la
\hat{S}_{(\mathcal{A})} \bigr\ra^\mathrm{out}$ and $\bigl\la
\hat{S}_{(\mathcal{A})} \bigr\ra^\mathrm{in}$:
\begin{equation}\label{q1130}
\bigl\la \hat{S}_{(\mathcal{A})} \bigr\ra^\mathrm{out} =
\sum_{\mathcal{B} =0}^3  M_{\mathcal{A} \mathcal{B} } \bigl\la
\hat{S}_{(\mathcal{B})} \bigr\ra^\mathrm{in},
%
\end{equation}
where
\begin{equation}\label{q1140}
\begin{array}{lcl}
 \displaystyle{M_{\mathcal{A} \mathcal{B} } }& \equiv & \displaystyle{\sum_{p,q,n}^{0, N-1} \sum_{\mathcal{C}=0 }^{3}
\Delta_{\mathcal{A} \mathcal{C}} (n)
 M_{\mathcal{C} \mathcal{B} }(n,n,p,q)
R_{pq}^{\, \mathrm{in}}}\\
& = & \displaystyle{\sum_{p,q,n}^{0, N-1} \sum_{\mathcal{C}=0
}^{3} \Delta_{\mathcal{A} \mathcal{C}} (n)
\sum_i \mathrm{Tr} \left\{ \sigma_{(\mathcal{C})} A_{i} (n,p)
\sigma_{(\mathcal{B})}
A_{i}^\dagger (q,n)
 \right\}
R_{pq}^{\, \mathrm{in}}}\\
& \cong & \displaystyle{\sum_{p,q,n}^{0, N-1}
\sum_i \mathrm{Tr} \left\{ \sigma_{(\mathcal{A})} A_{i} (n,p)
\sigma_{(\mathcal{B})} A_{i}^\dagger (q,n)
 \right\}
R_{pq}^{\, \mathrm{in}}},
\end{array}%
%
\end{equation}
where the last, approximate equality is valid only in the limit of
\emph{paraxial detection}.
\subsection{Two-photon scattering}
Let us consider now the case of two  photons, say $A$ and $B$,
that are scattered by two independent, spatially separated media.
We denote with $| a \alpha \ra$ and $| b \beta \ra$ the
single-photon basis states for photons $A$ and $B$ respectively,
where $a,b \in \{0, \ldots, N-1 \}$ and $\alpha, \beta \in \{0,1
\}$. A two-photon basis state will be indifferently written as
\begin{equation}\label{q1150}
|a \alpha\ra \otimes | b \beta\ra = |a \alpha\ra  | b \beta\ra =
|a \alpha, b \beta\ra= |A B\ra,
%
\end{equation}
where $A$ and $B$ are cumulative indices for the pairs of indices
$(a \alpha)$ and $( b \beta)$, respectively. Let
$\hat{\rho}^\mathrm{in}$ denotes the density operator describing
the input two-photon state:
\begin{equation}\label{q1160}
\begin{array}{lcl}
\displaystyle{ \hat{\rho}^\mathrm{in} } &  =  & \displaystyle{
\sum_{a,b \atop a', b'}^{0,N-1} \sum_{\alpha, \beta \atop \alpha',
\beta'}^{0,1} \rho_{a \alpha, b \beta ; a' \alpha', b' \beta'} |a
\alpha, b \beta \ra \la a' \alpha', b' \beta'| } \\
&  =  & \displaystyle{ \sum_{A, B} \sum_{A', B'} \rho_{AB, A'B'}
|A B \ra \la A' B'| } ,
\end{array}
%
\end{equation}
where
\begin{equation}\label{q1170}
\rho_{a \alpha, b \beta ; a' \alpha', b' \beta'} = \la a \alpha, b
\beta | \hat{\rho}^\mathrm{in} | a' \alpha', b' \beta'\ra =
\rho_{\alpha \beta , \alpha' \beta'}(ab,a'b') .
%
\end{equation}
A \emph{linear} scattering process due to two independent,
spatially separated media, transforms the input two-photon density
operator $\hat{\rho}^\mathrm{in}$ into the output two-photon
density operator $\hat{\rho}^\mathrm{out}$:
\begin{equation}\label{q1180}
\begin{array}{lcl}
\displaystyle{\hat{\rho}^\mathrm{out}} & = &
\displaystyle{ \mathcal{L}[\hat{\rho}^\mathrm{in}]}\\
& = & \displaystyle{ \sum_{i,j} \left( \hat{A}_i \otimes \hat{B}_j
\right) \hat{\rho}^\mathrm{in} \left( \hat{A}_i^\dagger \otimes
\hat{B}_j^\dagger \right) },
\end{array}%
%
\end{equation}
where
\begin{equation}\label{q1190}
\sum_i \hat{A}_i  \hat{A}_i^\dagger = \hat{I} = \sum_j \hat{B}_j
\hat{B}_j^\dagger.
%
\end{equation}
The relevant quantities to calculate are the transition amplitudes
\begin{equation}\label{q1200}
\begin{array}{c}
\la a \alpha | \hat{A}_i | a' \alpha' \ra = A_{i,\alpha \alpha'}
(a,a') = A_{i,A A'},
\\
\la b \beta | \hat{B}_j | b' \beta' \ra = B_{j,\beta \beta'}
(b,b')= B_{j,B B'},
\end{array}
%
\end{equation}
where with $ A_{i} (a,a') $ and $ B_{j} (b,b') $ we have denoted
the $2 \times 2$ matrices whose elements are $ A_{i,\alpha
\alpha'} (a,a')$ and $ B_{j,\beta \beta'} (b,b')$, respectively.
Then, we can rewrite Eq. (\ref{q1180}) as
\begin{equation}\label{q1210}
\begin{array}{l}
\displaystyle{ \rho_{a \alpha, b \beta ; a' \alpha', b'
\beta'}^\mathrm{out}} \\
%
%
\displaystyle{ \; \; \; =

\sum_{i,j} \sum_{a'',b'' \atop a''', b'''}^{0,N-1} \sum_{\alpha'',
\beta'' \atop \alpha''', \beta'''}^{0,1}
 A_{i,\alpha \alpha''} (a,a'')  B_{j,\beta \beta''} (b,b'')
\rho_{a'' \alpha'', b'' \beta'' ; a''' \alpha''', b'''
\beta'''}^\mathrm{in}
A_{i,\alpha''' \alpha'}^\dagger (a''',a') B_{j,\beta''' \beta'}^\dagger (b''',b')}\\
\displaystyle{ \; \; \; = \sum_{i,j} \sum_{A'',B'' \atop A''',
B'''}
 A_{i,A A''} B_{j,B B''}
\rho_{A'' B'' , A''' B'''}^\mathrm{in}
A_{i,A''' A'}^\dagger  B_{j,B''' B'}^\dagger }.
%
%
%
\end{array}
%
%
\end{equation}
From the algebra of the Pauli matrices it is easy to see that if
we define the $4 \times 4$ matrices $\Sigma_{(\mathcal{A}
\mathcal{B})}$ as
\begin{equation}\label{q1220}
\Sigma_{(\mathcal{A} \mathcal{B})} \equiv \sigma_{(\mathcal{A})}
\otimes \sigma_{(\mathcal{B})},
%
\end{equation}
they form a complete (by definition) and orthonormal set of basis
matrices in $\mathbb{C}^{4 \times 4}$:
\begin{equation}\label{q1230}
\begin{array}{rcl}
\displaystyle{ \mathrm{Tr} \left\{ \Sigma_{(\mathcal{A} \mathcal{B})}
 \Sigma_{(\mathcal{A'} \mathcal{B'})} \right\}} & = &
\displaystyle{ \mathrm{Tr} \left\{ \left( \sigma_{(\mathcal{A})}
\otimes \sigma_{( \mathcal{B})} \right) \left(
\sigma_{(\mathcal{A'})}
\otimes \sigma_{( \mathcal{B'})} \right) \right\}}\\
& = &
\displaystyle{ \mathrm{Tr} \left\{  \sigma_{(\mathcal{A})}
 \sigma_{( \mathcal{A'})} \right\} \mathrm{Tr} \left\{
\sigma_{(\mathcal{B})} \sigma_{( \mathcal{B'})} \right\}}\\
& = &
\displaystyle{ \delta_{\mathcal{A} \mathcal{A'}}
\delta_{\mathcal{B} \mathcal{B'}}}.
\end{array}
%
%
\end{equation}
Then, it is clear that it is always possible to write
\begin{equation}\label{q1240}
\rho_{a \alpha, b \beta ; a' \alpha', b' \beta'} =
\sum_{\mathcal{A} ,\mathcal{B}}^{0,3}
\mathcal{S}_{\mathcal{A}\mathcal{B}}(ab,a'b')
[\Sigma_{(\mathcal{A}\mathcal{B})}]_{\alpha \beta, \alpha'
\beta'},
%
\end{equation}
where we have defined
\begin{equation}\label{q1250}
\begin{array}{lcl}
\displaystyle{ \mathcal{S}_{\mathcal{A}\mathcal{B}}(ab,a'b')}
& = & \displaystyle{ \sum_{\alpha, \beta \atop \alpha',
\beta'}^{0,1} \rho_{a \alpha, b \beta ; a' \alpha', b' \beta'}
[\Sigma_{(\mathcal{A}\mathcal{B})}]_{\alpha' \beta', \alpha
\beta} } \\
& = & \displaystyle{ \sum_{\alpha, \beta \atop \alpha',
\beta'}^{0,1} \rho_{\alpha \beta , \alpha' \beta'}(ab,a'b')
[\Sigma_{(\mathcal{A}\mathcal{B})}]_{\alpha' \beta', \alpha
\beta}} \\
& = & \displaystyle{ \mathrm{Tr} \left\{ \rho (a b , a' b' )
\Sigma_{(\mathcal{A}\mathcal{B})} \right\}}.
\end{array}
%
\end{equation}
If we use Eq. (\ref{q1240}) in both sides of Eq. (\ref{q1210}), we
obtain, after a lenghty but straightforward calculation,
\begin{equation}\label{q1260}
 \mathcal{S}_{\mathcal{A}\mathcal{B}}^\mathrm{out}(ab,a'b') =
 \sum_{\mathcal{A'},
 \mathcal{B'}}^{0,3}\sum_{a'',b'' \atop a''', b'''}^{0,N-1}
 \left[ \mathbb{M}(ab,a'b'; a'' b'', a''' b''') \right]_{\mathcal{A}\mathcal{B},\mathcal{A'}\mathcal{B'}}
  \mathcal{S}_{\mathcal{A'}\mathcal{B'}}^\mathrm{in}(a''b'',a'''b'''),
%
%
\end{equation}
where  the $16 \times 16 $ matrix $ \mathbb{M}(ab,a'b'; a'' b'',
a''' b''')$ is defined as:
\begin{equation}\label{q1270}
 \mathbb{M}(ab,a'b'; a'' b'', a''' b''') \equiv
M^{(A)}(a a', a'' a''') \otimes M^{(B)}(b b', b'' b'''),
%
%
\end{equation}
and where, as in Eq. (\ref{q940}), we have defined the  $4 \times
4 $ matrices $M^{(A)}(a a', a'' a''')$ and  $M^{(B)}(b b', b''
b''')$ as
\begin{equation}\label{q1280}
\begin{array}{lcl}
\displaystyle{ M^{(A)}_{\mathcal{A} \mathcal{A'}}(a a', a'' a''')}
& = & \displaystyle{ \sum_i \mathrm{Tr} \left\{
\sigma_{(\mathcal{A})} A_{i} (a,a'') \sigma_{(\mathcal{A'})}
A_{i}^\dagger (a''',a') \right\}} ,\\
\displaystyle{ M^{(B)}_{\mathcal{B} \mathcal{B'}}(b b', b'' b''')}
& = & \displaystyle{ \sum_j \mathrm{Tr} \left\{
\sigma_{(\mathcal{B})} B_{j} (b,b'') \sigma_{(\mathcal{B'})}
B_{j}^\dagger (b''',b') \right\}}.
\end{array}
%
\end{equation}
Now we want to relate the quantities displayed in Eq.
(\ref{q1260}) with quantities that are actually measured which,
therefore, corresponds to mean values of Hermitian operators. To
this end, we calculate step by step the mean value of the
two-photon Stokes operator $\hat{S}_{(\mathcal{A})} \otimes
\hat{S}_{(\mathcal{B})}$:
\begin{equation}\label{q1290}
\begin{array}{lcl}
\displaystyle{\bigl\la \hat{S}_{(\mathcal{A})} \otimes
\hat{S}_{(\mathcal{B})} \bigr\ra} & = & \displaystyle{ \mathrm{Tr}
\bigl\{ \hat{\rho} \left( \hat{S}_{(\mathcal{A})} \otimes
\hat{S}_{(\mathcal{B})} \right) \bigr\}}\\
& = & \displaystyle{ \mathrm{Tr} \bigl\{ \sum_{A, B} \sum_{A', B'}
\rho_{AB, A'B'} |A B \ra \la A' B'| \left(\hat{S}_{(\mathcal{A})}
\otimes
\hat{S}_{(\mathcal{B})} \right) \bigr\}}\\
& = & \displaystyle{ \sum_{A, B} \sum_{A', B'} \rho_{AB, A'B'}
\mathrm{Tr} \bigl\{  |A B \ra \la A' B'|
\left(\hat{S}_{(\mathcal{A})} \otimes
\hat{S}_{(\mathcal{B})} \right) \bigr\}}\\
& = & \displaystyle{ \sum_{A, B} \sum_{A', B'} \rho_{AB, A'B'}
 \la A' B'| \hat{S}_{(\mathcal{A})} \otimes
\hat{S}_{(\mathcal{B})} |A B \ra }\\
& = & \displaystyle{ \sum_{A, B} \sum_{A', B'} \rho_{AB, A'B'}
 \la A'| \hat{S}_{(\mathcal{A})} |A \ra  \la B'|
\hat{S}_{(\mathcal{B})} |B \ra }\\
& = & \displaystyle{ \sum_{a,b \atop a', b'}^{0,N-1} \sum_{\alpha,
\beta \atop \alpha', \beta'}^{0,1} \rho_{a \alpha, b \beta ; a'
\alpha', b' \beta'}
 \la a' \alpha'| \hat{S}_{(\mathcal{A})} |a \alpha \ra  \la b' \beta'|
\hat{S}_{(\mathcal{B})} |b \beta \ra }\\
& = & \displaystyle{ \sum_{a,b \atop a', b'}^{0,N-1} \sum_{\alpha,
\beta \atop \alpha', \beta'}^{0,1} \rho_{a \alpha, b \beta ; a'
\alpha', b' \beta'}
 \delta_{a' a} \left( \varepsilon_{a' \alpha'}, \sigma_{(\mathcal{A})}\varepsilon_{a \alpha} \right)
  \delta_{b' b} \left( \varepsilon_{b' \beta'}, \sigma_{(\mathcal{B})}\varepsilon_{b \beta} \right)  }\\
& = & \displaystyle{ \sum_{a,b}^{0,N-1} \sum_{\alpha, \beta \atop
\alpha', \beta'}^{0,1} \rho_{a \alpha, b \beta ; a \alpha', b
\beta'}
 \left( \varepsilon_{a \alpha'}, \sigma_{(\mathcal{A})}\varepsilon_{a \alpha} \right)
 \left( \varepsilon_{b \beta'}, \sigma_{(\mathcal{B})}\varepsilon_{b \beta} \right)  }\\
& \equiv & \displaystyle{ \sum_{a,b}^{0,N-1} \sum_{\alpha, \beta
\atop \alpha', \beta'}^{0,1} \rho_{a \alpha, b \beta ; a \alpha',
b \beta'} \left[ \sigma_{a(\mathcal{A})} \right]_{\alpha' \alpha}
\left[ \sigma_{b(\mathcal{B})} \right]_{\beta' \beta} }\\
& \equiv & \displaystyle{ \sum_{a,b}^{0,N-1} \sum_{\alpha, \beta
\atop \alpha', \beta'}^{0,1} \rho_{\alpha \beta , \alpha' \beta'}
(ab,ab) \left[ \sigma_{a(\mathcal{A})} \otimes \sigma_{b(\mathcal{B})} \right]_{\alpha' \beta', \alpha \beta} }\\
& \equiv & \displaystyle{ \sum_{a,b}^{0,N-1} \mathrm{Tr} \bigl\{
\rho (ab,ab) \left( \sigma_{a(\mathcal{A})} \otimes
\sigma_{b(\mathcal{B})} \right) \bigr\}}.
\end{array}%
%
\end{equation}
From Eq. (\ref{q960}) it is easy to see that
\begin{equation}\label{q1300}
\begin{array}{rcl}
\displaystyle{\sigma_{f (\mathcal{F})}} & = & \displaystyle{\sum_{\mathcal{F}' = 0}^3
\Delta_{\mathcal{F} \mathcal{F'}}(f) \sigma_{(\mathcal{F}')}, }\\
\displaystyle{\Delta_{\mathcal{F} \mathcal{F'}}(f)} & = &
\displaystyle{\mathrm{Tr} \left\{\
 \sigma_{f (\mathcal{F})} \sigma_{(\mathcal{F}')} \right\} ,}\\
\end{array}%
%
\end{equation}
where $f \in \{a,b \}$, and $\mathcal{F} \in \{ \mathcal{A},
\mathcal{B}\}$. By using this result and Eq. (\ref{q1250}) we can
rewrite Eq. (\ref{q1290}) as
\begin{equation}\label{q1310}
\begin{array}{lcl}
\displaystyle{\bigl\la \hat{S}_{(\mathcal{A})} \otimes
\hat{S}_{(\mathcal{B})} \bigr\ra}
& = &
 \displaystyle{ \sum_{a,b}^{0,N-1} \mathrm{Tr} \bigl\{
\rho (ab,ab) \left( \sigma_{a(\mathcal{A})} \otimes
\sigma_{b(\mathcal{B})} \right) \bigr\}}\\
& = &
 \displaystyle{ \sum_{a,b}^{0,N-1} \sum_{\mathcal{A}',
 \mathcal{B}'}^{0,3}
 \Delta_{\mathcal{A} \mathcal{A'}}(a) \Delta_{\mathcal{B}
 \mathcal{B'}}(b)
 \mathrm{Tr} \bigl\{
\rho (ab,ab) \left( \sigma_{(\mathcal{A}')} \otimes
\sigma_{(\mathcal{B}')} \right) \bigr\}}\\
& = &
 \displaystyle{ \sum_{a,b}^{0,N-1} \sum_{\mathcal{A}',
 \mathcal{B}'}^{0,3}
 \Delta_{\mathcal{A} \mathcal{A'}}(a) \Delta_{\mathcal{B}
 \mathcal{B'}}(b)
 \mathrm{Tr} \bigl\{
\rho (ab,ab)  \Sigma_{(\mathcal{A}' \mathcal{B}')}  \bigr\} }\\
& = &
 \displaystyle{ \sum_{a,b}^{0,N-1} \sum_{\mathcal{A}',
 \mathcal{B}'}^{0,3}
 \Delta_{\mathcal{A} \mathcal{A'}}(a) \Delta_{\mathcal{B}
 \mathcal{B'}}(b)
 \mathcal{S}_{\mathcal{A'}\mathcal{B'}}(ab,ab) }\\
& = &
 \displaystyle{ \sum_{a,b}^{0,N-1} \sum_{\mathcal{A}',
 \mathcal{B}'}^{0,3} \left[
 \Delta (a) \otimes \Delta (b) \right]_{\mathcal{A}\mathcal{B},\mathcal{A'}
 \mathcal{B}'}
 \mathcal{S}_{\mathcal{A}' \mathcal{B}'}(ab,ab) }\\
& \cong &
 \displaystyle{ \sum_{a,b}^{0,N-1}
 \mathcal{S}_{\mathcal{A} \mathcal{B}}(ab,ab) },
\end{array}%
%
\end{equation}
where the last, approximate equality holds in the paraxial limit
where $ \Delta (a) \cong I_4 \cong \Delta (b)$. By using Eqs.
(\ref{q1260}) and (\ref{q1310}) it is easy to see that
\begin{equation}\label{q1320}
\begin{array}{lcl}
\displaystyle{\bigl\la \hat{S}_{(\mathcal{A})} \otimes
\hat{S}_{(\mathcal{B})} \bigr\ra^\mathrm{out}}
& = &
 \displaystyle{ \sum_{a,b}^{0,N-1} \sum_{\mathcal{A}',
 \mathcal{B}'}^{0,3} \left[
 \Delta (a) \otimes \Delta (b) \right]_{\mathcal{A}\mathcal{B},\mathcal{A'}
 \mathcal{B}'}
 \mathcal{S}_{\mathcal{A}' \mathcal{B}'}^\mathrm{out}(ab,ab) }\\
& = &
 \displaystyle{
 \sum_{a,b \atop { a',b' \atop a'', b''}}^{0,N-1}
 \sum_{\mathcal{A}',
 \mathcal{B}' \atop \mathcal{A}'',
 \mathcal{B}''}^{0,3} \Bigr\{ \Bigl.
  \left[
 \Delta (a) \otimes \Delta (b) \right]_{\mathcal{A} \mathcal{B}, \mathcal{A'}
 \mathcal{B}'} }\\
 & & \displaystyle{ \Bigl. \times
 \left[ \mathbb{M}(ab,ab; a' b', a'' b'')
\right]_{\mathcal{A'} \mathcal{B'},\mathcal{A''}\mathcal{B''} }
  \mathcal{S}_{\mathcal{A''}\mathcal{B''}}^\mathrm{in}(a'b',a''b'')
  \Bigr\} }\\
& = &
 \displaystyle{
 \sum_{a,b \atop { a',b' \atop a'', b''}}^{0,N-1}
 \sum_{\mathcal{A}',
 \mathcal{B}' \atop \mathcal{A}'',
 \mathcal{B}''}^{0,3} \Bigr\{ \Bigl.
  \left[
 \Delta (a) \otimes \Delta (b) \right]_{\mathcal{A} \mathcal{B}, \mathcal{A'}
 \mathcal{B}'} }\\
 & & \displaystyle{ \Bigl. \times
 \left[ M^{(A)}(a a, a' a'') \otimes M^{(B)}(b b, b' b'')
\right]_{\mathcal{A'} \mathcal{B'},\mathcal{A''}\mathcal{B''} }
  \mathcal{S}_{\mathcal{A''}\mathcal{B''}}^\mathrm{in}(a'b',a''b'')
  \Bigr\} }\\
& = &
 \displaystyle{
 \sum_{ { a',b' \atop a'', b''}}^{0,N-1}
 \sum_{\mathcal{A}'',
 \mathcal{B}''}^{0,3} \Bigr\{ \Bigl.
  \left[
  \sum_{a=0}^{N-1}\Delta (a) M^{(A)}(a a, a' a'') \otimes
   \sum_{b=0}^{N-1} \Delta (b) M^{(B)}(b b, b' b'')\right]_{\mathcal{A} \mathcal{B}, \mathcal{A''}
 \mathcal{B}''} }\\
 & & \displaystyle{ \Bigl. \times
  \mathcal{S}_{\mathcal{A''}\mathcal{B''}}^\mathrm{in}(a'b',a''b'')
  \Bigr\} }\\
& \equiv &
 \displaystyle{
 \sum_{ { a',b' \atop a'', b''}}^{0,N-1}
 \sum_{\mathcal{A}'',
 \mathcal{B}''}^{0,3}
M_{\mathcal{A} \mathcal{B}, \mathcal{A''}
 \mathcal{B}''} (a'b', a'' b'')
  \mathcal{S}_{\mathcal{A''}\mathcal{B''}}^\mathrm{in}(a'b',a''b'')},
\end{array}%
%
\end{equation}
where we have defined
\begin{equation}\label{q1330}
M_{\mathcal{A} \mathcal{B}, \mathcal{A''}
 \mathcal{B}''} (a'b', a'' b'') \equiv
  \left[
  \sum_{a=0}^{N-1}\Delta (a) M^{(A)}(a a, a' a'') \otimes
   \sum_{b=0}^{N-1} \Delta (b) M^{(B)}(b b, b' b'')\right]_{\mathcal{A} \mathcal{B}, \mathcal{A''}
 \mathcal{B}''},
%
\end{equation}
or, in compact matrix form:
\begin{equation}\label{q1340}
M(a'b', a'' b'') \equiv
  \sum_{a=0}^{N-1}\Delta (a) M^{(A)}(a a, a' a'') \otimes
   \sum_{b=0}^{N-1} \Delta (b) M^{(B)}(b b, b' b'').
%
\end{equation}
Then, we can rewrite Eq. (\ref{q1320}) as
\begin{equation}\label{q1350}
\bigl\la \hat{S}_{(\mathcal{A})} \otimes \hat{S}_{(\mathcal{B})}
\bigr\ra^\mathrm{out} =
 \sum_{ { a',b' \atop a'', b''}}^{0,N-1}
 \sum_{\mathcal{A}',
 \mathcal{B}'}^{0,3}
M_{\mathcal{A} \mathcal{B}, \mathcal{A'}
 \mathcal{B}'} (a'b', a'' b'')
  \mathcal{S}_{\mathcal{A'}\mathcal{B'}}^\mathrm{in}(a'b',a''b'').
%
\end{equation}
When the input state is \emph{not} hyperentangled, one can write
\begin{equation}\label{q1360}
\rho_{a \alpha, b \beta; a' \alpha', b' \beta'}^\mathrm{in} =
R_{ab,a'b'}^\mathrm{in} r_{\alpha \beta, \alpha'
\beta'}^\mathrm{in},
%
\end{equation}
where $R^\mathrm{in}$ and $r^\mathrm{in}$ are a $N^2 \times N^2$
and a $4 \times 4$ matrices, respectively, and $\mathrm{Tr} \{
R^\mathrm{in} \} = 1 = \mathrm{Tr} \{ r^\mathrm{in} \}$. In this
case, a straightforward calculation shows that
\begin{equation}\label{q1365}
\begin{array}{rcl}
\displaystyle{ \bigl\la \hat{S}_{(\mathcal{A})} \otimes
\hat{S}_{(\mathcal{B})} \bigr\ra^\mathrm{in} }
& = & \displaystyle{
 \sum_{\mathcal{A}',
 \mathcal{B}'}^{0,3}
\Delta_{\mathcal{A} \mathcal{B}, \mathcal{A'}
 \mathcal{B}'}
S_{\mathcal{A'}\mathcal{B'}}^\mathrm{in}},
\end{array}
%
\end{equation}
%
%
and
\begin{equation}\label{q1370}
\begin{array}{rcl}
\displaystyle{
\mathcal{S}_{\mathcal{A'}\mathcal{B'}}^\mathrm{in}(a'b',a''b'') }
& = & \displaystyle{\sum_{\alpha, \beta \atop \alpha',
\beta'}^{0,1} \rho_{a' \alpha, b' \beta ; a'' \alpha', b''
\beta'}^\mathrm{in} [\Sigma_{(\mathcal{A'}\mathcal{B'})}]_{\alpha'
\beta', \alpha
\beta}  }\\
& = & \displaystyle{\sum_{\alpha, \beta \atop \alpha',
\beta'}^{0,1} R_{a'b',a''b''}^\mathrm{in} r_{\alpha \beta, \alpha'
\beta'}^\mathrm{in} [\Sigma_{(\mathcal{A'}\mathcal{B'})}]_{\alpha'
\beta', \alpha
\beta}  }\\
& = & \displaystyle{ R_{a'b',a''b''}^\mathrm{in} \mathrm{Tr} \{
r^\mathrm{in} \Sigma_{(\mathcal{A'}\mathcal{B'})} \}  }\\
& \equiv & \displaystyle{ R_{a'b',a''b''}^\mathrm{in}
S_{\mathcal{A'}\mathcal{B'}}^\mathrm{in}  },
\end{array}
%
\end{equation}
where we have defined
\begin{equation}\label{q1380}
S_{\mathcal{A'}\mathcal{B'}}^\mathrm{in} = \mathrm{Tr} \{
r^\mathrm{in} \Sigma_{(\mathcal{A'} \mathcal{B'})} \},
%
\end{equation}
and
\begin{equation}\label{q1390}
\begin{array}{rcl}
\displaystyle{\Delta_{\mathcal{A} \mathcal{B}, \mathcal{A'}
 \mathcal{B}'}} & \equiv & \displaystyle{   \sum_{ a,b}^{0,N-1} R_{ab,ab}^\mathrm{in}
 \left[
\Delta(a) \otimes \Delta (b) \right]_{\mathcal{A} \mathcal{B},
\mathcal{A'}
 \mathcal{B}'}}\\
& \cong & \displaystyle{    \delta_{\mathcal{A} \mathcal{A'}}
\delta_{\mathcal{B} \mathcal{B'}}},
 \end{array}
%
\end{equation}
where the last, approximate equality holds in the paraxial limit
only. In this limit, we have
\begin{equation}\label{q1400}
\bigl\la \hat{S}_{(\mathcal{A})} \otimes \hat{S}_{(\mathcal{B})}
\bigr\ra^\mathrm{in} =  S_{\mathcal{A}\mathcal{B}}^\mathrm{in},
%
\end{equation}
and, by combining this result with Eq. (\ref{q1370}), we obtain
\begin{equation}\label{q1410}
\displaystyle{
\mathcal{S}_{\mathcal{A'}\mathcal{B'}}^\mathrm{in}(a'b',a''b'')  =
R_{a'b',a''b''}^\mathrm{in} \bigl\la \hat{S}_{(\mathcal{A'})}
\otimes \hat{S}_{(\mathcal{B'})} \bigr\ra^\mathrm{in} }.
%
\end{equation}
Finally, we substitute Eq. (\ref{q1410}) into Eq. (\ref{q1350}) to
obtain
\begin{equation}\label{q1420}
\begin{array}{rcl}
\displaystyle{ \bigl\la \hat{S}_{(\mathcal{A})} \otimes
\hat{S}_{(\mathcal{B})} \bigr\ra^\mathrm{out} }
& = & \displaystyle{
 \sum_{ { a',b' \atop a'', b''}}^{0,N-1}
 \sum_{\mathcal{A}',
 \mathcal{B}'}^{0,3}
M_{\mathcal{A} \mathcal{B}, \mathcal{A'}
 \mathcal{B}'} ( a'b', a'' b'')
R_{a'b',a''b''}^\mathrm{in} \bigl\la \hat{S}_{(\mathcal{A'})}
\otimes \hat{S}_{(\mathcal{B'})} \bigr\ra^\mathrm{in}}\\
& \equiv & \displaystyle{
 \sum_{\mathcal{A}',
 \mathcal{B}'}^{0,3}
\mathbf{M}_{\mathcal{A} \mathcal{B}, \mathcal{A'}
 \mathcal{B}'}
 \bigl\la \hat{S}_{(\mathcal{A'})}
\otimes \hat{S}_{(\mathcal{B'})} \bigr\ra^\mathrm{in}},
\end{array}
%
\end{equation}
where we have defined the field-dependent two-photon $16 \times
16$ Mueller matrix $\mathbf{M}$ as:
\begin{equation}\label{q1430}
\mathbf{M}_{\mathcal{A} \mathcal{B}, \mathcal{A'}
 \mathcal{B}'} \equiv
 \sum_{ { a',b' \atop a'', b''}}^{0,N-1}
M_{\mathcal{A} \mathcal{B}, \mathcal{A'}
 \mathcal{B}'} ( a'b', a'' b'')
R_{a'b',a''b''}^\mathrm{in}
%
\end{equation}
Equation (\ref{q1420}) is our final result: It represent the
linear relation between input and output \emph{measured}
quantities. This equation is the two-photon quantum analogue of
the classical Mueller-Stokes relation. This similarity can be made
manifest if we define the $16$ two-photon Stokes parameters as
\begin{equation}\label{q1440}
\bigl\la \hat{S}_{(\mathcal{A})} \otimes \hat{S}_{(\mathcal{B})}
\bigr\ra \equiv S_{\Phi},
%
\end{equation}
where we introduced the cumulative index $\Phi = (\mathcal{A}
\mathcal{B}) \in \{ 0, \ldots, 15\}$. Then, Eq. (\ref{q1420}) can
be rewritten as
\begin{equation}\label{q1450}
S_{\Phi} ^\mathrm{out} =
 \sum_{\Phi = 0}^{15}
\mathbf{M}_{\Phi \Phi'}
S_{\Phi'} ^\mathrm{in},
%
\end{equation}
which is formally equivalent to the classical one.

\newpage

\begin{figure}[h!]
\includegraphics[angle=0,width=12 truecm]{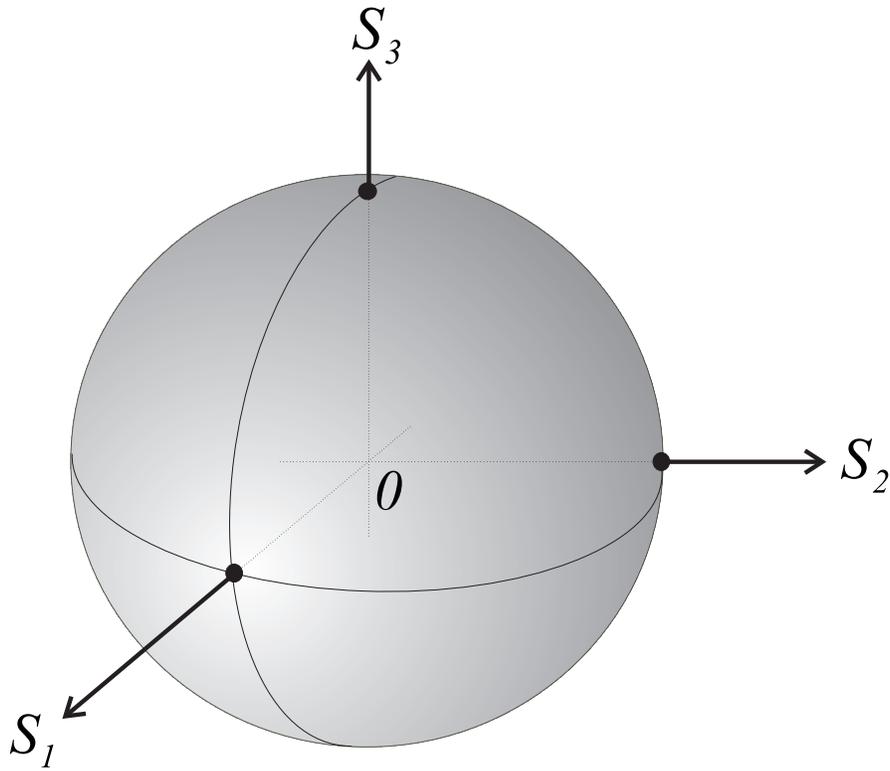}
\caption{\label{fig:1} The Poincar\'e sphere. To each point on the
sphere it is possible to associate a definite \emph{pure}
polarization state of the light. Moreover, internal points are
associate with \emph{mixed} (or partially polarized) states.}
\end{figure}

\end{document}